\documentclass[a4paper,11pt]{article}
\usepackage{jheppub} 
\usepackage[dvipsnames]{xcolor}
\usepackage{verbatim}
\usepackage{tikz}
\usepackage{tikz-cd}
\usepackage{tikz-feynman}
\usepackage{pgfplots}
\usepackage{pgfplotstable}
\usepackage{ytableau}
\ytableausetup{boxsize=0.7em}
\usepackage{amsmath}
\preprint{IPPP/25/42}

\title{\boldmath Charge quantisation, monopoles and emergent symmetry
in the Standard Model and its embeddings}

\author{Rodrigo Alonso, Despoina Dimakou, Yunji Ha, and Valentin V. Khoze}
\affiliation{Institute of Particle Physics Phenomenology, Department of Physics, Durham University, Durham DH1 3LE, U.K.}

\abstract{This work studies the connection of the global properties of the SM gauge group to 1-form discrete symmetries, the possible non-Abelian embeddings of the SM group, and electric and magnetic charge quantisation. Building on previous work, we introduce indexes to characterise the group choices, connect the concept of compositeness degree to emergent electric 1-form symmetry, introduce a new model to fill in the $p=1$ gap, and analyse the magnetic spectrum while connecting its UV and IR realisations.}

\begin{document}
\maketitle
\flushbottom

\section{Introduction}

Quantum mechanics postulates the universe is made up of quanta, i.e. particles, which we can classify given a number of their properties. Some of these properties are themselves quantised and at times conserved, e.g. spin or electric charge. We have come to understand the possible set of discrete values for these properties in terms of fundamental compact groups and their representations. This is one reason why one will find groups at the heart of our formulation of Nature: the Lorentz group and the SM gauge group stand alone with singular importance. On the other hand, one might hear a particle physicist murmur to themselves, \textit{but is electric charge really quantised?} If the Standard Model and hence its group have been put to the test and corroborated so many times, how can these two thoughts be held by the same mind\footnote{Assuming a physicist's mind cannot hold simultaneously contradictory views.}.

The answer is that we do not fully know what the Standard Model group is. We know enough to do perturbative computations but not enough to state what the smallest possible quantum of charge is. The extra knowledge required to determine the group in full is subtle and in practice overlooked, which is only reasonable given how one simply does not need it for `everyday life' computations. Schematically to fully determine the group, one needs to find out what its periodicity is since this determines the possible charge spectrum. Finding the periodicity requires taking a trip in group space that loops back to the beginning: {\it to circumnavigate the gauge group!}

While the different groups lead indeed to different theories and have potential distinct non-perturbative phenomena, see~\cite{Anber:2021upc} for a study of the SM case, in practice the known spectrum of the SM and its phenomenology does not offer an answer. A known possibility to tell the groups apart is the discovery of new particles or monopoles with distinct charges, given that the different possible groups have different electric and magnetic charge quantisation conditions. This is the line of work this paper aligns and builds upon with subject study the SM case, its embeddings and its unbroken group after symmetry breaking. A prominent issue when postulating new fractionally electrically or magnetically charged quanta are experimental constraints. In particular, cosmological fractionally-charged remnants could have been produced in the early universe, yet multiple searches, see~\cite{Langacker:2011db} for a collation of several of bounds, have found no evidence for them. This deserves its own dedicated and systematic study; for the purposes of this work, we simply note the viable possibility of an inflation scale below new fractional particle masses. This solution is the same one that dilutes monopoles away, but we note that the scale of inflation need not be way above the SM scale and could be as low as the tenths of MeV~\cite{Hannestad:2004px,deSalas:2015glj}, so as to produce Big Bang Nucleosynthesis (BBN) as we have measured it.  

\medskip

The paper is organised as follows. Sec.~\ref{sec:global} reviews the different possible Standard Model groups, introduces the charge operator $Q_6$ as well as electric and magnetic 1-form symmetries, discusses the topology of the different groups and possible monopoles in terms of $Q_6$, exemplifies the stairway structure characteristic of each group, discusses compositeness degree and the unbroken SM groups $SU(3)_c\times U(1)_{\textrm{em}}$, in a to identifying the emergent electric 1-form in terms of ($p,k$). Sec.~\ref{sec:2} illustrates how embedding in a group predicts the $p$ value for the unbroken group besides the appearance of compositeness degree. Sec~\ref{sec:ThePSpec} discusses known models from the perspective of minimal embeddings that lead to different $p$ and proposes a renormalisable, anomaly-free new model with a realistic mass spectrum to fill in the $p=1$ gap in \ref{sec:3.1}. Sec~\ref{sec:PSpeck} reviews the same models but with different representation assignment for matter showing how it leads to $k\neq 0$. 
All of the non-Abelian embeddings constructed in this paper necessarily include magnetic degrees of freedom. These
are manifested by dynamical magnetic monopoles present in the UV models, or by `t~Hooft lines representing non-dynamical magnetic probes from the perspective of the SM theory in the IR. 
The relation between the two and a study of the monopole spectrum 
is presented in sec.~\ref{sec:mono} for the same models as in sec.~\ref{sec:ThePSpec}, as well as connecting these with the monopole discussion of sec.~\ref{sec:global}. Sec.~\ref{sec:mono-comp} extends the monopole spectrum analysis to $k\neq0$. Conclusions are to be found in sec.~\ref{sec:Concl}.

\section{Global structure of the Standard Model}
\label{sec:global}

\subsection{The true Standard Model group}

Knowing the group perturbatively amounts to being in possession of the Lie algebra, obtainable from the gauge boson spectrum and self-interactions. Any ambiguities left should then be associated to groups with the same Lie algebra, yet still different. Systematically finding this set of groups is attained by considering which quotients the theory admits, i.e. which spectrum-compatible subgroups of the centre of the group\footnote{The centre of the group is the set of elements that commute with themselves and every other element.} can be removed.

The case of the Standard Model reads 
\begin{align}
\label{eq:Gpdef}
    G_p
    \,=\,\frac{SU(3)_c \times SU(2)_L\times U(1)_Y }{Z_p}\equiv\frac{\widetilde{G}_{SM}}{Z_6}\,\,, \quad p\,=\,1,2,3,6.
\end{align}
In this form, the group in the denominator, $Z_p$,  sets an equivalence relation, i.e. identifies elements of the numerator among themselves that differ by the quotient $g\sim z g=gz$ with $z\in Z_p$. The resulting possible spectrum has to be consistent with this equivalence relation and in particular, $z \Psi=\Psi $; that is, only representations in which the group $Z_p$ collapses into the identity are allowed. The possible quotients $Z_p$ were already found in~\cite{Hucks:1990nw} and revisited in the context of generalised symmetries in~\cite{Tong:2017oea}.
Here, we will specify the action of $Z_p$ in terms of the operator $Q_6$ defined as follows
\begin{align}
\label{eq:Q6def}
    Q_6&=
    2\tilde\lambda_8+3 \tilde T_{3L}+6Q_Y\,, \quad \text{with}\quad \tilde\lambda_8^F=\left(\begin{array}{ccc}
         1&  &\\
         &1 &\\
         &&-2
    \end{array}\right)\,,\quad \tilde T_{3L}^{F}=\left(\begin{array}{cc}
         1&  \\
         & -1
    \end{array}
    \right),
\end{align}
where $Q_Y$ is the $U(1)_Y$ hypercharge, and the $SU(3)_c$ and $SU(2)_L$ generators, $\tilde\lambda_8$ and $\tilde T_{3L}$, are specified by 
the matrices $\tilde\lambda_8^{F}$ and $\tilde T_{3L}^{F}$ when acting on the fundamental representation $F$. Here and throughout, the convention is that {\it a tilded non-Abelian generator has its minimal $|$eigenvalue$|$ defined to be one}, and hence the rest of eigenvalues are integers. Note that this is not the usual normalisation where the trace of pairs of generators in the fundamental is one half, but just as in that case, having specified the generator in the fundamental determines uniquely the generator in any other rep, e.g. for a symmetric colour rep $\mathbf{6}=(\Phi_{rr},\Phi_{bb},\Phi_{gg},\Phi_{rb},\Phi_{rg},\Phi_{bg})$, we have $\tilde\lambda_8=$Diag$(2,2,-4,2,-1,-1)$. The use of this convention is that the group is $2\pi$ periodic in this direction. Given this, the group $Z_p$ is defined by its action on group elements $g=e^{iX}$ as
\begin{align}
\label{eq:Zpdef}
    Z_p:&\quad  g\,\to\, e^{2\pi i\ell Q_6/p}\,g \,=\, g\, e^{2\pi i\ell Q_6/p}\,, &X&\to X+2\pi \ell\frac{ Q_6}{p}\,,\quad \ell=1,\ldots,p,
\end{align}
or equivalently, each group $Z_p$ is generated by its generating element $\xi_p$ defined by
\begin{equation}
\label{eq:Zpdef2}
    \xi_p = e^{2\pi i \frac{Q_6}{p}} \,, \qquad Z_p = \{ (\xi_p)^\ell\}_{\ell=1}^p.
\end{equation}
We will also sometimes find it convenient to use the generating element of $Z_6$ which we call $\xi$
\begin{equation}
\label{eq:xidef}
    \xi \,\equiv\, \xi_6\, =\, e^{2\pi i \frac{Q_6}{6}}.
\end{equation}
Note that in the definition eq.~\eqref{eq:Zpdef} it is assumed that the elements of $Z_p$ are contained in the centre of the group (i.e. the subgroup of elements commuting with every other element ).
To see that this is indeed the case, we note that
\begin{align}
\label{eq:phases}
e^{2\pi i \tilde \lambda_8 /3}&=e^{2\pi i n_c/3}\,, & 
e^{\pi i \tilde T_{3L}}&=e^{\pi i n_L},   
\end{align}
where $n_c$ is the triality and $n_L$ is the duality index of a given representation. We use the convention that $n_c$ is an integer mod 3 and $n_L$ is an integer mod 2, with the fundamental of $SU(3)_c$ corresponding to $n_c=1$
and the fundamental of $SU(2)_L$ to $n_L=1$.
Equations~\eqref{eq:phases} imply that the action of the generating element
$\xi_p$ is an overall phase factor and, hence, the elements of $Z_p$ commute with any group element $g$ in agreement with eq.~\eqref{eq:Zpdef}.

 We should also highlight that all $Z_p$ groups contain the element $(\xi_p)^p =e^{2\pi iQ_6}$ as the identity, in the case of $p=1$ even if the group is trivial this imposes a constraint: $Q_6$ having integer eigenvalues.
Let us call $q_6$ the possible eigenvalues of the operator $Q_6$, 
\begin{equation}
Q_6|q_6\rangle=q_6|q_6\rangle\,, \quad
e^{2\pi i Q_6}=1 \,,\forall\, p \quad \Rightarrow\quad q_6 \in \mathbb{Z}.
\end{equation}
The action of the group $Z_p$ on the eigenstates $|q_6\rangle$ is described by the action of the generating element 
$\xi_p=e^{2\pi i q_6/p}$. For all values of $p=1,2,3,6$ the set of $q_6$ can be divided into 6 equivalence classes labeled by $n_6$ as
\begin{align}
\label{eq:Zpn6}
    n_6&\equiv q_6 \mod 6\,, & Z_p:\,\, e^{2\pi i\ell Q_6/p}|q_6\rangle&=e^{2\pi i \ell n_6/p}|q_6\rangle\,, 
    & \xi_p=e^{2\pi i n_6/p}.
\end{align}

\begin{figure}
    \centering
    \begin{tikzpicture}[scale=0.85]
  \draw[->] (-0.5,0) -- (5.5,0) node[right] {$n_6$};
  \draw[->] (0,-0.5) -- (0,4) node[above] {$G_p$};
  \node[left] at (-0.1,0.2) {$G_6$};
  \node[left] at (-0.1,1) {$G_3$};
  \node[left] at (-0.1,2) {$G_2$};
  \node[left] at (-0.1,3) {$G_1$};
  \foreach \x in {1,...,5}
    \node[below] at (\x,-0.1) {\x};
  \node[below] at (0.2,-0.1) {0};
  \foreach \x in {0,...,5} {
    \foreach \y in {0,...,3} {
      \draw (\x,\y) circle (3pt);
    }
  }
  \fill (0,0) circle (3pt);
  \fill (0,1) circle (3pt);
  \fill (3,1) circle (3pt);
  \fill (0,2) circle (3pt);
  \fill (2,2) circle (3pt);
  \fill (4,2) circle (3pt);
  \foreach \x in {0,...,5}
    \fill (\x,3) circle (3pt);
\end{tikzpicture}
\hspace{0.5 cm}
\begin{tikzpicture}[scale=0.85]
  \draw[->] (-0.5,0) -- (5.5,0) node[right] {$n_6^m$};
  \draw[->] (0,-0.5) -- (0,4) node[above] {$G_p$};
  \node[left] at (-0.1,0.2) {$G_6$};
  \node[left] at (-0.1,1) {$G_3$};
  \node[left] at (-0.1,2) {$G_2$};
  \node[left] at (-0.1,3) {$G_1$};
  \foreach \x in {1,...,5}
    \node[below] at (\x,-0.1) {\x};
  \node[below] at (0.2,-0.1) {0};
  \foreach \x in {0,...,5} {
    \foreach \y in {0,...,3} {
      \draw (\x,\y) circle (3pt);
    }
  }
  \fill (0,3) circle (3pt) ;
  \fill (0,2) circle (3pt);
  \fill (3,2) circle (3pt);
  \fill (0,1) circle (3pt);
  \fill (2,1) circle (3pt);
  \fill (4,1) circle (3pt);
  \foreach \x in {0,...,5}
    \fill (\x,0) circle (3pt);
\end{tikzpicture}
    \caption{Left panel shows the allowed values of $n_6$ characterising all possible electric representations for each $G_p$ choice of the SM group as black filled-in nodes. On the right we show the corresponding magnetic spectrum $n_6^m$ for each $G_p$ theory.}
    \label{fig:n6n6m}
\end{figure}
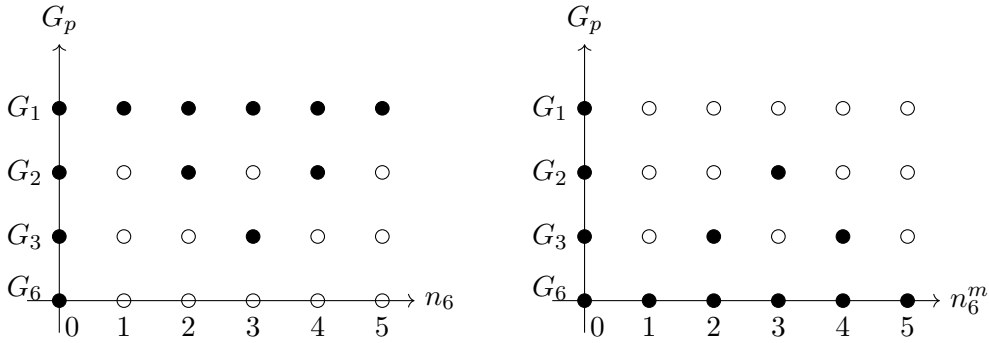

The Standard Model gauge group $G_p$ in eq.~\eqref{eq:Gpdef} is defined by taking the quotient by $Z_p$ which demands that $Z_p$ must act trivially on any representation in $G_p$ which further restricts the allowed values of $n_6$ for each $G_p$ as follows
\begin{align}
\label{eq:1.6}
    e^{2\pi i\frac{Q_6}{p}}\,=\,1 \quad \Rightarrow\quad
    n_6 \,=\, 6Q_Y+2n_c+3n_L \,\,\,\text{mod}\,6
    \,=\, p\,\mathbb{Z} \,\,\,\text{mod}\,6.
\end{align}
The resulting correlation between possible values of $n_6$ and the group $G_p$, from eq.~\eqref{eq:1.6}, is shown on the left panel in Fig.~\ref{fig:n6n6m}. 
Different values of $n_6$ in each given $G_p$ model label different classes of states transforming under the SM group. In particular, all currently known SM matter fields have $n_6=0$, which follows from their representations and hypercharge assignments in Table~\ref{Tab:SMn6} and 
the definition of $n_6$ in eq.~\eqref{eq:1.6}. For simplicity from now on we will refer to them as the \emph{SM} spectrum. Any other particle not in the $SM$ spectrum will be dubbed a $BSM$ state and while they might have $n_6\neq 0$, they need not
. To distinguish them from magnetically charged states or probes in $G_p$, they will be called \emph{electric} states. 
\begin{table}[h]
    \centering
\begin{tabular}{c|c|c|c|c|c|c}
     & $q_L$& $u_R$& $d_R$ & $\ell_L$ & $e_R$ & $H$\\ \hline
    $SU(3)_c$ & $\boldsymbol3$ & $\boldsymbol3$ & $\boldsymbol3$ & $\boldsymbol1$ &$\boldsymbol1$& $\boldsymbol1$\\
    $SU(2)_L$ &$\boldsymbol2$ & $\boldsymbol1$ & $\boldsymbol1$& $\boldsymbol2$ & $\boldsymbol1$&$\boldsymbol2$\\
    $ Q_Y$ & $\frac{1}{6}$ & $\frac{2}{3}$ &$-\frac{1}{3}$ &$-\frac{1}{2}$ &$-1$&$\frac{1}{2}$ 
    \\\hline
    $n_6$ & 0 & 0 & 0 & 0 & 0 & 0
\end{tabular}
\caption{All known SM fields have $n_6=0$ as follows from their hypercharge and non-Abelian representations assignments and the definition of $n_6 \,=\, 6Q_Y+2n_c+3n_L \mod 6$.}
    \label{Tab:SMn6}
\end{table}

States with different values of $n_6$ in Fig.~\ref{fig:n6n6m} can be formally distinguished from one another by the action of a leftover electric discrete symmetry in each given $G_p$ theory. For example for $G_3$ one can have $n_6=0,3$ and we can tell the two apart by a parity defined as $e^{2\pi i n_6/6}$. This is generalised to all $p$ 
by specifying the action of the discrete electric symmetry as
\begin{equation}
\label{Z0el}
 Z_{6/p}^{\textrm{el},(0)}\,=\, \left\{e^{2\pi i \ell_e Q_6/6}\right\}\,\,,\quad \ell_e =1,\dots,\frac{6}{p},
\end{equation}
 where the superscript $(0)$ marks it as a $0$-form or ordinary symmetry. Defining the `electric parity' or hexality of a given state $\Psi$ as the eigenvalue of the generating element of $Z_{6/p}^{\textrm{el},(0)}$
in eq.~\eqref{Z0el}:
\begin{align}
\label{eq:el-parity}
e^{2\pi i  Q_6/6}\, \Psi\,=\,e^{2\pi i  n_6/6}\, \Psi\, \equiv\, \textrm{`hexality'}\cdot\Psi.
\end{align}
We have no electric discrete symmetry left for $p=6$ when we eliminate the centre $Z_6$ in eq.~\eqref{eq:Gpdef}, whereas if we keep it, i.e. take $p=1$, the electric group in eq.~\eqref{Z0el} is itself $Z_6$.

The index $n_6$ is in fact related to electric charge $Q_{\textrm{em}}$. In our convention,
\begin{equation}
\label{eq:Qemdef}
Q_{\textrm{em}}\,=\, Q_Y + \frac{1}{2}\tilde T_{3L},
\end{equation}
which we use to re-write $Q_6$ in eq.~\eqref{eq:Q6def} in the form
$Q_6 \,=\, 2\tilde\lambda_{8}+6 Q_{\textrm{em}}$.
For the electric charge this gives
\begin{equation}
\label{eq:Qem}
Q_{\textrm{em}}\,=\,\frac{n_6}{6}-\frac{n_c}{3} + \mathbb{Z}.
\end{equation}
The presence of a non-zero $n_6$ signals new fractional electric charges, including fractionally charged colour-singlets, while the conservation of $n_6$ (given its form in terms of triality and electromagnetic charge) implies that the lightest of such new particles cannot decay into known SM particles that all have $n_6=0$.

\medskip

The discussion so far has left the gauge nature of the symmetry under discussion aside, yet a question brings it to the fore. The electric discrete symmetry can be used to tell new particles apart
by assigning to them conserved electric hexality in eq.~\eqref{eq:el-parity},
and hence can be used as a physical discriminant.
Yet, these transformations are a subgroup of the SM gauge group, so how can there be any physical consequence to a gauge transformation? Let us answer this in two ways.

The effect of elements of $Z_p$ on representations is a discrete overall phase and not e.g. mixing among colours. In fact, the information encoded in $Z_p$ could be traded for Casimirs; i.e. in $SU(2)_L$, $n_L$ would measure if a state has even or odd dimension. This is to say that we are considering, in fact, a global symmetry and therefore one with observable consequences; this is what has led us to $Z^{\textrm{el},(0)}_{p/6}$ in eq.~(\ref{Z0el}).

However, this answer does not quite align with our quest to find the periodicity of the group, one would need some object that explores the group along a trajectory.
This object cannot be an ordinary local operator as those in the path integral action, we need instead an operator, evaluated along a trajectory in space that maps to a trajectory in the group with (possibly but not necessarily) two endpoints. This leads us to line operators, Wilson and `t~Hooft lines~\cite{Wilson:1974sk,tHooft:1977nqb}, and 1-form symmetries acting upon them~\cite{Aharony:2013hda,Gaiotto:2014kfa}, which we turn to now.

\subsection{1-form symmetries and monopoles}
\label{sec:1.2}

Let us start by rewriting the very same SM group of eq.~\eqref{eq:Gpdef} in a different form especially useful for topology and monopole construction
\begin{align}
  G_p=\frac{SU(3)_c \times SU(2)_L\times \mathbb{R}_Y }{K_p}  \label{GpKp},
\end{align}
where despite the numerator containing a non-compact direction, the quotient is accordingly enlarged so that the ratio is still a compact group.
Let us define the group $K_p$ by its action on the exponent $X=\alpha_i T_i$ of a group element $e^{iX}$
\begin{align}
    K_p=\mathbb{Z}&\quad :X\to X+\left\{\dots\,,\,-4\pi\frac{Q_6}{p}\,,\,-2\pi  \frac{Q_6
    }{p}\,,\,0\,,\,2\pi  \frac{Q_6}{p}\,,\,4\pi\frac{Q_6}{p}\,\dots \right\}.
\end{align}
\begin{figure}[h]
    \centering
\begin{tikzpicture}
\filldraw [gray!5] (-0.3,2) -- (6.3,2) -- (6.3,0)-- (-0.3,0);
    \draw[fill=blue!5] (2,0) -- (4,2) -- (6,2)-- (4,0)--(2,0);
    \draw (0,2.5) node{$p=1$};
    \draw [<->,thick] (-0.3,2)--(3,2);
    \draw [thick,->]  (0,0)--(0,2);
    \draw (-.1,1.8)node[anchor=north east] {$y$};
    \draw [thick] (3,2)--(3.2,2);
    \draw [>->,thick] (3.1,2)--(6.3,2);
    \draw [<->,thick] (-0.3,0)--(3,0);
    \draw [thick] (3,0)--(3.2,0);
    \draw [>->,thick] (3.1,0)--(6.3,0)  node[anchor=north west] {$x$};
    \draw [thick,red,->] (0,0) -- (2,2);
    \draw [thick,red,->,dashed] (0,0) -- (2,0);
    \draw [thick,blue] (2,0) -- (4,2);
    \draw [thick,blue,->] (4,0) -- (6,2);
    \draw[thick,dashed,blue] (4,0) -- (2,0) -- (2,2)-- (4,2)--(4,0);
    \end{tikzpicture}\,\,
\begin{tikzpicture}
  \filldraw [gray!5] (-0.3,2) -- (6.3,2) -- (6.3,0)-- (-0.3,0);
  \draw[fill=blue!5] (2,0) -- (3,1) -- (5,1)-- (4,0)--(2,0);
    \draw (-.1,1.8)node[anchor=north east] {$y$};
        \draw [thick,->]  (0,0)--(0,2);
    \draw (0,2.5) node{$p=2$};
        \draw [<->,thick] (-0.3,2)--(3,2);
    \draw [thick] (3,2)--(3.2,2);
    \draw [>->,thick] (3.1,2)--(6.3,2);
    \draw [<->,thick] (-0.3,0)--(3,0);
    \draw [thick] (3,0)--(3.2,0);
    \draw [>->,thick] (3.1,0)--(6.3,0) node[anchor=north west] {$x$};
    \draw[thick,blue] (2,0) -- (3,1) -- (5,1)-- (4,0)--(2,0);
    \draw[thick,blue,->](2,0) -- (2+1/2,1/2);
    \draw [thick,blue,->] (4,0)--(4+1/2,1/2);
    \draw [thick,blue,>->] (4-.16,1)--(4+.16,1);
    \draw [thick,red,->] (0+0.5,0+0.5) -- (1+0.5,1+0.5);
\end{tikzpicture}
\begin{tikzpicture}
\filldraw [gray!5] (-0.3,2) -- (6.3,2) -- (6.3,0)-- (-0.3,0);
  \draw[fill=blue!5] (2,0) -- (2+2/3,2/3) -- (4+2/3,2/3)-- (4,0)--(2,0);
      \draw [thick,->]  (0,0)--(0,2);
    \draw (0,2.5) node{$p=3$};
       \draw [<->,thick] (-0.3,2)--(3,2);
    \draw [thick] (3,2)--(3.2,2);
    \draw (-.1,1.8)node[anchor=north east] {$y$};
    \draw [>->,thick] (3.1,2)--(6.3,2);
    \draw [<->,thick] (-0.3,0)--(3,0);
    \draw [thick] (3,0)--(3.2,0);
    \draw [>->,thick] (3.1,0)--(6.3,0)node[anchor=north west] {$x$};
    \draw[thick,blue] (2,0) -- (2+2/3,2/3) -- (4+2/3,2/3)-- (4,0)--(2,0);
    \draw [thick,blue,>->] (3+2/3-.16,2/3)--(3+2/3+.16,2/3);
    \draw [thick,red,->] (0+0.43,0+0.43) -- (2/3+0.43,2/3+0.43);
\end{tikzpicture}\,\,
\begin{tikzpicture}
\filldraw [gray!5] (-0.3,2) -- (6.3,2) -- (6.3,0)-- (-0.3,0);
  \draw (-.1,1.8)node[anchor=north east] {$y$};
 \draw[fill=blue!5] (2,0) -- (2+2/6,2/6) -- (4+2/6,2/6)-- (4,0)--(2,0);    
    \draw (0,2.5) node{$p=6$};
    \draw [thick,->]  (0,0)--(0,2);
    \draw [<->,thick] (-0.3,2)--(3,2);
    \draw [thick] (3,2)--(3.2,2);
    \draw [>->,thick] (3.1,2)--(6.3,2);
    \draw [<->,thick] (-0.3,0)--(3,0);
    \draw [thick] (3,0)--(3.2,0);
    \draw [>->,thick] (3.1,0)--(6.3,0)node[anchor=north west] {$x$};
     \draw[thick,blue] (2,0) -- (2+2/6,2/6) -- (4+2/6,2/6)-- (4,0)--(2,0);    
     \draw [thick,blue,>->] (3+2/6-.16,2/6)--(3+2/6+.16,2/6);
    \draw [thick,red,->] (0+0.5,0+0.5)  -- (2/6+0.5,2/6+0.5);
  
\end{tikzpicture}
    \caption{The action of the quotient group, i.e. the equivalence relation we impose on the group, in the plane $(x,y)$ where $X=2\pi x(6Q_Y)+2\pi y (2\tilde\lambda_8+3 \tilde T_{3L})$ and a group element is $g=\exp(iX)$. We have $x\in\mathbb{R}_Y$ while $y$ is periodic (by definition) with period 1, the red arrow represents the action of the generating element of $K_p$ so any two points differing by an integer times this vector are identified. The blue-filled rhomboid is the minimum domain all points in $x,y$ can be mapped to, one can think of it as a torus with its interior (a non-Abelian off-diagonal direction) filled in.}
    \label{fig:Kp}
\end{figure}
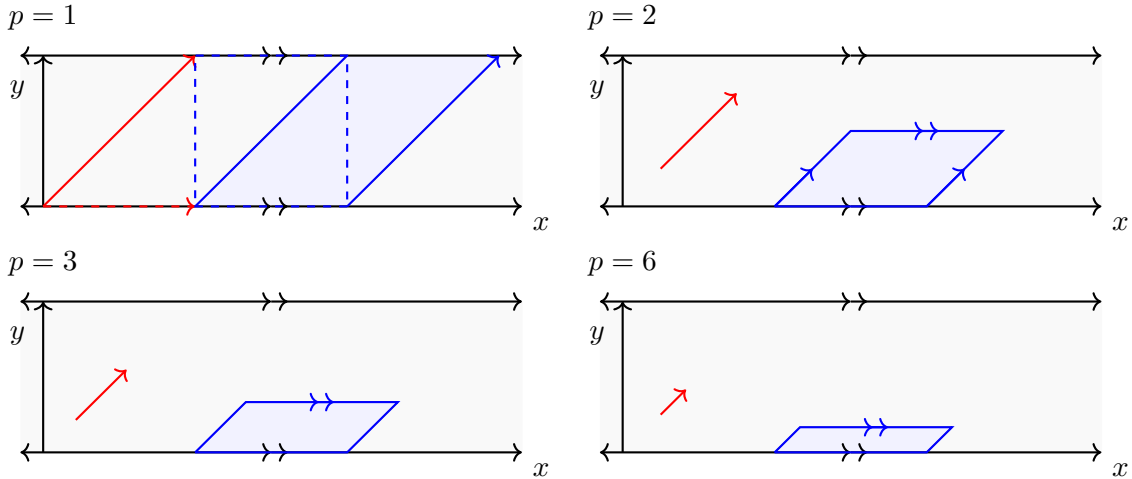

 One has that $K_p$ is isomorphic to $\mathbb{Z}$ regardless of $p$, i.e. it  is always the group of translations by an integer multiple of an elementary vector $2\pi Q_6/p$. The vector itself is of different length for different $p$; in fact the generating element of $K_6$ added to itself $6/p$ times returns the generating element for $K_p$. One would be tempted to say that $K_6$ is larger than the rest, yet they are all the same dimension by virtue of the infiniteness of $\mathbb{Z}$ as can be shown with a Hilbert-hotel-like reasoning\footnote{Hilbert's hotel is a thought experiment that highlights a surprising feature of infinite sets. It can be shown that a fully booked hotel with an infinite number of rooms can still make space for more guests -even an infinite number of them- and that is a process that can be repeated infinitely many times.}. Given $K_p$ is novel we expand on how it acts in Fig.~\ref{fig:Kp}. 
 
The use of eq.~(\ref{GpKp}) is in that the numerator in this form has trivial fundamental group and so one has the following exact sequence
\begin{align}
    \pi_1(G_p)=\pi_1\left(\frac{SU(3)_c \times SU(2)_L\times \mathbb{R}_Y }{K_p}\right)=\pi_0(K_p)=\mathbb{Z}.
\end{align}
Topology is hence blind to $p$; to go beyond it, let us introduce two 1-form discrete symmetries that do depend on $p$
\begin{align}
\label{eq:mag1f}
    Z^{\textrm{mag}(1)}_{p}&=\left\{ U_{\ell_m}(V_3)\right\} \quad \ell_m =1,\dots,p,\\
\label{eq:el1f}    
    Z^{\textrm{elec}(1)}_{6/p}&=\left\{ \tilde U_{\ell_e}(\tilde V_3)\right\} \quad \ell_e =1,\dots,6/p.
\end{align}
The rest of this section discusses how these symmetries arise, are defined, and allow us to formulate the phenomenology of different quotients in a gauge-invariant form. 

\medskip

We start with `t~Hooft lines, which are built analogously to the Wilson lines by replacing the gauge field $A_\mu$ by its magnetic dual $\tilde A_\mu$. 
A `t~Hooft line is a line operator that describes a magnetic monopole charge travelling along its world-line. The monopole here is treated as an external non-dynamical magnetic probe of the $G_p$ theory and in its rest frame it sources the gauge field ($D_\mu=\partial_\mu-iA_\mu$)
\begin{align}
    A^\pm=A^{\pm}_\mu dx^\mu\,=\, \frac{\vec g_m \cdot \vec T}{2}\left(\cos\theta\pm 1\right) d\phi,
\label{eq:WuYangdef}
\end{align}
with $\theta,\phi$ being polar and azimuthal angles in spherical polar coordinates.
The linear combination $\vec g_m \cdot \vec T$ implements an embedding of the familiar $U(1)$ Dirac monopole gauge configuration~\cite{Dirac:1931kp}
into the non-Abelian $G_p$ theory where $\vec T$ are the generators of its Lie algebra (or more precisely the Cartan subalgebra).
Equation~\eqref{eq:WuYangdef} assumes the Wu-Yang picture~\cite{Wu:1975es} where two distinct gauge potentials, $A_\mu^+$ and 
$A_\mu^-$, describing the monopole on the Southern and the Northern hemispheres, are wielded together on the equator by the 
gauge transformation,\footnote{This picture is physically equivalent to the Dirac string picture which uses a single gauge field for the monopole and requires that the apparent singularity along the Dirac string direction is unobservable. In this case, the requirement of the periodicity of the gauge transformation along the equator in eq.~\eqref{eq:Uwuyang} is interpreted as the unobservability of the Aharonov-Bohm phase around the Dirac string.}
\begin{equation} 
U(\phi)=e^{i\phi \vec g_m \cdot \vec T}\,, \qquad \text{so that at }\theta={\pi}/{2}:\,\,
A_\mu^-\,=\, U(\phi)(A_\mu^+ + i\partial_\mu) U(\phi)^\dagger.
\label{eq:Uwuyang}
\end{equation}
For consistency $U(\phi)=U(\phi+2\pi)$, which results in the Dirac quantisation condition for the monopole magnetic charges $\vec g_m$ in eq.~\eqref{eq:WuYangdef}.

This $U(\phi)$ is the foretold trajectory, a mapping from spacetime to the group that probes the true global structure, {\it this is our circumnavigation}. Indeed, imposing a quotient produces new ways of satisfying  $U(\phi)=U(\phi+2\pi)$, for given $p$ one has $\exp(2\pi iQ_6/p)= 1$. In particular, let us consider the monopoles eq.~\eqref{eq:WuYangdef} with magnetic charges $\vec g_m$ chosen such that
\begin{align}
\label{eq:n6mdefExp}
   \exp\left[2\pi i\,  \vec g_m \cdot\vec T\right] \,=\, \exp\left[2\pi i\,\frac{n_6^m Q_6}{6}\right].
\end{align}
In section~\ref{sec:mono} we will show that {\it all} monopoles or, equivalently all `t~Hooft line operators, that are consistent with the $G_p$ Standard Model gauge theory satisfy this equation. 
This implies that without loss of generality the monopoles in a $G_p$ theory for any given $p$ are characterised by an integer-mod(6)-valued index $n_6^m$ whose possible values depend on $p$ as shown on the right panel in Fig.~\ref{fig:n6n6m}. 

To simplify our notation, we can formally write 
\begin{align}
\label{eq:n6mdef}
    \vec g_m \cdot\vec T=\frac{n_6^m Q_6}{6},
\end{align}
while keeping in mind that for a generic monopole this equation is expected to hold only in the exponentiated form in the sense of eq.~\eqref{eq:n6mdefExp}.
We now take said monopole, and for concreteness, assume it is traveling at 3-speed $(0,0,\beta)$ along the $z$ axis such that it crosses the origin at $t=0$. This will be our `t~Hooft line $T_{n_6^m}(C)$, where $C$ denotes the world-line, and the associated magnetic current is ({\it cf.}~eqs.~\eqref{eq:WuYangdef}, \eqref{eq:n6mdef})
\begin{equation}
\label{eq:Jmgdef}
\tilde J_\mu\,=\,\frac{2\pi \, Q_6 n_6^m}{6}\,\gamma(1,0,0,\beta)\, \delta(x)\,\delta(y)\,\delta(z-\gamma\beta t).
\end{equation}

Next, consider the operator enacting a new magnetic symmetry $Z_p^{\textrm{mag}(1)}$ which will tell the different monopoles apart. A $d$-form symmetry action is defined on a $4-d$ volume; thus for the 1-form symmetry the volume is $V_3$, which, for concreteness, we take to be defined as 
\begin{align}
\label{eq:Vol3}
    &V_3(x,y,t): \quad t_-\leq t\leq t_+\,,\quad
    -\infty<x,y<+\infty,\quad z=0,
    \\
 \label{eq:Sigpm}   
    &(\partial V_3)_{\pm} \,\equiv\, \Sigma_{\pm}(x,y):\quad z=0, \quad t=t_{\pm},
    \\  
    &\int d^4x\, \delta_{V_3}\equiv\int d^4x\, \Theta(t-t_-)\Theta(t_+-t)\delta(z)=\int_{t_-}^{t_+}dt \int dx dy=V_3.
\end{align}
Then the $\ell_m$'th element of the 1-form magnetic symmetry eq.~\eqref{eq:mag1f} 
is defined on the boundary of $V_3$ as 
\begin{align}
\label{eq:UlmS}
    U_{\ell_m}((\partial V_3)_\pm)\,=\,\exp\left(i \frac{\ell_m}{p}\int d\Sigma_{\pm} n_\mu^\perp n'_\nu \tilde F^{\nu\mu}\right)\,,
    \quad \ell_m=1,\ldots, p ,
\end{align}
where $n^\perp,n'$ are the two normal vectors to a 2d-surface in 4d and $\tilde F$ is the dual field strength. In particular, for our choice of the boundary surface components $\Sigma_{\pm}(x,y)$ in eq.~\eqref{eq:Sigpm}, the normal vectors are $(n^\perp)^\mu=(0,0,0,1)$, $(n')^\mu=(1,0,0,0)$. Let the operator in eq.~\eqref{eq:UlmS} act on the `t~Hooft line $T_{n_6^m}(C)$ as 
\begin{equation}
    \label{eq:mgsymlong}
Z^{\textrm{mag}(1)}_{p} \,:\, T_{n_6^m}(C) \,\to\, U_{\ell_m}(\Sigma_+)\,T_{n_6^m}(C)\,U_{\ell_m}^\dagger(\Sigma_-),
\end{equation}
where
    \begin{align}
    &U_{\ell_m}(\Sigma_+)\, T_{n_6^m}(C)\,U_{\ell_m}^\dagger(\Sigma_-)\,
    \\ 
\label{eq:longfirst}    
    &=\,\textrm{exp}\left(i \frac{\ell_m}{p}\int d\Sigma_+ n_\nu^\perp n'_\mu \tilde F^{\mu\nu}_m\right)T_{n_6^m}(C)\,\textrm{exp}\left(-i \frac{\ell_m}{p}\int d\Sigma_- n_\nu^\perp n'_\mu \tilde F^{\mu\nu}\right)
    \\
    &=\,\textrm{exp}\left(i \frac{\ell_m}{p}\int dxdy( \tilde F^{t\nu}_m(t_+)-\tilde F^{t\nu}_m(t_-) )n^\perp_\nu) \right)T_{n_6^m}(C)
    \\
\label{eq:longthird} 
   & =\,\textrm{exp}\left(i \frac{\ell_m}{p}\int dx dydt \,\partial_t \tilde F^{t\nu}_m n_\nu^\perp  \right)T_{n_6^m}(C)
    \,=\,\textrm{exp}\left(i \frac{\ell_m}{p}\int dV_3 \partial_\rho \tilde F^{\rho\nu}_m n_\nu^\perp  \right)T_{n_6^m}(C)
    \\
 \label{eq:longfourth}   
    &=\,\exp\left(
    \frac{i\ell_m }{p}\int dV_3 dz\delta(z) \frac{2\pi n_6^mQ_6}{6}\delta(x)\delta(y)\delta(z-\gamma\beta t)\gamma\beta\right)T_{n_6^m}(C) \\
    &=\,\exp\left(\frac{2\pi\ell_m i n_6^m}{6}\frac{Q_6}{p}\int_{t_-}^{t^{+}} dt\delta(t) \right)T_{n_6^m}(C)\,=\, \exp\left(\frac{2\pi\ell_m i n_6^m}{6}\frac{Q_6}{p}\textrm{Link}(V_3,C) \right)T_{n_6^m}(C)
\label{LinkN}
    \\
    &=\,\exp\left(\frac{2\pi\ell_m i n_6^m}{6}\frac{Q_6}{p}\right)T_{n_6^m}(C)
    \,=\, \left[\exp\left(\frac{2\pi\ell_m i n_6^m}{6}\right)\right]^{Q_6/p}T_{n_6^m}(C).
\label{eq:longfinal}    
\end{align}
For clarity of exposition, we have provided an explicit step-by-step derivation, let us now go through the steps in eqs.~\eqref{eq:longfirst}-\eqref{eq:longfinal} with relevant commentary. We denote the field strength evaluated in the monopole solution with the subscript $m$ (e.g. $\tilde F_m^{\mu\nu}$ is the Hodge dual of the field strength evaluated on the monopole configuration eq.~\eqref{eq:WuYangdef}) and, in particular, operators in eq.~\eqref{eq:longfirst} to the left of the `t~Hooft line $T_{n_6^m}(C)$ have the monopole background turned on. On line eq.~\eqref{eq:longthird} we re-write the field strength integral as a 3-volume integral of its time derivative which is then promoted to the full divergence by adding $\partial_x \tilde F^{xz}+\partial_y \tilde F^{yz}$. This step is justified in our case given that we have sent the space boundary to the circle at infinity and assumed that all fields go to zero fast enough there. The divergence of $\tilde F$ is given by the magnetic current $\tilde J$ in eq.~\eqref{eq:Jmgdef} which contains a 3-Dirac delta function marking the path of the monopole in eq.~\eqref{eq:longfourth}. The combination of the $V_3$ volume integral and this Dirac delta function returns the Link number, here equal to 1 since in line eq.~\eqref{LinkN} $t_-\leq 0\leq t_+$. In general, the Link number is 1 or 0 depending on whether the line intersects the 3-volume -- which is diffeomorphism invariant\footnote{Note that Link$(C,V_3)=0,1$ because $V_3$ extends to spatial infinity, the reader might be aware that for finite surfaces this number can be a integer as the line can `coil' and intersect the surface many times yet these would not lead to a well-defined group element.}. 
We have thus derived the action of the magnetic 1-form symmetry on the `t~Hooft line in the form
\begin{equation}
    \label{eq:mgsymexpl}
Z^{\textrm{mag}(1)}_{p} \,:\, T_{n_6^m}(C) \,\to\, 
\exp\left(2\pi i \ell_m  \frac{n_6^m}{6}\frac{Q_6}{p}\textrm{Link}(V_3,C) \right)T_{n_6^m}(C)
\,,
    \quad \ell_m=1,\ldots, p.
\end{equation}

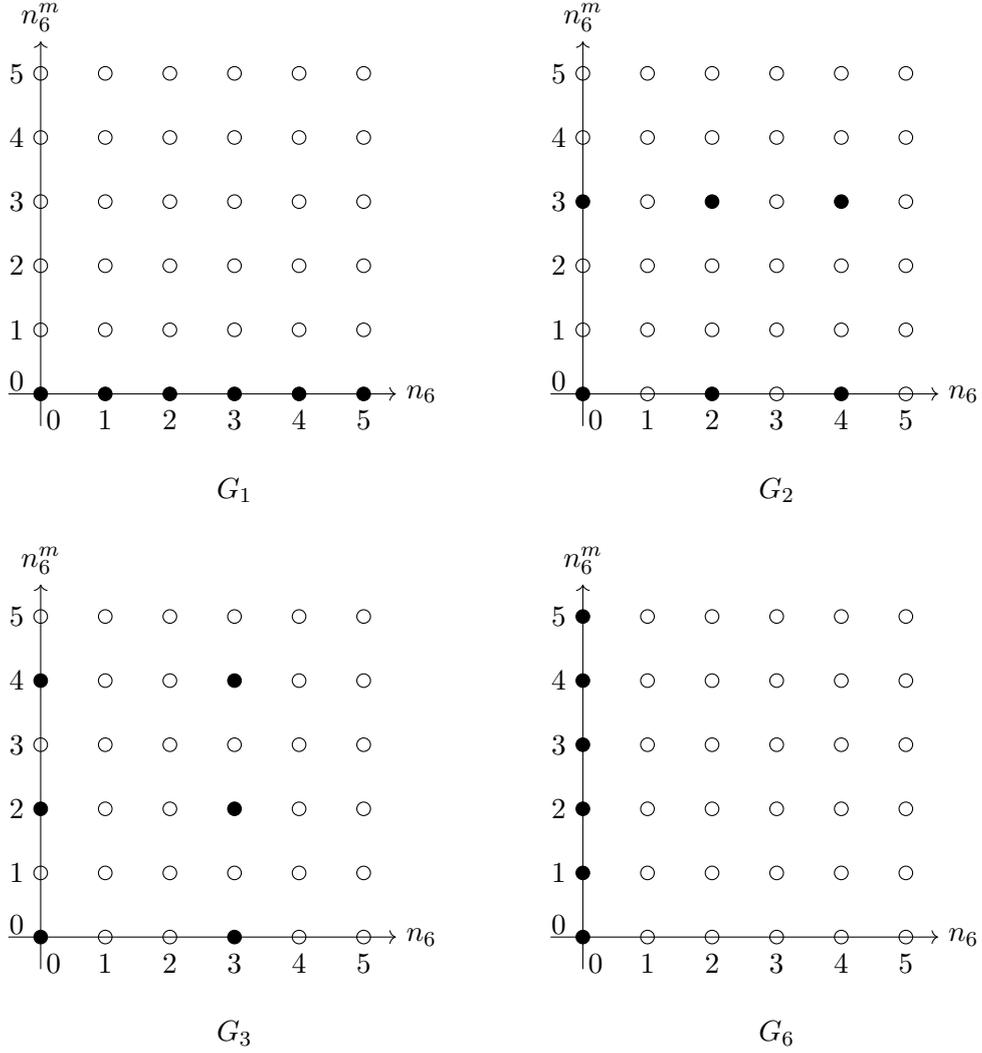
\begin{figure}
    \centering
     \begin{tikzpicture}[scale=0.85]
  \draw[->] (-0.5,0) -- (5.5,0) node[right] {$n_6$};
  \draw[->] (0,-0.5) -- (0,5.5) node[above] {$n^m_6$};
  \node[left] at (-0.1,0.2) {$0$};
  \node[left] at (-0.1,1) {$1$};
  \node[left] at (-0.1,2) {$2$};
  \node[left] at (-0.1,3) {$3$};
  \node[left] at (-0.1,4) {$4$};
  \node[left] at (-0.1,5) {$5$};
  \foreach \x in {1,...,5}
    \node[below] at (\x,-0.1) {\x};
  \node[below] at (0.2,-0.1) {0};
  \foreach \x in {0,...,5} {
    \foreach \y in {0,...,5} {
      \draw (\x,\y) circle (3pt);
    }
  }
  \foreach \x in {0,...,5}
    \fill (\x,0) circle (3pt);

    \node at (3,-1.5) {$G_1$};
\end{tikzpicture}
\hspace{1cm}
\begin{tikzpicture}[scale=0.85]
  \draw[->] (-0.5,0) -- (5.5,0) node[right] {$n_6$};
  \draw[->] (0,-0.5) -- (0,5.5) node[above] {$n^m_6$};
  \node[left] at (-0.1,0.2) {$0$};
  \node[left] at (-0.1,1) {$1$};
  \node[left] at (-0.1,2) {$2$};
  \node[left] at (-0.1,3) {$3$};
  \node[left] at (-0.1,4) {$4$};
  \node[left] at (-0.1,5) {$5$};
  \foreach \x in {1,...,5}
    \node[below] at (\x,-0.1) {\x};
  \node[below] at (0.2,-0.1) {0};
  \foreach \x in {0,...,5} {
    \foreach \y in {0,...,5} {
      \draw (\x,\y) circle (3pt);
    }
  }
  \fill (0,0) circle (3pt);
  \fill (2,0) circle (3pt);
  \fill (4,0) circle (3pt);

\fill (0,3) circle (3pt);
  \fill (2,3) circle (3pt);
  \fill (4,3) circle (3pt);
     

  \node at (3,-1.5) {$G_2$};
\end{tikzpicture}
\\
\vspace{0.3 cm}
\begin{tikzpicture}[scale=0.85]
  \draw[->] (-0.5,0) -- (5.5,0) node[right] {$n_6$};
  \draw[->] (0,-0.5) -- (0,5.5) node[above] {$n^m_6$};
  \node[left] at (-0.1,0.2) {$0$};
  \node[left] at (-0.1,1) {$1$};
  \node[left] at (-0.1,2) {$2$};
  \node[left] at (-0.1,3) {$3$};
  \node[left] at (-0.1,4) {$4$};
  \node[left] at (-0.1,5) {$5$};
  \foreach \x in {1,...,5}
    \node[below] at (\x,-0.1) {\x};
  \node[below] at (0.2,-0.1) {0};
  \foreach \x in {0,...,5} {
    \foreach \y in {0,...,5} {
      \draw (\x,\y) circle (3pt);
    }
  }
 \fill (0,0) circle (3pt);
  \fill (3,0) circle (3pt);
  
 \fill (0,2) circle (3pt);
  \fill (3,2) circle (3pt);
     
    \fill (0,4) circle (3pt);
    \fill (3,4) circle (3pt);
\node at (3,-1.5) {$G_3$};
\end{tikzpicture}
\hspace{1cm}
\begin{tikzpicture}[scale=0.85]
  \draw[->] (-0.5,0) -- (5.5,0) node[right] {$n_6$};
  \draw[->] (0,-0.5) -- (0,5.5) node[above] {$n^m_6$};
  \node[left] at (-0.1,0.2) {$0$};
  \node[left] at (-0.1,1) {$1$};
  \node[left] at (-0.1,2) {$2$};
  \node[left] at (-0.1,3) {$3$};
  \node[left] at (-0.1,4) {$4$};
  \node[left] at (-0.1,5) {$5$};
  \foreach \x in {1,...,5}
    \node[below] at (\x,-0.1) {\x};
  \node[below] at (0.2,-0.1) {0};
  \foreach \x in {0,...,5} {
    \foreach \y in {0,...,5} {
      \draw (\x,\y) circle (3pt);
    }
  }
  \fill (0,0) circle (3pt);
  \fill (0,1) circle (3pt);
  \fill (0,2) circle (3pt);
\fill (0,3) circle (3pt);
     
    \fill (0,4) circle (3pt);
  \fill (0,5) circle (3pt);
  \node at (3,-1.5) {$G_6$};
\end{tikzpicture}
    \caption{The figure shows the spectrum of allowed electric $n_6$ and magnetic $n^m_6$ states for each quotient SM group $G_1, G_2,G_3,G_6$ as black filled-in nodes.}
    \label{fig:emlattice}
\end{figure}

The final expression in eq.~\eqref{eq:longfinal} is meant to illustrate that given the possible magnetic spectrum $n_6^{m}$, for the magnetic group to be well defined we need $Q_6/p=\mathbb{Z}$. This is the condition previously found for each $p$ but now it is derived from the existence of certain monopoles as allowed by the group's global structure. This correlation between electric and magnetic charges is usually derived through the commutation of 't Hooft and Wilson line operators, see~\cite{Tong:2017oea} for the SM case; the novelty in this work is the introduction of indexes $n_6, n_6^m$ which display this correlation clearly as shown in Fig.~\ref{fig:emlattice}.

The magnetic symmetry transformation $U_{\ell_m} \in Z^{\textrm{mag}(1)}_{p}$ in eq.~\eqref{eq:mgsymlong} is, in summary, sensitive to the monopole flux and its defining trajectory in $U(\phi)$ hence marking a true measure of the group globally while it also goes beyond topology which itself is insensitive to the index $n_6^m$, which is a remarkable fact since at times (in non-Abelian gauge theories) the two are considered to be equivalent.

The electric 1-form symmetry $Z^{\textrm{elec}(1)}_{6/p}$ in eq.~\eqref{eq:el1f} is a 
dual of the magnetic symmetry $Z^{\textrm{mag}(1)}_{p}$ above.
It can be defined as the exponential of the electric flux, yet the equations of motion now require a gauge-coupling-dependent prefactor. Let us instead define the action of the electric group on Wilson lines $W$ characterised by its representation, which here we collapse to $n_6$ and path $\tilde C$
\begin{align}
    W_{n_6}(\tilde C)\equiv\, &\textrm{Tr}\left(\mathcal P\exp\left[ i\int_{\tilde C} dx^\mu A_\mu \right]\right)=\textrm{Tr}\left(\mathcal P\exp\left[ i\int_{\tilde C}  A \right] \right),
    \end{align}
    by a shift in the gauge bosons
    \begin{align}
   \tilde  U_{\ell_e}\,:\,\,& A_\mu \,\to\, A_\mu+n^\perp_\mu \delta_{\tilde V_3} \frac{2\pi \ell_e }{6}Q_6,
\end{align}
where we note that the spacetime dependence is entirely fixed by the volume $\tilde V_3$ which one can take to be the same as $V_3$ for definiteness. Given this action, the Wilson loop transforms under $Z^{\textrm{elec}(1)}_{6/p}$ as
\begin{align}
   &\tilde U_{\ell_e}(\tilde V_3) W_{n_6}(\tilde C)\tilde U^\dagger_{\ell_e}(\tilde V_3)\\r=&\mbox{Tr}\,\mathcal P\exp\left[ \int_{\tilde C} A+\frac{2\pi \ell_e i Q_6}{6}\int_{\tilde C} dx^\mu n_\mu^\perp\, \delta_{\tilde V_3}(x(\tilde C))\right]\\
   =&\mbox{Tr}\,\mathcal P\exp\left[ \int_{\tilde C} A+\frac{2\pi \ell_e i Q_6}{6} \textrm{Link}(\tilde V_3,\tilde C)\right]\,=\,
   e^{\frac{2\pi \ell_e i n_6}{6}\textrm{Link}(\tilde V_3,\tilde C)} W_{n_6}(\tilde C),
\end{align}
and one can see that if for example $\tilde V_3=V_3\,,\,\tilde C=C$ the integrand reduces to 
\newline $dz \delta(z)\Theta(t_+-t(z))\Theta(t(z)-t_-)$ and equals one provided that $t_-\leq t(z=0)\leq t_+$. 

\begin{figure}
    \centering
    \begin{tikzpicture}
        \filldraw[color=blue!60, fill=blue!5, thick] (0,0) ellipse (2.5 and 1.5);
        \draw [thick,green] (0,-3)--(0,-1.5);
        \draw [thick,green,dashed] (0,-1.5)--(0,0);
        \draw [thick,green] (0,0)--(0,3);
        \draw [blue,->] (2.25,0) -- (2.75,0);
        \draw [blue,->] (-2.25,0) -- (-2.75,0);
        \draw [blue,->] (1.25,.95) -- (1.75,1.4);
        \draw [blue,->] (-1.25,.95) -- (-1.75,1.4);
        \draw [blue,->] (-1.25,-.95) -- (-1.75,-1.4);
        \draw [blue,->] (1.25,-.95) -- (1.75,-1.4);
        \draw (1,0) node {$\int d\Sigma \,\tilde F$};
        \draw (1.5,3.5) node {Magnetic line, $T_{n_6^m}(C)$};
    \end{tikzpicture} \qquad\qquad
    \begin{tikzpicture}
        \filldraw[color=green!60, fill=green!5, thick] (0,0) ellipse (2.5 and 1.5);
        \draw [thick,blue] (0,-3)--(0,-1.5);
        \draw [thick,blue,dashed] (0,-1.5)--(0,0);
        \draw [thick,blue] (0,0)--(0,3);
        \draw [green,->] (2.25,0) -- (2.75,0);
        \draw [green,->] (-2.25,0) -- (-2.75,0);
        \draw [green,->] (1.25,.95) -- (1.75,1.4);
        \draw [green,->] (-1.25,.95) -- (-1.75,1.4);
        \draw [green,->] (-1.25,-.95) -- (-1.75,-1.4);
        \draw [green,->] (1.25,-.95) -- (1.75,-1.4);
        \draw (1,0) node {$\int d\Sigma \,F$};
        \draw (1.5,3.5) node {Electric line, $W_{n_6}(\tilde C)$};
    \end{tikzpicture}
    \caption{This figure depicts the trajectory $C$, ($\tilde C$) in which the 't Hooft (Wilson) line is defined as a green (blue) line for a time interval $t_-\leq t\leq t_+$ as well as the projection into space of the 3-volume $V_3$, ($\tilde V_3$) in which the operator implementing the 1-form magnetic (electric) transformation is defined as the blue (green) surface for the same time interval. Both configurations displayed have linking numbers equal to one since the lines intersect the surfaces.}
    \label{The1-formsymmetries}
\end{figure}
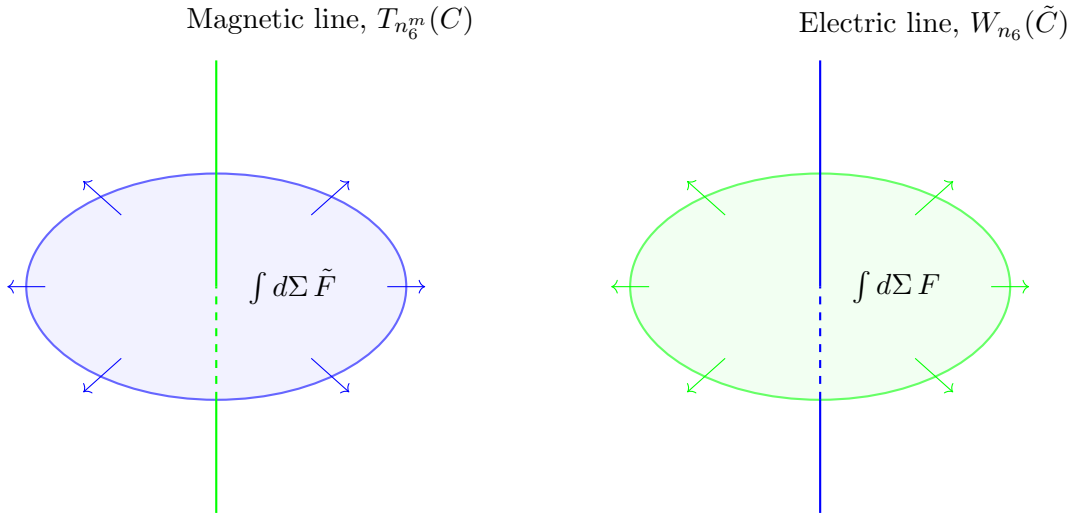

Thus, we have the action of the electric 1-form symmetry on Wilson lines (in analogy with its magnetic counterpart eq.~\eqref{eq:mgsymexpl})
\begin{equation}
    \label{eq:elsymexpl}
Z^{\,\textrm{el}(1)}_{6/p} \,:\, W_{n_6}(\tilde C) \,\to\, 
\exp\left(2\pi i \ell_e  \frac{n_6}{6}\textrm{Link}(\tilde V_3,\tilde C) \right)W_{n_6}(\tilde C)
\,,
    \quad \ell_e=1,\ldots, 6/p.
\end{equation}
This definition of the action of the 1-form electric symmetry best suits our discussion since it is independent of the gauge coupling. For definitions in terms of the electric flux see e.g.~\cite{Brennan:2023mmt}. Fig.~\ref{The1-formsymmetries} illustrates the action of electric and magnetic 1-form symmetries on Wilson and `t~Hooft lines.

\medskip

It will be useful to take a step back and compare 1-form with ordinary 0-form symmetries. One finds that the conservation equation for each 0- and 1-forms relies, respectively, on the relation 
\begin{align}
\label{eq:divless}
     &\int_{\partial V_4} d^3x J^0=\int_{V_4} \partial J \,,
     \qquad\,\,
     \int_{\partial V_3} d \vec \Sigma \cdot \vec E=\int_{ V_3}\partial F\cdot n^\perp(V_3).
\end{align} 
The charge in the equation for the 0-form case is replaced by the flux in the equation for the 1-form case. The condition for either of these transformations to be a \emph{symmetry} is for the integrals to vanish. This depends on whether the corresponding 1- and 2-forms on the RHS of eqs.~\eqref{eq:divless} have vanishing divergence. 
This leads to a very notable difference between the two -- the introduction of dynamical charged particles does not break a 0-form symmetry as long as they have charge-preserving interactions in the Lagrangian and their fields do not develop a vev. Indeed, for the 0-form we have
\begin{align}
    \int_{\partial V_4} d^3x J^0=\int d^3x\left[ J^0(t_+)-J^0(t_-)\right]=Q(t_+)-Q(t_-)=\int_{V_4} \partial_\mu J^\mu=0,
\end{align}
where we note that the presence of charged particles does not change the fact that $\partial_\mu J^\mu=0$.
On the other hand, for the 1-form case we have
\begin{align}
    \int_{\partial V_3} d\vec\Sigma\cdot \vec E =\int d\vec\Sigma\cdot\left[ \vec E(t_+)-\vec E(t_-)\right]=\textrm{Flux}(t_+)-\textrm{Flux}(t_-)=\int_{V_3} \partial_\mu F^{\mu\nu}n^\perp_\nu,
\end{align}
where in the presence of dynamical charged particles the field strength is no longer divergenceless but instead equals $\partial_\mu F^{\mu\nu}=j^\nu_{\textrm{el}}\neq 0$ with $j_{\textrm{el}}$ the electric current as dictated by Maxwell's equations.

\medskip

When applied to our case, given the conclusion above applies to both electric and magnetic 1-form symmetries, continuous as well as discrete, we have 
\begin{align}
    \textrm{dynamical particle with } n_6\neq0& & \partial F&\neq 0\,\Rightarrow\, \textrm{Broken}\,\,   Z^{\textrm{elec}(1)},\\
    \textrm{dynamical particle with } n_6^m\neq 0& & \partial \tilde F&\neq 0\,\Rightarrow\, \textrm{Broken}\,\, Z^{\textrm{mag}(1)}.
\end{align}
Hence, the very particles that would transform and tell us about the symmetry in the first place, unavoidably break it. The settings in which the 1-form symmetries would still be preserved involve Wilson and `t~Hooft lines 
with non-trivial $n_6$ and $n_6^m$ charges, and with no such charges present in the dynamical spectrum. These settings can be relevant for considerations of the phases of the theory and its vacuum structure. It is notable that for $p\neq 6$ (i.e. in the Standard Model with a $G_1$, $G_2$ or $G_3$ gauge group) and at the currently testable energy scales, the 1-form symmetry $Z^{\textrm{elec}(1)}_{6/p}$ is preserved. This is because all known SM particles have $n_6=0$. Similarly, given the absence of monopoles at current energy scales, there is a conserved magnetic 1-form symmetry in the Standard Model for any $p$ (at least up to an appropriately high energy below the monopole mass).
We shall come back to these points in the next subsection, while for further discussion of generalised symmetries we refer the interested reader to Refs.~\cite{Gaiotto:2014kfa,Cordova:2022ruw,McGreevy:2022oyu,Brennan:2023mmt,Bhardwaj:2023kri}.

\subsection{Compositeness degree and emergent 1-form symmetries}

The discussion so far has left the very question first posed unanswered: a simple $U(1)$ theory with say a charge $Q_{\textrm{exp}}$ particle does not contain enough information to discard the possibility that it is not $U(1)$ but $\mathbb{R}$ that is the true group. Let us refine this approach to put it to use by first noting that if one has $\mathbb{R}$, one can indeed famously have irrational charges, but also arbitrarily small ones.

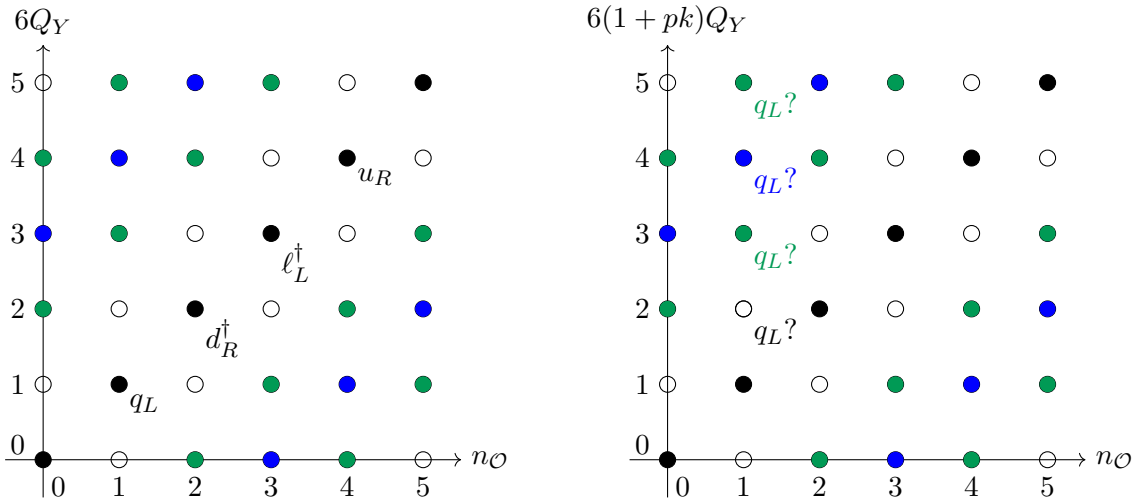
\begin{figure}
    \centering
    \begin{tikzpicture}[scale=1]
  \draw[->] (-0.5,0) -- (5.5,0) node[right] {$n_{\mathcal{O}}$};
  \draw[->] (0,-0.5) -- (0,5.5) node[above] {$6Q_Y$};
  \node[left] at (-0.1,0.2) {$0$};
  \node[left] at (-0.1,1) {$1$};
  \node[left] at (-0.1,2) {$2$};
  \node[left] at (-0.1,3) {$3$};  
  \node[left] at (-0.1,4) {$4$};
  \node[left] at (-0.1,5) {$5$};
  \foreach \x in {1,...,5}
    \node[below] at (\x,-0.1) {\x};
  \node[below] at (0.2,-0.1) {0};
  \foreach \x in {0,...,5} {
    \foreach \y in {0,...,5} {
      \draw (\x,\y) circle (3pt);
    }
  }
  \fill (0,0) circle (3pt);
  \fill (1,1)  node [anchor=north west] {$q_L$} circle (3pt);
  \fill (2,2)  node [anchor=north west] {$d_R^\dagger$}circle (3pt);
  \fill (3,3)  node [anchor=north west] {$\ell_L^\dagger$} circle (3pt);
  \fill (4,4)  node [anchor=north west] {$u_R$} circle (3pt);
  \fill (5,5) circle (3pt);
  \fill [blue](0,3) circle (3pt);
  \fill [blue](4,1) circle (3pt);
  \fill [blue](5,2) circle (3pt);
  \fill [blue](3,0) circle (3pt); 
  \fill [blue](1,4) circle (3pt);
  \fill [blue](2,5) circle (3pt);
  \fill [ForestGreen] (0,2) circle (3pt);
  \fill [ForestGreen] (1,3) circle (3pt);
  \fill [ForestGreen] (0,4) circle (3pt);
  \fill [ForestGreen] (1,5) circle (3pt);
  \fill [ForestGreen] (2,0) circle (3pt);
  \fill [ForestGreen] (3,1) circle (3pt);
  \fill [ForestGreen] (2,4) circle (3pt);
  \fill [ForestGreen] (3,5) circle (3pt);
  \fill [ForestGreen] (4,0) circle (3pt);
  \fill [ForestGreen] (5,1) circle (3pt);
  \fill [ForestGreen] (4,2) circle (3pt);
  \fill [ForestGreen] (5,3) circle (3pt);
\end{tikzpicture}\qquad
\begin{tikzpicture}[scale=1]
  \draw[->] (-0.5,0) -- (5.5,0) node[right] {$n_{\mathcal{O}}$};
  \draw[->] (0,-0.5) -- (0,5.5) node[above] {$6(1+pk) Q_Y$};
    \node[left] at (-0.1,0.2) {$0$};
  \node[left] at (-0.1,1) {$1$};
  \node[left] at (-0.1,2) {$2$};
  \node[left] at (-0.1,3) {$3$};  
  \node[left] at (-0.1,4) {$4$};
  \node[left] at (-0.1,5) {$5$};
  \foreach \x in {1,...,5}
    \node[below] at (\x,-0.1) {\x};
  \node[below] at (0.2,-0.1) {0};
  \foreach \x in {0,...,5} {
    \foreach \y in {0,...,5} {
      \draw (\x,\y) circle (3pt);
    }
  }
  \draw (1,2) node [anchor=north west] {$q_L$?}  circle (3pt);
  \fill (0,0) circle (3pt);
  \fill (1,1)  circle (3pt);
  \fill (2,2) circle (3pt);
  \fill (3,3) circle (3pt);
  \fill (4,4) circle (3pt);
  \fill (5,5) circle (3pt);
  \fill [blue](0,3) circle (3pt);
  \fill [blue](4,1) circle (3pt);
  \fill [blue](5,2) circle (3pt);
  \fill [blue](3,0) circle (3pt); 
  \fill [blue](1,4) node [anchor=north west] {$q_L$?} circle (3pt);
  \fill [blue](2,5) circle (3pt);
  \fill [ForestGreen] (0,2) circle (3pt);
  \fill [ForestGreen] (1,3) node [anchor=north west] {$q_L$?}  circle (3pt);
  \fill [ForestGreen] (0,4) circle (3pt);
  \fill [ForestGreen] (1,5) node [anchor=north west] {$q_L$?} circle (3pt);
  \fill [ForestGreen] (2,0) circle (3pt);
  \fill [ForestGreen] (3,1) circle (3pt);
  \fill [ForestGreen] (2,4) circle (3pt);
  \fill [ForestGreen] (3,5) circle (3pt);
  \fill [ForestGreen] (4,0) circle (3pt);
  \fill [ForestGreen] (5,1) circle (3pt);
  \fill [ForestGreen] (4,2) circle (3pt);
  \fill [ForestGreen] (5,3) circle (3pt);
\end{tikzpicture}
    \caption{Spectrum stairway structure of each group $G_p$ with compositeness degree $k=0$ on the LHS and $k\neq 0$ on the RHS. Black-filled points are common to all $G_p$ while in addition $p=3$ ($p=2$) allows for blue (green) and all points displayed are possible in $p=1$. On the RHS we display the same lattice structure but now place $q_L$ in different lattice sites which would correspond in ascending order to $(p,k)$ equal to $(1,1)$, $(2,1)$, $(3,1)$, and $(2,2)$.}
    \label{fig:novsQY}
\end{figure}

To visualize the spectrum of $U(1)_Y$ charges in the SM it is useful to introduce $n_{\mathcal O}$ as 
\begin{align}
    n_{\mathcal O}\equiv -2n_c-3n_L \mod 6,
\end{align}
and write the selection rule eq.~\eqref{eq:1.6} for the allowed values of $n_6$ in each of the $G_p$ models as
\begin{align}
    n_6\,=\, 6Q_Y - n_{\mathcal O} \,=\, p\mathbb{Z}.
\end{align}
The spectrum of different $G_p$ groups can then be plotted in the ($n_\mathcal{O}$, $Q_Y$) plane as shown in on the left panel  in Fig.~\ref{fig:novsQY}.

 The stairway structure, and in particular its periodicity, is what determines the quotient $Z_p$, yet placing the SM quark doublet $q_L$ in the first step of the stairway is a choice made unconsciously. It is rather possible to place $q_L$ not in the first, but in the $k$'th flight of stairs. This amounts to a rescaling of $Q_Y$ in $Q_6$ by a $p$-dependent amount, e.g. if $p=3$ and $q_L$ is placed on the next flight of stairs up, its charge is 4 times the smallest $n_\mathcal{O}=1$ charge. The generalisation and constraint on the spectum read
\begin{align}
\label{eq:Q6kdef}
    Q_6(k)\,=&\,2\tilde\lambda_{8}+3\tilde T_{3L}+6(1+k \,p)Q_Y\,, & k&\in \mathbb{Z},\\
\label{eq:n6kdef} 
    n_6(k)\,= &\,2n_c+3n_{L}+6\,(1+k \,p)Q_Y=\,p\mathbb{Z}.
\end{align}
The integer $k$, first introduced in \cite{Alonso:2024pmq}, is referred to as the {\it compositeness} degree.
Our investigation up to now was focused on the $k=0$ case, but from now on we will also explore implications of non-trivial compositeness degree $k\neq 0$.

\medskip

A study of models with non-vanishing $k$ might appear to be purely academic particularly in the light of well-known minimal examples of UV completion of the SM provided by $SU(5)$ and other GUT models, or Pati--Salam models, all yielding $k=0$. 
In our view such a general conclusion would be premature since the approach based on non-trivial degree of compositeness offers valuable insights:  
$(i)$ it embodies the $U(1)$ vs $\mathbb{R}$ problem, $(ii)$ in a comprehensive bottom-up approach compositeness cannot be discarded and, $(iii)$ compositeness appears naturally at low energy. To illustrate the last point consider the theory after electroweak symmetry breaking (EWSB),
\begin{align}
   G_p\,=\, \frac{SU(3)_c \times SU(2)_L\times U(1)_Y }{Z_p}\,\rightarrow \,\frac{SU(3)_c\times U(1)_{\textrm{em}}}{Z_{p'}}\,\equiv\,\frac{ G_{Q(C+E)D}}{Z_{p'}}\,,
\end{align}
where $p'$ is either $1$ or $3$. That implies that four possible groups $G_p$ reduce to two, yet we know there are physical differences between all four so there must be another handle to distinguish between them. One starts with the stand-alone analysis of $G_{p'}$, which would introduce $Q_3$ and the constraint on the spectrum, {\it cf.}~eqs~\eqref{eq:Q6kdef}-\eqref{eq:n6kdef}
\begin{align}
    Q_3&\equiv 3(1+p'k')Q_{\textrm{em}}+\tilde\lambda_{8},&  3(1+p'k')Q_{\textrm{em}}+\tilde\lambda_{8}&=p'\mathbb{Z}.
\end{align}
Note that $Q_6$ and its quantisation condition read
\begin{align}
6(1+pk)\,Q_Y+2\tilde\lambda_{8}+3\tilde T_{3L}\,=\, &\,p\mathbb{Z},
\\
6(1+p k)\,Q_{em}+2\tilde \lambda_8-3p\cdot k\tilde T_{3L}\,=\,&\,p\mathbb{Z},
\\
\quad    6(1+p k)\,Q_{em}+2 n_c\,=\,&\,p\mathbb{Z}.  \label{eq:UVmatch}
\end{align}
Now one can match the UV and IR descriptions. 
Let us first comment that for $p=1,2$ colour triality drops out, i.e. it can be swallowed up in the integers or even integers on the RHS of eq.~\eqref{eq:UVmatch}. 
\begin{align}
\label{eq:p2result}
    p=2& & 
    &\left.\begin{array}{c}
    6(1+2k)\,Q_{\textrm{em}}=2\mathbb{Z}\\
     3(1+p'k')\,Q_{\textrm{em}}=p'\mathbb{Z}
     \end{array}\right\} \, \Rightarrow\, p'=1\,,\, k'=2k,
     \\
    p=1& & 
    &\left.\begin{array}{c}
    6(1+k)\,Q_{\textrm{em}}=\mathbb{Z}\\
    3(1+p'k')\,Q_{\textrm{em}}=p'\mathbb{Z}
    \end{array}\right\}  \,\Rightarrow\, p'=1\,,\, k'=2k+1.
\end{align}
For the other two values of $p$ we have, were we use eq.~\eqref{eq:phases} to substitute $\tilde \lambda_8$ for $n_c$ in $Q_3$,
\begin{align}
    p=6& & 
    &\left.\begin{array}{c}
    6(1+6k)\,Q_{\textrm{em}}+2n_c=6\mathbb{Z}\\
     3(1+p'k')\,Q_{\textrm{em}}+n_c=p'\mathbb{Z}
     \end{array}\right\}\, \Rightarrow\, p'=3\,,\, k'=2k,
\label{eq:p6result}       
     \\
    p=3& & 
    &\left.\begin{array}{c}
    3(1+3k)\,Q_{\textrm{em}}+2n_c=3\mathbb{Z}\\
    3(1+p'k')\,Q_{\textrm{em}}+n_c=p'\mathbb{Z}
    \end{array}\right\} \, \Rightarrow\, p'=3\,,\, k'=-2k-1.
\label{eq:p3result}        
\end{align}
We conclude that the group correspondence is
\begin{align}
\label{eq:UVIR2pairs}
     \frac{\widetilde{G}_{SM}}{Z_6,Z_3} \, \longrightarrow\,  \frac{G_{Q(C+E)D}}{Z_3} 
    \,\,, \qquad
     \frac{\widetilde{G}_{SM}}{Z_2,Z_1}\, \longrightarrow\,  \frac{G_{Q(C+E)D}}{Z_1} \,\,,
\end{align}
and, less trivially, we can now also see how to tell the two IR theories with the same $p'$ apart. In the first equation of eq.~\eqref{eq:UVIR2pairs} the IR theory with $p'=3$ can arise either from a $p=6$ or from a $p=3$ UV model. Assume that both of these UV models have trivial degree of compositeness, $k=0$. Then the correspondence in eqs.~\eqref{eq:p6result}-\eqref{eq:p3result}  tells us that the resulting $p'=3$ theories in the IR are distinguished by their emerging IR compositeness degree, in the first case it is trivial, $k'=0$, and for the second it is $k'=-1$;
\begin{align}
\label{eq:SU30-1}
    &p=6\,,\,k=0 \, \longrightarrow\, p'=3\,,\, k'=0
\,\,, 
\quad 
&p=3\,,\,k=0 \, \longrightarrow\, p'=3\,,\, k'=-1.   
\end{align}
In the same way we can distinguish between the pair of $p'=1$ IR models in the second equation of eq.~\eqref{eq:UVIR2pairs},
\begin{align}
\label{eq:p1mods}
    &p=2\,,\,k=0 \, \longrightarrow\, p'=1\,,\, k'=0
\,\,, 
\quad 
    &p=1\,,\,k=0 \, \longrightarrow\, p'=1\,,\, k'=1.
\end{align}

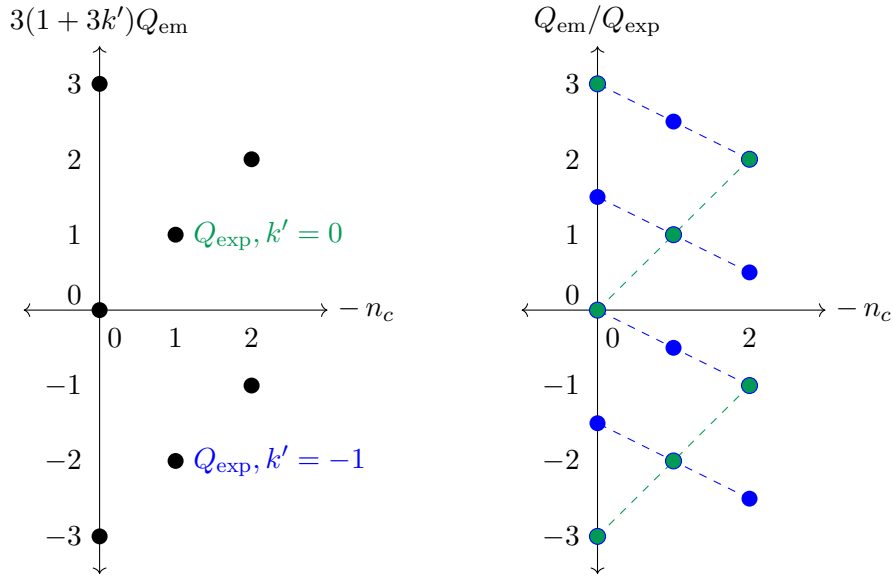
\begin{figure}
    \centering
\begin{tikzpicture}[scale=1]
  \draw[<->] (-1,0) -- (3,0) node[right] {$-\,n_c$};
  \draw[<->] (0,-3.5) -- (0,3.5) node[above] {$3(1+3k')Q_{\textrm{em}}$};
  \node[left] at (-0.1,0.2) {$0$};
  \node[left] at (-0.1,1) {$1$};
  \node[left] at (-0.1,2) {$2$};
  \node[left] at (-0.1,3) {$3$};
  \node[left] at (-0.1,-1) {$-1$};
  \node[left] at (-0.1,-2) {$-2$};
  \node[left] at (-0.1,-3) {$-3$};
  \foreach \x in {1,...,2}
    \node[below] at (\x,-0.1) {\x};
  \node[below] at (0.2,-0.1) {0};
  \fill (2,-1) circle (3pt);
  \fill (1,-2) circle (3pt);
  \node[right] at (1.1,-2) {{\color{blue} $Q_{\textrm{exp}}, k'=-1$}};
  \fill (0,-3) circle (3pt);
  \fill (0,0) circle (3pt);
  \fill (1,1) circle (3pt);
  \node[right] at (1.1,1) {\color{ForestGreen} $Q_{\textrm{exp}},k'=0$};
  \fill (2,2) circle (3pt);
  \fill (0,3) circle (3pt);
\end{tikzpicture}\qquad \qquad
\begin{tikzpicture}[scale=1]
  \draw[<->] (-1,0) -- (3,0) node[right] {$-\,n_{c}$};
  \draw[<->] (0,-3.5) -- (0,3.5) node[above] {$Q_{\textrm{em}}/Q_{\textrm{exp}}$};
  \node[left] at (-0.1,0.2) {$0$};
  \node[left] at (-0.1,1) {$1$};
  \node[left] at (-0.1,2) {$2$};
  \node[left] at (-0.1,3) {$3$};
  \node[left] at (-0.1,-1) {$-1$};
  \node[left] at (-0.1,-2) {$-2$};
  \node[left] at (-0.1,-3) {$-3$};
  \node[below] at (0.2,-0.1) {0};
  \node[below] at (2,-0.1) {2};
    \draw [ForestGreen,dashed] (0,0) --(2,2);
    \draw [ForestGreen,dashed] (0,-3) --(2,-1);
  \draw[blue,dashed] (2,-5/2) -- (0,-3/2);
  \draw[blue,dashed] (2,-1) -- (0,0);
  \draw[blue,dashed] (2,1/2) -- (0,3/2);
  \draw[blue,dashed] (2,2) -- (0,3);
  \fill [ForestGreen] (2,-1) circle (3pt);
  \fill [ForestGreen] (1,-2) circle (3pt);
  \fill [ForestGreen] (0,-3) circle (3pt);
  \fill [ForestGreen] (0,0) circle (3pt);
  \fill [ForestGreen] (1,1) circle (3pt);
  \fill [ForestGreen] (2,2) circle (3pt);
  \fill [ForestGreen] (0,3) circle (3pt);
    \fill [blue] (2,1/2) circle (3pt);
   \draw [blue](1,1) circle (3pt);
  \fill [blue](0,3/2) circle (3pt);
    \draw [blue](2,4/2) circle (3pt);
  \fill [blue](1,5/2) circle (3pt);
  \draw [blue](0,6/2) circle (3pt);
  \draw [blue](0,0) circle (3pt);
  \fill [blue](1,-1/2) circle (3pt);
  \draw [blue](2,-2/2) circle (3pt);
  \fill [blue](0,-3/2) circle (3pt);
  \draw[blue] (1,-4/2) circle (3pt);
  \fill [blue](2,-5/2) circle (3pt);
   \draw [blue](0,-3) circle (3pt);
\end{tikzpicture}
    \caption{Example of non-zero compositeness degree for $U(3)$ model eq.~\eqref{eq:U3mod} in the original lattice and the lattice normalised to the experimentally found fundamental of charge $Q_{\textrm{exp}}$. Green dots in the plot on the right give the $k'=0$ spectrum, while blue and blue-lined-green-filled dots indicate the spectrum of the $k'=-1$ theory. Each set has been connected with dashed lines to highlight the stairway structure.}
    \label{fig:kne0SU3}
\end{figure}

Next we turn to discuss the spectrum and normalisation of possible values of the electric charge in models with non-trivial compositeness degree. For concreteness consider the $U(3)$ model in the IR with compositeness degree $k'$
\begin{equation}
\label{eq:U3mod}
    U(3)\,=\, \frac{SU(3)_c\times U(1)_{\textrm{em}}}{Z_3} \,,
    \quad {k'}\in \mathbb{Z}\,.
\end{equation}
This model is characterised by charge $Q_3$ with eigenvalues $n_3$ given by
\begin{equation}
\label{eq:Q3def1}
n_3\,=\,n_c + 3Q_{\textrm{em}} (1+3k')  \,=\, 3\mathbb{Z} \mod 3\,.
\end{equation}
The condition above is satisfied when $3Q_{\textrm{em}} (1+3k')= -n_c$ mod 3, which defines the allowed spectrum. This spectrum is displayed in the LHS of Fig.~\ref{fig:kne0SU3}, where one can identify $p'$, regardless of $k'$, as the periodicity of the stairway structure, i.e. how many steps each flight of stairs has. The next question is, assuming a state with a value of electric charge  $Q_{\textrm{em}}=Q_{\textrm{exp}}$ and $n_c=1$ is observed at an experiment, how do we know which flight of stairs it belongs in? The answer is that we do not know with the information given, so we should consider placing it in each of the stairs that we labeled $k'$. The first two values of $k'$ are marked in the LHS of Fig.~\ref{fig:kne0SU3}; $k'=0$ would imply we have found the smallest quantum of charge  while for $k'=-1$ there exists another state with $-1/2$ times our $Q_{\textrm{exp}}$. We do, however, use $Q_{\textrm{exp}}$ itself as our measure of charge, and the RHS of Fig.~\ref{fig:kne0SU3} displays the same spectrum with this normalization. The purpose of these two diagrams shown side by side is to show how in the latter how the stairway structure of negative $k'$ ascends in the opposite direction as positive $k'$ yet they both belong to the same group (same $p'$).

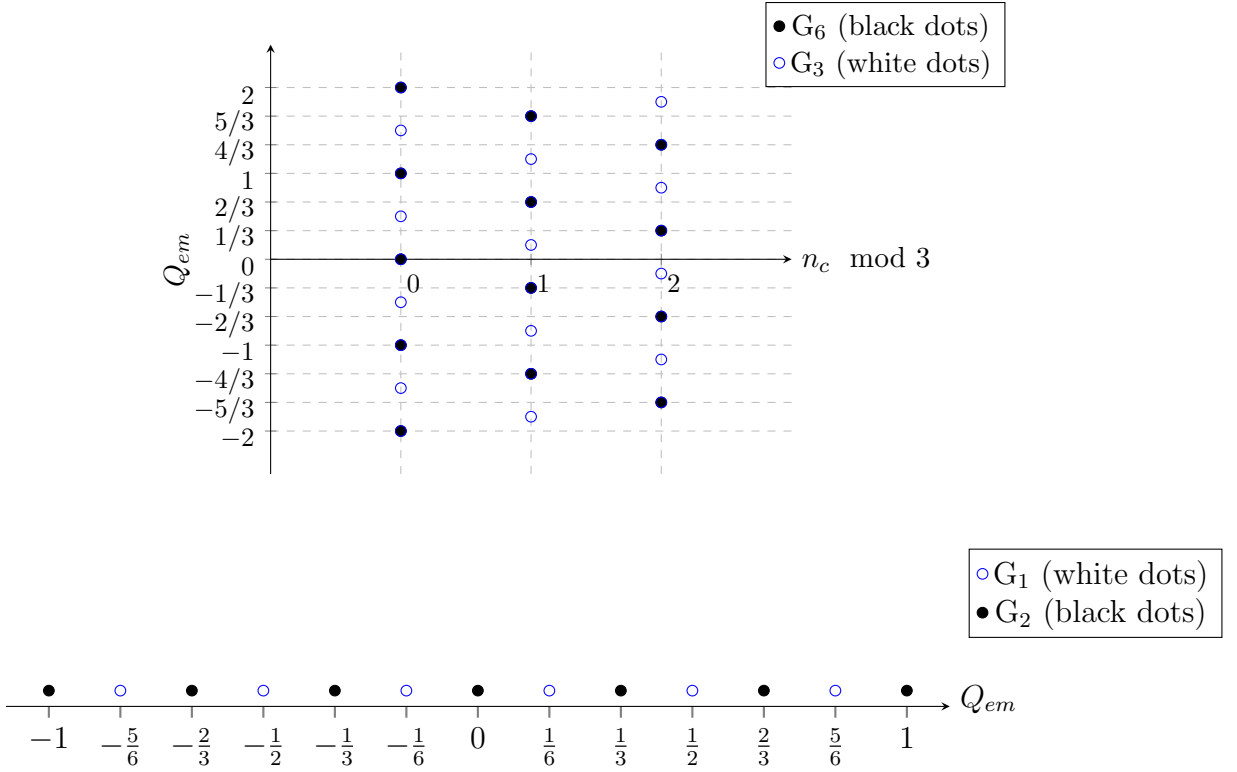
\begin{figure}[t]
\begin{center}
\begin{tikzpicture}
    \begin{axis}[
        axis x line=middle,
        axis y line=left,
        xlabel={$n_c \mod 3$},
        ylabel={$Q_{em}$},
        xlabel style={at={(axis description cs:1,0.5)}, anchor=west},
        xtick={0.5,1.5,2.5}, 
        xticklabels={0,1,2},
        ytick={-2,-5/3,-4/3,-1,-2/3,-1/3,0,1/3,2/3,1,4/3,5/3,2},
        yticklabels={$-2$, $-5/3$, $-4/3$, $-1$, 
                     $-2/3$, $-1/3$, $0$, $1/3$, 
                     $2/3$, $1$, $4/3$, $5/3$, $2$},
        ymin=-2, ymax=2,
        xmin=0, xmax=3,
        grid=major,
        grid style=dashed,
        enlargelimits={abs=0.5},
        scaled ticks=false,
        tick label style={font=\small},
        every y tick label/.style={yshift=-0.6ex},
        every x tick label/.style={xshift=1 ex},
        legend pos=north east,
        legend style={at={(.95,1)}, anchor=west}
    ]
    \addplot[only marks, mark=*, color=black] coordinates {
        (1.5,-4/3) (1.5,-1/3) (1.5,2/3) (1.5,5/3)
        (2.5,-5/3) (2.5,-2/3) (2.5,1/3) (2.5,4/3)
        (0.5,-2) (0.5,-1) (0.5,0) (0.5,1) (0.5,2)
    };
    \addplot[only marks, mark=o, color=blue] coordinates {
        (1.5,-11/6) (1.5,-4/3) (1.5,-5/6) (1.5,-1/3) (1.5,1/6) (1.5,2/3) (1.5,7/6) (1.5,5/3)
        (2.5,-5/3) (2.5,-7/6) (2.5,-2/3) (2.5,-1/6) (2.5,1/3) (2.5,5/6) (2.5,4/3) (2.5,11/6)
        (0.5,-2) (0.5,-3/2) (0.5,-1) (0.5,-1/2) (0.5,0) (0.5,1/2) (0.5,1) (0.5,3/2) (0.5,2)
    };
    \legend{G$_6$ (black dots), G$_3$ (white dots)}
    \end{axis}
\end{tikzpicture}
\end{center}
\begin{center}
\begin{tikzpicture}
    \begin{axis}[
        axis x line=bottom,
        axis y line=none,
        xlabel={$Q_{em}$},
        xlabel style={at={(axis description cs:1,1.5)}, anchor=west},
        xmin=-1.1, xmax=1.1,
        ymin=-0.1, ymax=0.1,
        xtick={-1,-5/6,-2/3,-1/2,-1/3,-1/6,0,1/6,1/3,1/2,2/3,5/6,1},
        xticklabels={
            $-1$, $-\frac{5}{6}$, $-\frac{2}{3}$, $-\frac{1}{2}$, $-\frac{1}{3}$, $-\frac{1}{6}$,
            $0$, $\frac{1}{6}$, $\frac{1}{3}$, $\frac{1}{2}$, $\frac{2}{3}$, $\frac{5}{6}$, $1$
        },
        height=2cm,
        width=14cm,
        enlargelimits=false,
        tick align=outside,
        tick style={thick},
        tick label style={font=\large},
        xlabel style={font=\large},
        legend style={font=\large, at={(1.02,5)}, anchor=north west}
    ]
    \addplot[
        only marks,
        mark=o,
        mark size=2pt,
        color=blue
    ] coordinates {
        (-5/6,0) (-1/2,0) (-1/6,0)
        (1/6,0) (1/2,0) (5/6,0)
    };
    \addplot[
        only marks,
        mark=*,
        mark size=2pt,
        color=black
    ] coordinates {
        (-1,0) (-2/3,0) (-1/3,0) (0,0) (1/3,0) (2/3,0) (1,0)
    };
    \legend{G$_1$ (white dots), G$_2$ (black dots)}
    \end{axis}
\end{tikzpicture}
\end{center}
\caption{Spectrum after EWSB for $G_p$, $p=1,2,3,6$ and $k=0$.\label{fig:SMk1m1}}
\end{figure}

Fig.~\ref{fig:SMk1m1} uses these patterns to recover the spectrum for all $G_p$ group realisations of the Standard Model with $k=0$ on the $(n_c\,,\, Q_{\textrm{em}})$ lattice.
Remaining results in eqs.~\eqref{eq:p2result}-\eqref{eq:p3result}
will be assembled at the end of the section in Table~\ref{tab:Matching}. 

In fact, the very same matching can be repeated at even lower energies; below $\Lambda_{\textrm{QCD}}$ the group is $U(1)_{\textrm{em}}$ so there is no quotient (i.e. $p''$ is always $1$) and all possibilities map into compositeness degree $k''$ as also shown in Table~\ref{tab:Matching}.

\medskip

Let us now discuss how the discrete symmetries act in this descent to the IR.

First, for the 0-form electric symmetry 
 $Z^{\textrm{el},(0)}_{6/p}$ defined in eq.~\eqref{Z0el}, a notable consequence of $k\neq 0$ is the $p$- and $k$-dependent assignment of $n_6$, an example of which can be seen in Table~\ref{n6kneq0}. In particular, the Higgs doublet itself might have a non-zero $n_6$ which would also imply the discrete symmetry is spontaneously broken. To examine this it will be useful to rewrite the operator $Q_6$ and its eigenvalues $n_6$ in eqs.~\eqref{eq:Q6kdef}-\eqref{eq:n6kdef}
in the following form
\begin{align}
    Q_6(k) &=2\tilde \lambda_8+6(1+p k)\,Q_{em}-3p \,k\,\tilde T_{3L}, \\
    n_6(k) &=2 n_c+6(1+p k)\,Q_{em}-3p\, k\,\tilde T_{3L}\,\,\,\textrm{mod}\, 6.
\end{align}
For $k=0$, one has that this is a conserved charge, given that triality and electric charge are conserved. For $k\neq 0$, however, the non-vanishing component along isospin introduces the possibility of breaking -- the only saving grace being, if it so happens, that $3\,p k\mod 6=0$, in which case the Higgs doublet is not charged and the electric symmetry stays unbroken. 
In summary one finds,
\begin{align}
    p=&6,2  && \,Z^{\textrm{el},(0)}_{6/p} \textrm{unbroken},\\
    p=&3    &&\left\{\begin{array}{cc}
         k\,\,\textrm{even}:\, & Z^{\textrm{el},(0)}_2\,\,\textrm{unbroken} \\
         k\,\,\textrm{odd} \,:\, & Z^{\textrm{el},(0)}_2\,\to 0
    \end{array}\right.,\\
    p=&1    &&\left\{\begin{array}{cc}
         k\,\,\textrm{even}\,:\, & Z^{\textrm{el},(0)}_6 \,\,\textrm{unbroken} \\
         k\,\,\textrm{odd}\,:\, & Z^{\textrm{el},(0)}_6\to Z^{\textrm{el},(0)}_3
    \end{array}\right..
\end{align}

\begin{table}[]
    \centering
\begin{tabular}{c|c|c|c|c|c|c}
     $G_3^{k=-1}$& $q_L$& $u_R$& $d_R$ & $\ell_L$ & $e_R$ & $H$\\ \hline
    $SU(3)_c$ & $\boldsymbol3$ & $\boldsymbol3$ & $\boldsymbol3$ & $\boldsymbol1$ &$\boldsymbol1$& $\boldsymbol1$\\
    $SU(2)_L$ &$\boldsymbol2$ & $\boldsymbol1$ & $\boldsymbol1$& $\boldsymbol2$ & $\boldsymbol1$&$\boldsymbol2$\\
    $ Q_Y$ & $\frac{1}{6}$ & $\frac{2}{3}$ &$-\frac{1}{3}$ &$-\frac{1}{2}$ &$-1$&$\frac{1}{2}$ 
    \\\hline
    $n_6$ & 3 & 0 & 0 & 3 & 0 & 3
\end{tabular}
\hspace{0.4cm}
\begin{tabular}{c|c|c|c|c|c|c}
     $G_2^{k=1}$& $q_L$& $u_R$& $d_R$ & $\ell_L$ & $e_R$ & $H$\\ \hline
    $SU(3)_c$ & $\boldsymbol3$ & $\boldsymbol3$ & $\boldsymbol3$ & $\boldsymbol1$ &$\boldsymbol1$& $\boldsymbol1$\\
    $SU(2)_L$ &$\boldsymbol2$ & $\boldsymbol1$ & $\boldsymbol1$& $\boldsymbol2$ & $\boldsymbol1$&$\boldsymbol2$\\
    $ Q_Y$ & $\frac{1}{6}$ & $\frac{2}{3}$ &$-\frac{1}{3}$ &$-\frac{1}{2}$ &$-1$&$\frac{1}{2}$ 
    \\\hline
    $n_6$ & 2 & 2 & 2 & 0 & 0 & 0
\end{tabular}
    \caption{LHS (RHS) values of $n_6$ for the $p=3$ ($p=2$) group with compositeness degree $k=-1$ ($k=1$) which, unlike the $k=0$ case, can be non-zero yet still within the spectrum allowed by the choice of $p$, see LHS of Fig.~\ref{fig:n6n6m} }
    \label{n6kneq0}
\end{table}

Remarkably, even when part of what one would naturally think of as the extension of the electric symmetry for $k\neq 0$ is broken, a new electric discrete symmetry emerges in its place. Consider the element
\begin{align}
    \eta=&\exp\left[\frac{2\pi}{6} i(6Q_Y+2n_c+3n_L)\right], 
\end{align}
and note that
\begin{align}
    \eta^{(6/p)(1+pk)}=\exp\left[2\pi i\frac{Q_6(k)}{p}+2\pi i k(2n_c+3n_L)\right]=1,
\end{align}
where we used the spectrum defining condition $n_6(k)=p\mathbb{Z}$. As such, $\eta$ generates the group $Z_{|(6/p)(1+pk)|}$ and moreover, $\eta$ acting on SM fields returns the identity, which is to say they are all neutral. This means that this emergent parity is a good IR quantum number, but it can further be promoted to an emergent 1-form symmetry in the IR since there are no particles charged under it. The 0-form charge is an expression that we have already made the acquaintance of
\begin{align}\label{eq:emergdef}
    \eta&=\exp\left(\frac{2\pi i}{6}(6Q_{\textrm{em}}+2n_c)\right), &
    \eta\, \Psi_{\hat n}&=\exp\left(2\pi i\frac{\hat n}{6/p+6k}\right)\Psi_{\hat n}.
\end{align}
The index $\hat n$ which ranges between $0$ and $6/p+6k$ is the phenomenologically useful generalisation of $n_6$ for $k\neq 0$ and provides a quantum number that cannot be changed by SM particle emission.

Let us give its 1-form realisation explicitly here for a given $\tilde V_3$ manifold
\begin{align}
    Z_{|(6/p)(1+pk)|}^{emg,(1)}\quad\tilde U_{\ell_e}^{emg}:\qquad A\to A+\frac{2\pi i}{6}(6Q_Y+2\tilde \lambda_8+3\tilde T_{3L})n_\mu^\perp dx^\mu \delta_{\tilde V_3},
\end{align}
so 
\begin{align}
   \tilde U_{\ell_e}^{emg}(\tilde V_3)W_{\hat n}(\tilde C)\tilde U_{\ell_e}^{emg\dagger}(\tilde V_3)=& \mbox{Tr}_R\,\mathcal P\exp\left[ \int A+\frac{2\pi \ell_e i (6Q_{\textrm{em}}+2n_c)}{6}\textrm{Link}(\tilde V_3,\tilde C)\, \right] \\
   \nonumber=&\exp\left(\frac{ 2\pi i\ell_e\hat n}{6/p+6k)}\textrm{Link}(\tilde V_3,\tilde C) \right)  W_{\hat n}(\tilde C).
\end{align}

The emergent symmetries for QCD+QED and QED are more straightfoward to obtain and are listed in Table~\ref{tab:Matching}. Let us highlight the connection with \cite{Koren:2024xof}, where the emergence of electric 1-form symmetry  all the way in the IR was spelled out. The discovery of a particle of charge $q$ would imply the emergence of a $Z_{(p/6)(1+pk)}$ 1-form symmetry with 
$q=\frac{p}{6(1+pk)}$. Remarkably, there is a one-to-one correspondence; for each $p$, $k$ there exists a unique value of $6/p+6k =q^{-1}$ which in turn means:
\begin{align}
    \boxed{\textrm{For each value of $p,k$ the minimal colour-singlet charge is unique and equal to } \frac{1}{6/p+6 k}.}
\end{align}

\begin{table}[]
    \centering
    \begin{tabular}{c|c|c|c}
        stage &SM& QCD+QED & QED \\ \hline
        & & &\\
        quotient &$p$& $p=6,3\to p'=3$& $1\,\,\forall$\\
        &&$p=2,1\to p'=1$&\\
        compositeness &$k$& $2k,-2k-1,1+2k,2+2k$ & $(6/p)(1+pk)-1$\\
        dim($Z^{(1)}_{emg}$) & $|6/p(1+pk)|$ & $|(p'/3) (6/p)(1+p k)| $& $|(6/p)(1+pk)|$
        \\ & & &
    \end{tabular}
    \begin{tikzpicture}
        \draw [thick,->](-7,0)--(7,0);
        \draw  (-1.4,0) node [anchor=north] {$v^{-1}$};
        \draw  (3.5,0) node [anchor=north] {$\Lambda_{QCD}^{-1}$};
    \end{tikzpicture}
    \caption{Matching of group quotients, compositeness degree, and emergent electric symmetries across energy scales. Here, \textit{emg} denotes “emergent”.}

    \label{tab:Matching}
\end{table}

\section{Embedding the SM group, general considerations}
\label{sec:2}

The classification above is fully general; a given UV completion may correspond to specific values of $p,k$ in the language used here. This also means that we can use these $p,k$ indexes as classifiers answering questions such as: Which completions lead to $p=6$? Which to $p=3$? Given the structure we discussed, these labels are exclusive -- i.e. if a model falls in the $p=6$ category, it will not belong to the $p=3$ one. {\it Thus, excluding certain $p$ has direct consequences for GUTs}. 

In this spirit, we will look for larger non-Abelian groups
\begin{align}
    \mathcal{G}\to G_{p},
\end{align}
not with the aim of providing an extensive list, but rather to examine the simplest or most well-known models and comprehend how the various low energy rules are realised in the high energy regime.

\subsection{Illustrative example}

Let us first illustrate how an embedding predicts the value of $p$; consider as an example $(SU(2)\times U(1))/Z_p$ so $p=1,2$ as our IR group. In this theory, the centre is (assuming a fundamental rep with charge one is in the spectrum)
\begin{align}
    Z_2=\{1 \,,\, \xi\equiv\exp(\pi i(Q-\tilde T_3)) \}.
\end{align}
Consider next the embedding of the IR group into $SU(3)$ in the UV. Given this choice only one of the two IR groups follows but which one? That is $SU(3)\to (SU(2)\times U(1))/Z_p$ where we want to solve for $p$. To do this, one has to connect the generators to the embedding group, so in our case
\begin{align}
    \tilde T_3\,=\,\left(\begin{array}{ccc}
        1 &  &  \\
         & -1 &  \\
         &  & 0
    \end{array}\right),\qquad Q\,=\,\tilde\lambda_8=\left(\begin{array}{ccc}
        1 &  &  \\
         & 1 &  \\
         &  & -2
    \end{array}\right).
\end{align}
One can simply now check what $\xi=\exp(\pi i(Q-\tilde T_3))$ looks like
\begin{align}
    \exp[\pi i(Q-\tilde T_3)]\,=\,\exp\left[\pi i\left(\left(\begin{array}{ccc}
        1 &  &  \\
         & 1 &  \\
         &  & -2
    \end{array}\right)-\left(\begin{array}{ccc}
        1 &  &  \\
         & -1 &  \\
         &  & 0
    \end{array}\right) \right)\right]\,=\,
    \exp \left[\pi i\left(\begin{array}{ccc}
        0 &  &  \\
         & 2 &  \\
         &  & -2
    \end{array}\right)\right],
\end{align}
with $\xi$ being the identity,
the condition $\xi\psi=\psi$ is trivially realised, and given that this is the condition of the spectrum for $p=2$, we conclude that $SU(3)\to (SU(2)\times U(1))/Z_2$. As opposed to our bottom-up approach where taking a quotient meant setting constraints in the spectrum, our top-down embedding group has no constraint imposed in its UV spectrum, the $Z_2$ quotient only appearing in the IR. We have carried out the discussion in fundamental-representation-space and indeed we assumed our IR doublet came from a fundamental of $SU(3)$, i.e. $\mathbf{3}\to \mathbf{2}\oplus\mathbf{1}$.

The case we have discussed is indeed that of the embedding group being the universal cover and the low energy fermion being contained in a fundamental, but one can consider quotients of the embedding group itself: here there is just one possibility $SU(3)/Z_3$. This group however, forbids $\mathbf{3}$, so the doublet would instead arise from, say, an adjoint 
\begin{align}
    \textrm{Adj} (\mathbf{8})=\left(\begin{array}{cc}
         \mathbf{3}_0& \mathbf{2}_{3} \\
         \mathbf{2}_{-3}& 0 
    \end{array}
    \right)+\tilde \lambda_8 \mathbf{1}_0,
\end{align}
where the $3\times 3$ matrix representation for $\textrm{Adj} (\mathbf{8})$ is decomposed into the $2\times 2$ block $\mathbf{3}_0$, and $1\times 2$ and $2\times 1$ blocks $\mathbf{2}_{3}$ and $\mathbf{2}_{-3}$ plus a singlet times $\tilde \lambda_8 $; and the subscripts on $\mathbf{3}_0$, $\mathbf{2}_{3}$ and $\mathbf{2}_{-3}$ give the  charges of unbroken $U(1)$.
There is a charged doublet, $\mathbf{2}_{3}$, but it has 3 times the unit charge, so one should revisit the relation between $Q$ and $\tilde\lambda_8$
\begin{align}\label{eq:newYSU3Z3}
    Q&=\frac{1}{3}\tilde\lambda_8 \,\,, \qquad  \mathbf2\in \mathbf{8},
\end{align}
and
\begin{align}\label{eq:xiZ3}
    \xi&=\exp\left(\pi i\left[\frac{\tilde\lambda_8}{3}-\tilde T_3\right]\right) 
 \,\,, \qquad   
    \left[\frac{\tilde\lambda_8^F}{3}-\tilde T_3^F\right]=\left(\begin{array}{ccc}
         -\frac{2}{3}&  &\\
         &  \frac{4}{3}&\\
         &&-\frac{2}{3}
    \end{array}\right),
\end{align}
which is still the identity but not as clearly as before. The expression above returns (minus) the triality of the representation as $\xi=\exp(-2\pi i n_c/3)$ but the spectrum constraint forbids all but those with $n_c=0$ (in fact the fundamental representation in which we wrote eq.~\eqref{eq:xiZ3} is itself forbidden), so $\xi$ is indeed the identity and again we have $SU(3)/Z_3\to SU(2)\times U(1)/Z_2$.

One last remark is in order: assume we knew that $\mathbf{2}$ indeed came from a $SU(3)$ adjoint.  Then we could have $SU(3)/Z_3$ or $SU(3)$ itself, since the latter contains the spectrum of the former. However, this would lead to
\begin{align}
    \mathbf 2_{Q}\in \mathbf{8}\,\, \textrm{of}\,\, SU(3) \left\{\begin{array}{cc}
         SU(3)/Z_3 \to SU(2)\times U(1)/Z_2,  & k=0\\
         SU(3)\to SU(2)\times U(1)/Z_2, &k=1
    \end{array}\right..
\end{align}
That is, if the group is $SU(3)/Z_3$ our doublet is in the lowest-dimensional possible representation,
corresponding to $k=0$. On the other hand, if the group is $SU(3)$ and the doublet comes from the adjoint representation, there exists another smaller-charge representation that is not part of the IR spectrum, and in this case, one has $k=1$.

The take away message is that embedding in a non-Abelian universal cover group leads to a quotient in the unbroken group. Further quotienting the UV group can result in possibly an even larger quotient at low energies and (or) to a higher compositeness degree.

\subsection{The SM case}

For a first hint at which embeddings would lead to which $G_p$ group let us rewrite the 4 choices, given $\frac{SU(N)\times U(1)}{Z_N}=U(N)$, one has
\begin{align}
    G_6&=S[U(3)_c\times U(2)_L], \\  G_3&=SU(2)_L\times U(3)_c,\\
    G_2&=U(2)_L\times SU(3)_c, \\ G_1&=SU(3)_c\times SU(2)_L\times \frac{U(1)}{Z_1}.
\end{align}
So we see that $G_2$ correlates weak isospin and hypercharge; if we embed both within a simple group, we can expect to reproduce this correlation. Similarly, $G_3$ correlates colour and hypercharge, and $G_6$ correlates all three SM sectors. The question then boils down to what is the smallest non-Abelian group that contains the following groups with their respective {\it p} values? 
$(i)\, p=6 : SU(3)\times SU(2)\times U(1),$ $(ii)\, p=3 : SU(3)\times U(1),$ $(iii)\, p=2 : SU(2)\times U(1),$ $(iv)\, p=1 : U(1)$. Dynkin diagrams are helpful in addressing this question since the cartan sub-algebra of the embedding group should have dimension of at least the cartan subalgebra of the sector of the SM group it is embedding. We will briefly discuss each possibility in the following sections.

The embedding is characterised by the relation between the generators of $\widetilde{G}_{SM}$ and those of the larger group. This is straightforward for colour and weak isospin, where one can identify the corresponding generators {\it and} determine the proportionality constant between them -- for example, by requiring that they obey the same Lie algebra (which is not invariant under rescaling). This is not the case for $Q_Y$ due to the triviality of its Lie algebra. However, one essential constraint the relation must obey is a consistent periodicity. Suppose we have a candidate generator $\tilde{T}_Y$, its periodicity can be determined by finding the distance $\alpha$ that takes $e^{i\alpha \tilde{T}_Y}$ back to the origin {\it for all} representations. This amounts to finding the smallest eigenvalue of $\tilde{T}_Y$, as the rest will be multiples of it by construction (and note that this might not always be in the fundamental). Once this eigenvalue is found, it sets the normalisation of $\tilde{T}_Y$. Provided that the smallest eigenvalue\footnote{Note that the smallest eigenvalue is a function of whether the group $\mathcal G$ has a quotient itself with an instance given in eq.~\eqref{eq:newYSU3Z3} for $SU(3)/Z_3$} is where we place the quark doublet, then
\begin{align}
    6\,Q_Y&\,=\,\tilde{T}_Y,  &\textrm{smallest $|$eigenvalue$|$ of}\, \,\tilde{T}_Y\,\textrm{defined as 1}.
\label{eq:6QYnorm}    
\end{align}

This yields a compatible relation, but it is not unique -- we could have a proportionality constant equal to an integer. In fact, the form of this constant can be put in one-to-one correspondence with the compositeness degree introduced earlier.
\begin{align}
    6(1+p k)\,Q_Y&\,=\,\tilde{T}_Y, &\textrm{smallest $|$eigenvalue$|$ of}\,\, \tilde{T}_Y\,\textrm{defined as 1}.
\end{align}
This result however goes beyond saying that $6 \mathbb{Z}Q_Y=\tilde T_Y$ to correlate the proportionality constant with $p$; we shall see this at play in the next section.

Once the $\tilde T_Y-Q_Y$ connection is specified, everything else can be inferred; in particular, the value of $p$ can be deduced by solving for the \emph{maximal} $p$ that satisfies the equation ({\it cf.}~eqs.~\eqref{eq:xidef}, \eqref{eq:1.6})
\begin{align}\label{eq:embfindp}
    \xi^{6/p}=e^{\frac{2\pi i Q_6(k)}{p}}=\exp\left( \frac{2\pi i}{p}\left(2\tilde\lambda_{8}+3\tilde T_{3L}+\tilde T_{Y}\right)\right)=1,
\end{align}
where $\tilde\lambda_{8}$ and $\tilde T_{3L}$ should be taken as the $\mathcal G$-embedded generators which however are straigthfoward to find, unlike for $\tilde T_Y$.
gives the identity, by construction.

Let us close this section with a number of lessons to be used in the following
\begin{itemize}
    \item The $n_6$ spectrum determines $G_p$ (for $k=0$), see Fig.~\ref{fig:n6n6m}.
    \item Any embedding group one introduces must be fully specified, i.e. if the embedding group admits quotients, one shall pick one.
    \item Once the group is given, the resulting monopole spectrum can be derived and compared with the predictions outlined above.
\end{itemize}

\section{The particle spectrum}\label{sec:ThePSpec}

The best known and studied embeddings are those that `pack' matter into the smallest number of representations. While these models and some novel ones are the case studies of this section, here we shall approach it from the group structure, in particular, starting with the Cartan subalgebra and Dynkin diagrams. The aim of this section is to give explicit examples for all $p$ discussing also possible quotients of $\mathcal G$. For previous work along these lines see~\cite{Cordova:2023}.

\subsection{Embedding hypercharge}
\label{sec:3.1}

A straightforward gauge group that presents the $Z_1$ quotient is the universal cover $SU(3)_c \times SU(2)_L\times U(1)_Y$, which is what the average particle physicist will picture in their minds if the words Standard Model group are uttered. However, the extension of the $p>1$ cases with $k=0$ to $p=1$ also has $k=0$ which is to say the minimum allowed hypercharge is $1/6$ and this assumption is a step beyond stating our group is $G_1$. We have encountered none other than the problem of finding the smallest quantum of charge in a $U(1)$ group; a known solution to this problem is the external constraint of Dirac magnetic monopoles. It can also be solved by embedding hypercharge in a non-Abelian group, which is in fact a solution that neccesarily brings along monopoles\footnote{In fact for the monopoles to be dynamical finite-energy states, they are expected to be realised as `t~Hooft--Polyakov solutions~\cite{tHooft:1974kcl,Polyakov:1974ek}, which again requires an embedding of $U(1)$ into a non-Abelian group.}. The simplest embedding group for $U(1)_Y$ is $SU(2)_Y$, a statement that hardly needs explanation but which we find useful in reframing with a modification of Dynkin diagrams. Take a dashed node to be a $U(1)$ generator as opposed to a regular node which neccesarily has roots and brings along at least an $SU(2)$ subalgebra, one can diagrammatically and mathematially characterise this embedding starting from $SU(3)_c \times SU(2)_L\times SU(2)_Y$ as in Fig.~\ref{tab:DynkinY} and as
\begin{align}
\label{eq:2to1}
     SU(3)_c& \times SU(2)_L \times (SU(2)_Y \to U(1)_Y), & 6Q_Y = \tilde{T}_{Y}.
\end{align}

\begin{figure}[h]
    \centering    
\begin{tikzpicture}
    \draw [thick,dashed] (0,0) circle (3pt);
    \draw [thick,blue,<-] (0.5,0)-- (2,0);
    \draw [thick] (2.75,0) circle (3pt);
    \draw (0.25,-0.75) node {$U(1)_Y$};
    \draw (2.75,-.75) node {$SU(2)_Y$};
\end{tikzpicture}
\caption{The RHS shows the smallest group $SU(2)$ that can embed the $U(1)_Y$ hypercharge of the SM that can further break into $U(1)_{em}$.}
\label{tab:DynkinY}
\end{figure}
\noindent Equation~\eqref{eq:2to1} fully determines the group embedding and is the basis for the new anomaly-free 
$SU(3)_c \times SU(2)_L \times SU(2)_Y$ model that we propose next, dubbed $\mathbf{SU2Y}$ for reference.
\medskip

\noindent$\bullet$ {\bf The value of} $p$. The value of $p$ which determines which $G_p$ theory we are in can be derived by finding the \emph{maximal} value of $p$ for which 
$\xi_p=e^{2\pi iQ_6/p} =1$, see eqs.~\eqref{eq:Zpdef2}, \eqref{eq:1.6}, \eqref{eq:embfindp}. There is no correlation apriori between colour, weak-isospin and hypercharge so our equation shall involve $Q_Y$ only and one has (using eq.~\eqref{eq:6QYnorm})
\begin{align}
    \xi_1 =  \exp\left(2\pi i\tilde{T}_{Y} \right) =\exp{\left(12\pi iQ_Y\right)} = \exp\left(2\pi i\left(\begin{array}{ccc}
        1 &   \\
         & -1  
    \end{array}\right)\right),
\end{align}
where the RHS implies that the maximal value of $p$ for which $\xi_p=1$ is $p=1$ and we confirm that the model is a $G_1$ theory
\begin{align}
  G_1 \,\,\, : \,\,\, e^{2\pi i(6Q_6)}= e^{12\pi i Q_Y} = 1.
\end{align}
As we shall see and anticipated in sec.~\ref{sec:2}, this value of $p$ could be increased if the embedding group admits itself a quotient, yet to discuss this one needs the matter content so let us postpone this question till the end of this subsection.

The constraint imposed on the SM by the $Z_1$ quotient allows hypercharge values to be independent of $n_L$ and $n_c$, with the minimum $Q_Y$ being $1/6$. The hypercharge of this group comes from the third component, $\tilde{T}_Y$ of $SU(2)_Y$ group so the electric charge is
\begin{align}
    Q_{em} \,=\, \frac{1}{2}\,\tilde{T}_{3L} + Q_Y \,=\, 
    \frac{1}{2}\,\tilde{T}_{3L}+ \frac{1}{6}\,\tilde{T}_Y \,\,\, ; \quad \tilde{T}_Y^{(F)} = \text{diag} (1, -1).
    \label{Q_em for 322}
\end{align}

\noindent$\bullet$ \textbf{Fermionic matter content.} Given the smallest hypercharge is $1/6$ corresponding to the fundamental of $SU(2)_Y$, one has $q_L$ $\in \mathbf{2}$ of $SU(2)_Y$. To form hypercharge representations for other SM fields we can group together particles that belong in the same $SU(3)_c, \, SU(2)_L$ representation. For example $u_R$ and $d_R$ are both left-handed (LH) singlets under $SU(2)_L$ and colour triplets under $SU(3)_c$ so they can be placed in the quintet ($\mathbf{5}$) representation of $SU(2)_Y$. Similarly, $e_R$ is a single under both $SU(3)_c, \, SU(2)_L$ and its high hypercharge means it should be placed in a septet ($\mathbf{7}$) representation for $SU(2)_Y$.

Strongly interacting LH matter fermions -- i.e. LH quarks -- are then grouped into the fundamental representation $(\mathbf{3},\mathbf{2},\mathbf{2})$
of our $SU(3)_c \times SU(2)_L\times SU(2)_Y$ model as follows
\begin{align}
\psi_{2, \, L} = (Q_L, q_L) = 
\begin{pmatrix}
D^{i} & u^i \\
U^{i} & d^i
\end{pmatrix}_L, \,\, \quad Q_L \equiv
\begin{pmatrix}
D^i\\ U^i
\end{pmatrix}_L,
\end{align}
where $SU(2)_L$ acts on the $\psi_{2, \, L}$ matrix from the left and $SU(2)_Y$ acts from the right.
Here $i=r,b,g$ is the colour index, $u^i$ and $d^i$ are the SM up and down quarks,
the remaining fermions $U^{i}$ and $D^{i}$ are new `BSM' quarks and here and in the following we shall give all our representations in the \emph{diagonal} $\tilde T_{Y}$ basis, in particular, $\tilde T_{Y}(\psi_{2,L})=(-Q_L,+q_L)$.

Similarly, we can place LH leptons into the $(\mathbf{1},\mathbf{2},\mathbf{4})$ representation
\begin{align}
\psi_{4, \, L} = (\ell_L, L_L) = 
    \begin{pmatrix}
\nu & N' & N'' & N''' \\
e & E' & E'' & \nu'
\end{pmatrix}_L,
\end{align}
where $SU(2)_L$ acts on the left and $SU(2)_Y$ on the right and each column has $\tilde T_{Y}$ eigenvalue, from left to right, $(-3,-1,1,3)$; so there is an $SU(2)_Y$-quartet of LH lepton doublets under $SU(2)_L$. The first of these doublets $\ell_L$ is composed of the standard $\nu_L$ and $e_L$, and the remaining three doublets are new leptons 
\begin{align}
\quad L_L' \equiv 
\begin{pmatrix}
N' \\ E'
\end{pmatrix}_L \,\, , \,\, L_L'' \equiv 
\begin{pmatrix}
N'' \\ E''
\end{pmatrix}_L \,\, , \,\, L_L''' \equiv 
\begin{pmatrix}
N''' \\ \nu'
\end{pmatrix}_L.
\end{align}

The right-handed (RH) quarks are assembled into the $\psi_{5,\,R}$ under $SU(2)_Y$ (which is a singlet of $SU(2)_L$)
\begin{align}
\psi_{5,\,R}= 
\begin{pmatrix}
U_R^i & d_R^i & X_R^i & D_R^i & u_R^i
\end{pmatrix},
\end{align}
where $U_R^i$, $X_R^i$ and $D_R^i$ are new RH quarks. In addition to the fields listed so far,  we also introduce a new LH fermion
$X^i_L$ $\in \bold 3$ to make the model a vector-like theory in the $SU(3)_c$ sector and cancel the colour anomaly.

The RH leptons are put into the septet representation of $SU(2)_Y$ 
\begin{align}
\psi_{7,\,R}= 
\begin{pmatrix}
e_R & E'_R & E''_R & \nu'_R & N'_R & N''_R & N'''_R  \\
\end{pmatrix},
\end{align}
and we also introduce a total singlet $\nu_R \in \bold1$ which we will identify with the RH partner of the active SM neutrinos\footnote{The representation $\psi_{7,R}$ contains $\nu'_R$ which has the same quantum numbers after EWSB as $\nu$ and the two will mix in a generic scenario.}.

\textbf{Particle count:} 19 more chiral fermions (times 3 generations) counting only chirality of states and not colour.

\begin{figure}[h!]
\centering
\begin{tikzpicture}
    \begin{axis}[
        axis x line=bottom,
        axis y line=none,
        xlabel={\,$Q_{em}$},
        xlabel style={at={(axis description cs:1,1.5)}, anchor=west},
        xmin=-1.1, xmax=1.1,
        ymin=-0.1, ymax=0.1,
        xtick={-1,-2/3,-1/3,0,1/3,2/3,1},
        xticklabels={
            $-1$, $-\frac{2}{3}$, $-\frac{1}{3}$,
            $0$, $\frac{1}{3}$, $\frac{2}{3}$, $1$
        },
        height=2cm,
        width=14cm,
        enlargelimits=false,
        tick align=outside,
        tick style={thick},
        tick label style={font=\large},
        xlabel style={font=\large},
        clip=false 
    ]
    \node at (axis cs:-2/3,0.3) {\begin{tabular}{c}$U$\\$E'$\end{tabular}};
    \node at (axis cs:-1/3,0.15) {\begin{tabular}{c}$E''$\end{tabular}};
    \node at (axis cs:0,0.5) {\begin{tabular}{c}$X$\\$\nu_R$\\$\nu'$\end{tabular}};
    \node at (axis cs:1/3,0.3) {\begin{tabular}{c}$D$\\$N'$\end{tabular}};
    \node at (axis cs:2/3,0.15) {\begin{tabular}{c}$N''$\end{tabular}};
    \node at (axis cs:1,0.15) {\begin{tabular}{c}$N'''$\end{tabular}};
    \node at (axis cs:-1,-0.75) {\begin{tabular}{c}\\$H'_-$\end{tabular}};
    \node at (axis cs:-2/3,-0.75) {\begin{tabular}{c}\\$H''_-$\end{tabular}};
    \node at (axis cs:-1/3,-1) {\begin{tabular}{c}$H'''_-$\\$B^-(T^-)$\end{tabular}};
    \node at (axis cs:0,-1) {\begin{tabular}{c}$H'_0$\\$T^0$\end{tabular}};
    \node at (axis cs:1/3,-1) {\begin{tabular}{c}$H''_+$\\$B^+(T^+)$\end{tabular}};
    \node at (axis cs:2/3,-0.85) {\begin{tabular}{c}$H'''_+$\end{tabular}};
    \end{axis}
\end{tikzpicture}
\caption{Electromagnetic spectrum of BSM particles. New fermions have been named so that LH and RH have the same symbol and form Dirac pairs, $\nu_R$ is the Dirac partner of the SM active neutrinos $\nu_L$, and the massive $B^\pm$ bosons have in parenthesis the scalar they incorporate as their longitudinal component.}
\label{fig:QemSU2Y}
\end{figure}
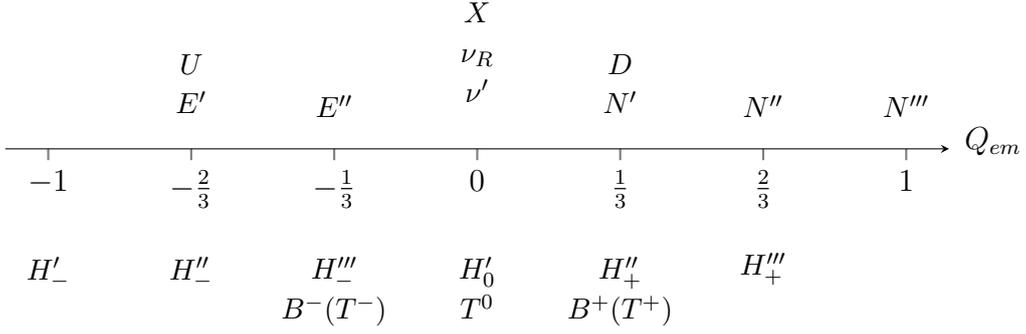

The classification of the fermionic matter content and the corresponding $n_6$ number of this model can be found in Table \ref{tab:Ysimple} whereas Fig.~\ref{fig:QemSU2Y} gives the electric charges of new states.

\begin{table}[h]
    \centering
    \begin{tabular}{c||c||c||c||c}
                $SU(3)_c\times SU(2)_L \times SU(2)_Y$  & $(\mathbf{3},\mathbf{2},\mathbf{2})$ & $(\mathbf{1},\mathbf{2},\mathbf{4})$     &  $(\mathbf{3},\mathbf{1}, \mathbf{5})$  & ($\mathbf{1}$,$\mathbf{1}$,$\mathbf{7}$)  \\ \hline
        SM fields  & $ q_L$  & $ \ell_L$    & $ u_R, d_R$  & $e_R$ \\
        $n_6$   & 0   & 0   &  0 &  0 \\ 
        \hline
        BSM fields  & $ Q_L$  & $ L_L', L''_L, L'''_L$    & $ U_R, X_R$,$D_R$  & $E_R',E''_R, \nu'_R, N_R' ,N''_R,N'''_R$ \\
         $n_6$  & 4 &  2,\, 4,\, 0  & 4,\, 2,\, 4  & 2,\,\,\, 4,\,\,\, 0,\,\,\, 2,\,\,\, 4,\,\,\, 0 \\
    \end{tabular}\\
    \vspace{1cm}
     \begin{tabular}{c||c||c}
                $SU(3)_c\times SU(2)_L \times SU(2)_Y$  & $(\mathbf{3},\mathbf{1},\mathbf{1})$ & $(\mathbf{1},\mathbf{1},\mathbf{1})$      \\ \hline
        BSM fields  & $ X_L$  & $ \nu_R$ \\
         $n_6$  & 4 &  0 \\ 
    \end{tabular}
    \caption{Fermionic matter content of the $SU(3)_c\times SU(2)_L \times SU(2)_Y$  model. The  $n_6$ values for each particle match the numbers below the corresponding fields. For example, $n_6$ of $L'_L$ is 2, and 4 for $L''_L$. All conventional SM particles have $n_6 = 0$.}
    \label{tab:Ysimple}
\end{table}

\noindent$\bullet$ \textbf{Scalar matter content and symmetry breaking}.
Encompassing the $U(1)_Y$ group into the larger $SU(2)_Y$ group means the $B_\mu$ boson is now recast as the third of the three gauge boson $B^i$ of $SU(2)_Y$, i.e. $B_\mu=B_\mu^3$. The two extra gauge boson fields we group as usual and have electric charge
\begin{align}
    B^{\pm}_\mu=&\frac{1}{\sqrt{2}}(B^1_\mu\pm iB^2_\mu), &
    Q_{em}&(B^i\sigma_i)=B^i[\tilde T_{Y},\sigma_i]/6, &Q_{em}B^\pm &= \pm \frac{1}{3}B^{\pm},
\end{align} 
where $\sigma^i$ are Pauli matrices and  we note that the transformation from the $B^i\sigma_i$ basis to $B^\pm$ is in keeping in our convention to write representations in the $\tilde T_{Y}$-diagonal basis as
  \begin{align} 
  \label{eq:Bs}
    {\cal B}_\mu \,&=\, \left(B^-_\mu\,\, B_\mu\, B_\mu^+\right)\,\quad\in\, (\mathbf{1},\mathbf{1},\mathbf{3}),\qquad \tilde T_Y({\cal B})=(-2 B^-, 0, 2B^+).
\end{align}

The SM Higgs boson $H$ is also upgraded from a single $SU(2)_L$ doublet in the SM to a quartet representation $(H,H',H'',H''')$ of $SU(2)_Y$ which includes three additional $SU(2)_L$ doublets as
\begin{align}
    \label{eq:Hs}    
    {\cal H} \,=\, \left(H',H'',H''',H\,\right)\,\in\, (\mathbf{1},\mathbf{2},\mathbf{4})\,\,\,;\quad (H',H'',H''')=\left(\begin{array}{ccc}
         H'_0 & H''_+& H'''_+\\
          H'_-& H''_-& H'''_-
    \end{array}\right) \,\,\,,\,\,\, H =\begin{pmatrix}
        H_+ \\
        H_0
    \end{pmatrix}.
\end{align}

To trigger the $SU(2)_Y \rightarrow U(1)_Y$ spontaneous symmetry breaking (SSB) we
also need to add a new Higgs field  that in the minimal settings can be taken to be in the adjoint representation of $SU(2)_Y$. We refer to this as the Higgs triplet ${\cal T}$ with components $T^0,\,T^{\pm}$ that have the same electromagnetic charge as those of the gauge bosons $B_\mu$ and a non-zero VEV $\langle T^0 \rangle \equiv t\neq 0$,
\begin{align}
\label{eq:Ts}    
    {\cal T} \,&=\, \left(T^-\,,\,T^0\,,\,T^+\right)\quad\qquad\in\, 
    (\mathbf{1},\mathbf{1},\mathbf{3}).
\end{align}
The bosonic fields and their hexality are collected in Table ~\ref{tab:Yboson}.
\begin{table}[h]
    \centering
    \begin{tabular}{c||c||c||c}
                $SU(3)_c\times SU(2)_L \times SU(2)_Y$  & $(\mathbf{1},\mathbf{2},\mathbf{4})$ & $(\mathbf{1},\mathbf{1},\mathbf{3})$  &  $(\mathbf{1},\mathbf{1},\mathbf{3})$ \\ \hline
        SM fields  & 
        $H$  & 
         & $B^0_\mu$\\
        $n_6$   & 0   &    &  0 \\ 
        \hline
        BSM fields  & $H', H'',H'''$  & $T^+,T^0,T^-$  & 
       $B^+_\mu,B^-_\mu$  \\
         $n_6$  & 0,\,\,\,2,\,\,\,4 &  
         2,\,\,\,0,\,\,\,4  & 2,\,\,4 
         \\ 
    \end{tabular}
    \caption{Higgs and vector boson fields of the $SU(3)_c\times SU(2)_L \times SU(2)_Y$ model eqs.~\eqref{eq:Hs}-\eqref{eq:Bs} and their $n_6$ values.}
    \label{tab:Yboson}
\end{table}

\textbf{Particle count:} 2 new massive gauge bosons, with 15 new scalar d.o.f., two of them ($T^{\pm}$) are the longitudinal components of $B^{\pm}$ and the remaining scalars split into 3 $Q_{\textrm{em}}=0$ real scalars and out of which 10 are charged.
\medskip

 \noindent$\bullet$~{{\textbf{Anomaly cancellation}}.}
The anomaly cancellation condition for the conservation of a general gauge current $J_c^\mu$ requires that~\cite{Gross:1972pv,Weinberg:1996kr}  
\begin{equation}
\sum_{\text{ferm.reps}} \mathrm{Tr}[T_c\{T_a,T_b\}] (-1)^{L,R}
\,=\, 0,
\end{equation}
where we sum over all representations of chiral fermions propagating in the loop in Fig.~\ref{fig:anomaly} with the LH and the RH fermions contributing with the opposite sign.

The cancellation of $SU(3)_c$ anomaly is verified by counting and matching the number of LH fermions charged under $SU(3)_c$ to the number of the RH ones in Table~\ref{tab:Ysimple}
\begin{align}
  (\psi_{2,L})  2\times 2  + (X_{L})  \,=\, (\psi_{\mathbf{5},R}) 5,
\end{align}
where in parenthesis we indicated the representation whose colour-perpendicular d.o.f. are being counted and all coloured fermions are in the same representation, i.e. {\bf3}.

\begin{figure}
    \centering
    \begin{tikzpicture}
    \begin{feynman}
    \vertex (a);
    \vertex [right=1.2cm of a] (v1);
    \vertex [above right=1cm and 1.5cm of v1] (v2);
    \vertex [below right=1cm and 1.5cm of v1] (v3);
    \vertex [right=1.8cm of v2] (g1);
    \vertex [right=1.8cm of v3] (g2);
    \diagram* {
      (a) (v1),
      (v1) -- (v2) -- (v3) -- (v1),
      (v2) -- [photon] (g1),
      (v3) -- [photon] (g2)
    };
    \node[above right=2pt] at (v2) {\(T_a\)};
    \node[left=2pt] at (v1) {\(T_c\)};
    \node[below right=2pt] at (v3) {\(T_b\)};

    \node at (v1) {\(\mathbf{\otimes}\)};
    \hspace{2cm}
    \node[right] at ($ (g1)!0.5!(g2) $) {\(\partial_\mu J^{\mu}\propto \ \displaystyle\sum\limits_{\text{ferm.reps}} \mathrm{Tr}[T_c\{T_a,T_b\}] (-1)^{L,R}\)};
    \end{feynman}
    \end{tikzpicture}
    \caption{Triangle anomaly for the current $J{\mu}_c$. Trace is summed over all chiral fermion representations propagating in the loop and $L,R$ are their chiralities.}
    \label{fig:anomaly}
\end{figure}
Mixed anomalies of $Q_Y$ with colour and $Q_Y$ with weak isospin now cancel for the same reason that the colour-weak anomaly does -- traceless $T_Y$ generators ensure that
$\mathrm{Tr}[T^a_Y\{T^b_L,T^c_L\}] \propto
\mathrm{Tr} (T_Y^i)=0$. In fact this also holds for the mixed anomaly with gravity. The remaining anomaly is $SU(2)_Y^3$, however $SU(2)_Y$ is pseudo-real which makes both $SU(2)_L$ and $SU(2)_Y$ naturally anomaly free. For a theory to be free from gauge anomalies, the cubic Casimir invariant has to vanish
\begin{align}
    d^{abc} = \mathrm{Tr}[T^a\{T^b,T^c\}] = 0,
\end{align}
where $T^i$ are the generators of $SU(2)$. Real representations of the local $SU(2)$ gauge group are automatically anomaly-free due to the symmetry properties of the trace under cyclic permutations. For pseudo-real representations, the similarity transformation of their generators $T^a_{\bar{R}}$
\begin{align}
    T^a_{\bar{R}} = S^{-1}T^a_R S,
\end{align}
ensures that all real and pseudo-real representations of SU(2) are anomaly-free~\cite{Weinberg:1996kr}.

Finally, we can show that \emph{global} $SU(2)$ anomalies~\cite{Witten:1982fp,Wang:2018qoy,Davighi:2019rcd} also cancel in our SU2Y model.
To ensure that these global anomalies are cancelled it is sufficient to check that that there is always an \emph{even} number of chiral fermions transforming in representations of $SU(2)$ with a half-integer isospin.
For $SU(2)_L$ this is the case since there is an even number (6 counting colour) of quark doublets in $\psi_{2,L}$ and 4 lepton doublets $\psi_{4,L}$. It is also easy to check that these fields contribute an even number of chiral fermions transforming under $SU(2)_Y$. For $\psi_{5,R}$ and $\psi_{7,R}$ this is irrelevant since they furnish integer (2 and 3) rather than half-integer isospin representations of $SU(2)_Y$.

\medskip

 \noindent$\bullet$~{{\textbf{Mass terms}}.}
Masses of new quarks and leptons are generated by coupling to the new Higgs-like BSM vector bosons or the SM Higgs bosons. Finding the invariant operators will require combining the different representations and looking for a resulting singlet. The nature of $SU(2)$ means one can build invariants with the square of a representation or alternatively that the complex conjugate of a representation can be manipulated to transform just as the representation itself. For weak isospin this takes the form
\begin{align}
    \ell_L^T \,\epsilon^T \,\ell_L= \ell_L^T \,\begin{pmatrix}
        0&-1\\
        1&0
    \end{pmatrix} \,\ell_L\equiv \tilde \ell_L^\dagger \ell_L,\qquad \tilde \ell_L=\epsilon \ell_L^*=\begin{pmatrix}
        0&1\\
        -1&0
    \end{pmatrix} \ell_L^*,
\end{align}
and one would identify this invariant as $\mathbf{2}\otimes\mathbf{2}=\mathbf{1}\oplus\dots$ while this `tilding' operation is what allows the SM Higgs to give mass to both up and down-type particles. The extension to $SU(2)_Y$ and a higher-representations R reads
\begin{align}
    \textrm{R}\,\varepsilon^T \,\textrm{R}^T=\textrm{R}\,\begin{pmatrix}
         &&&1\\
         &&-1\\
         &1&&\\
         \dots &&&
    \end{pmatrix} \,\textrm{R}^T\equiv R\tilde R^\dagger,\qquad \tilde R=R^*\varepsilon,\label{eq:vareps}
\end{align}
where we have kept with the convention of $SU(2)_Y$ representations as row-arrays (hence the transpose) and we note that $\varepsilon$ is an anti-symmetric (symmetric) tensor\footnote{The matrix $\varepsilon$ can be found by solving for eq.~\eqref{eq:vareps} to be an invariant, i.e. $(T_Y^{i})^T\varepsilon+\varepsilon T_Y^i=0$ with $T_Y^i$ the $SU(2)_Y$ generators.} for even (odd) dimensional representations. The operation of making a complex conjugate transform as the original representation when the field has both $SU(2)_L\times SU(2)_Y$ indexes is, for the Higgs multiplet,
\begin{align}
    \widetilde{\mathcal H}\equiv(\tilde H,-\tilde H''',\tilde H'',-\tilde H'),\qquad \tilde H=\begin{pmatrix}
        0&1\\
        -1&0
    \end{pmatrix} H^*.
\end{align}

The quark and lepton sectors can be discussed separately. First for coloured particles,  the exercise in group theory  to find invariant reads (where on the RHS we count the number of independent operators)
\begin{align}
    \mathcal{H}\bar{\psi}_{2,L}  \psi_{5,R} & : \mathbf{4} \otimes (\mathbf{2} \otimes \mathbf{5} )= \mathbf{4}\otimes (\mathbf{4}\oplus \mathbf{6})   & 1& \textrm{\,Op\,} (\mathbf{4}^2),\\
     \widetilde{\mathcal{H}}\bar{\psi}_{2,L} \psi_{5,R}&: \mathbf{4} \otimes (\mathbf{2} \otimes \mathbf{5 }) = \mathbf{4} \otimes (\mathbf{4} \oplus \mathbf{6}) &  1&\textrm{\,Op\,} (\mathbf{4}^4),\\
    \mathcal{T} \mathcal{H}\bar{\psi}_{2,L}  \psi_{5,R}& : \mathbf{3} \otimes \mathbf{4} \otimes \mathbf{2} \otimes \mathbf{5} = (\mathbf{2} \oplus \mathbf{4} \oplus \mathbf{6}) \otimes ( \mathbf{4} \oplus \mathbf{6} \oplus \mathbf{8}) & 2& \textrm{\,Ops\,} (\mathbf{4}^2,\mathbf{6}^2),
     \\
    \mathcal T \widetilde{\mathcal{H}}\bar{\psi}_{2,L}\mathcal  \psi_{5,R}&: \mathbf{3} \otimes \mathbf{4} \otimes \mathbf{2} \otimes \mathbf{5} = (\mathbf{2} \oplus \mathbf{4} \oplus \mathbf{6}) \otimes ( \mathbf{4} \oplus \mathbf{6} \oplus \mathbf{8}) & 2& \textrm{\,Ops\,} (\mathbf{4}^2,\mathbf{6}^2),\\
     \mathcal T^2\bar{\psi}_{1,L} \mathcal \psi_{5,R}&: \mathbf{3}^2 \otimes \mathbf{1} \otimes \mathbf{5} = (\mathbf{1}  \oplus \mathbf{5}) \otimes \mathbf{5} & 1 &\textrm{\,Op\,} (\mathbf{5}^2 ),
\end{align}
where we have taken into account that $\mathcal T$ is a real field and in $\mathbf{3}^2$ only the symmetric representations are non-vanishing. This amounts to 7 operators at dimension 5 and hence 7 free parameters, on the other hand we have a total of 5 massive quarks to account for, $u,d,U,D,X$.

For the explicit construction of any of these operators one can combine at every stage any two fields with the corresponding Clebsch Gordan and repeat until the singlet is obtained. In order to simplify this procedure it is simpler to treat $\bar\psi_L\psi_R$ as the conjugate of a representation, $R^\dagger_\psi$, and the scalar combination of the same dimension $R'_S$ so that the invariant reads $R^\dagger_\psi R'_S$. The fermionic representation is, borrowing spin-group notation
\begin{align}
    R_{\psi,m}^J= \sum _{m_i}C_{m}^{m_1,m_2}(J,j_1,j_2) \bar\psi_{j_1,R,m_3}\varepsilon_{m_3,m_1} \psi_{j_2,L,m_2} ,
\end{align}
where we first tilded $\bar\psi_R$ back to a regular representation and then combined with $\psi_L$. This is to be combined with the scalar fields to finally form a singlet in the form of eq.~\eqref{eq:vareps}, as an example
\begin{align}
   \bar\psi_{\mathbf{2},L}\mathcal H\psi_{\mathbf{5},R}\equiv\sum_{m_i} (C^{m_4}_{m1,m_2})^*\bar\psi_{\mathbf{2},L,m_2}\psi_{\mathbf{5},R,m_3}\varepsilon_{m_3,m_1}\mathcal H_{m_4}.
\end{align}
The explicit combination of Clesbsch Gordan for the fermionic half of the operator, in the case of a resulting $\mathbf{4}$ representation, dubbed $R_{\psi, m}^\mathbf{4}$, reads
\begin{align}\label{CGC4-1}
    m= \frac{3}{2} & : -\sqrt{\frac{4}{5}}\underbrace{(\bar{\psi}_{5,R}\varepsilon)_2(\psi_{2,L})_{-\frac{1}{2}}}_{\bar{U}_R \, Q_L} + \sqrt{\frac{1}{5}}\underbrace{(\bar{\psi}_{5,R}\varepsilon)_1({\psi}_{2,L})_{\frac{1}{2}}}_{-\bar{d}_R \, q_L}, \\
  m= \frac{1}{2} & : -\sqrt{\frac{3}{5}}\underbrace{ (\bar{\psi}_{5,R}\varepsilon)_1({\psi}_{2,L})_{-\frac{1}{2}}}_{-\bar{d}_R \, Q_L} + \sqrt{\frac{2}{5}}\underbrace{(\bar{\psi}_{5,R}\varepsilon)_0({\psi}_{2,L})_{\frac{1}{2}} }_{\bar{X}_R \, q_L},\\
  m=- \frac{1}{2} & : \sqrt{\frac{3}{5}}\underbrace{ (\bar{\psi}_{5,R}\varepsilon)_{-1}({\psi}_{2,L})_{\frac{1}{2}}}_{{-\bar{D}_R \, q_L}}- \sqrt{\frac{2}{5}}\underbrace{(\bar{\psi}_{5,R}\varepsilon)_0({\psi}_{2,L})_{-\frac{1}{2}} }_{{\bar{X}_R \, Q_L}}, \\
 m=- \frac{3}{2} & : \sqrt{\frac{4}{5}}\underbrace{(\bar{\psi}_{5,R}\varepsilon)_{-2}({\psi}_{2,L})_{\frac{1}{2}} }_{\bar{u}_R \, q_L}- \sqrt{\frac{1}{5}}\underbrace{(\bar{\psi}_{5,R}\varepsilon)_{-1}({\psi}_{2,L})_{-\frac{1}{2}}}_{ -\bar{D}_R \, Q_L},
\label{CGC4-4}
\end{align}
where on the LHS we give $+\frac{3}{2},+\frac{1}{2},-\frac{1}{2},-\frac{3}{2}$, the eigenvalues of the Cartan generator $\tilde T_{Y}/2$, of $SU(2)_Y$ and the RH  field $\bar \psi_R$ is tilded so as to transform as a $\mathbf{5}$ hence the sings arising from $\varepsilon$ in the underbraces. The invariant Yukawa term is then finally produced by adding the complex conjugate of each $m$ entry with the scalar representation of the same $m$.

 After both the triplet $\mathcal{T}$ and $\mathcal H$ take a vev
\begin{align}
    \langle \mathcal T\rangle \equiv (0, t,0), \qquad\langle \mathcal H\rangle=\begin{pmatrix}
        0&0&0&0\\
        0&0&0&v/\sqrt{2}
    \end{pmatrix},
\end{align} one has that the following dimension 4 operator produces a mass term
\begin{align}
     \bar{\psi}_{2,L}\langle \mathcal{H} \rangle\psi_{5,R} =-\frac{2}{\sqrt{5}}\bar Q_L\langle H\rangle U_R-\frac{1}{\sqrt{5}}\bar q_L\langle H\rangle d_R=-\frac{v}{\sqrt{10}}(2\bar U_L U_R+\bar d_L d_R)\,.\label{eq:RenMass}
\end{align}
This term alone gives the new particles $U$ twice the mass of $d$, this factor being set by the Clebsch-Gordan coefficients. This mass correlation is not viable phenomenologically which is the reason why one has to go to the non-renormalisable level to obtain a realistic model or, as shown later, modify the scalar spectrum for a renormalisable theory. Of the different dimension-5 operators one finds that the $\mathbf{4}$ rep that one builds out of the combination of $\mathcal{T}$ and $\mathcal{H}$, let us call it $(\mathcal{T}\mathcal{H})_\mathbf{4}$, combines with the same ferminon bilinear $\mathbf{4}$ of eqs.~\eqref{CGC4-1}-\eqref{CGC4-4} and hence after symmetry breaking gives mass to the very same linear combination in eq.~\eqref{eq:RenMass}.
 In practice then the operator $\bar \psi \mathcal H\psi$ and $\bar \psi(\mathcal{T}\mathcal{H})_4\psi$ become degenerate in their contribution to masses. The $\mathbf{6}$ combination will break the degeneracy, one has the fermion bilinear $R^\mathbf{6}_{\psi,m}$ is
\begin{align}
    m=\frac{5}{2} &: \underbrace{(\bar\psi_{5,R}\varepsilon)_{2}({\psi}_{2,L})_{\frac{1}{2}}}_{\bar{U}_Rq_L}, \\
       m=\frac{3}{2} & : \sqrt{\frac{1}{5}}\underbrace{(\bar{\psi}_{5,R}\varepsilon)_2(\psi_{2,L})_{-\frac{1}{2}}}_{\bar{U}_R \, Q_L} + \sqrt{\frac{4}{5}}\underbrace{(\bar{\psi}_{5,R}\varepsilon)_1({\psi}_{2,L})_{\frac{1}{2}}}_{{-}\bar{d}_R \, q_L}, \\
  m=\frac{1}{2} & : \sqrt{\frac{2}{5}}\underbrace{ (\bar{\psi}_{5,R}\varepsilon)_1({\psi}_{2,L})_{-\frac{1}{2}}}_{{-}\bar{d}_R \, Q_L} + \sqrt{\frac{3}{5}}\underbrace{(\bar{\psi}_{5,R}\varepsilon)_0({\psi}_{2,L})_{\frac{1}{2}} }_{\bar{X}_R \, q_L},\\
  m=- \frac{1}{2} & : \sqrt{\frac{2}{5}}\underbrace{ (\bar{\psi}_{5,R}\varepsilon)_{-1}({\psi}_{2,L})_{\frac{1}{2}}}_{{{-}\bar{D}_R \, q_L}}+ \sqrt{\frac{3}{5}}\underbrace{(\bar{\psi}_{5,R}\varepsilon)_0({\psi}_{2,L})_{-\frac{1}{2}} }_{{\bar{X}_R \, Q_L}}, \\
 m=- \frac{3}{2} & : \sqrt{\frac{1}{5}}\underbrace{(\bar{\psi}_{5,R}\varepsilon)_{-2}({\psi}_{2,L})_{\frac{1}{2}} }_{\bar{u}_R \, q_L}+ \sqrt{\frac{4}{5}}\underbrace{(\bar{\psi}_{5,R}\varepsilon)_{-1}({\psi}_{2,L})_{-\frac{1}{2}}}_{ {-}\bar{D}_R \, Q_L}, \\
   m= -\frac{5}{2} &: \underbrace{(\bar\psi_{5,R}\varepsilon)_{-2}({\psi}_{2,L})_{-\frac{1}{2}}}_{\bar{u}_R Q_L}, 
\end{align}
combined with $(\mathcal{T}\mathcal{H})_\mathbf{6}$ it provides an independent linear combination of $\bar U_L U_R$ and $\bar d_L d_R$ and a de-correlation of masses. Let us then give the mass Lagrangian for coloured particles
\begin{align}
    -\mathcal{L}_{\textrm{Yuk},Q} =& y_q \bar{\psi}_{2,L} \mathcal{H} \psi_{5,R} + \sum_{n=\mathbf{4,6}}\frac{y'_{q,n}}{\Lambda}\bar{\psi}_{2,L} (\mathcal{T} \mathcal{H})_{n} \psi_{5,R}+ \frac{y_X}{\Lambda}\bar X_L \mathcal{T}^2\psi_{5,R}\\ \nonumber
    &+\tilde y_q \bar{\psi}_{2,L}\widetilde{\mathcal{H}} \psi_{5,R} + \sum_{n=\mathbf{4,6}}\frac{\tilde y'_{q,n}}{\Lambda}\bar{\psi}_{2,L} (\mathcal{T} \widetilde{\mathcal{H}})_{n} \psi_{5,R}+h.c. \,,
\end{align}
which leads to masses as
 \begin{align}
    \begin{pmatrix}
        m_d\\
        m_{U}
    \end{pmatrix}&=\frac{v}{\sqrt{2}}\frac{1}{\sqrt{5}}
    \begin{pmatrix}
        -1& -2  \\
        -2& 1
    \end{pmatrix}
    \begin{pmatrix}
        \underline{y}_{q,\mathbf{4}}\\
        \underline{y}_{q,\mathbf{6}}
    \end{pmatrix}\,, & &\begin{array}{l}
         \underline{y}_{q,\mathbf{4}}\equiv y_q+\sqrt{\frac{3}{5}}\frac{t}{\Lambda}y'_{q,\mathbf{4}}  \\
          \underline{y}_{q,\mathbf{6}}\equiv \sqrt{\frac{2}{5}}\frac{t}{\Lambda}y'_{q,\mathbf{6}}
    \end{array}\,,\\
        \begin{pmatrix}
        m_u\\
        m_{D}
    \end{pmatrix}&=\frac{v^*}{\sqrt{2}}\frac{1}{\sqrt{5}}
    \begin{pmatrix}
        2& 1  \\
        1& -2
    \end{pmatrix}
    \begin{pmatrix}
        \underline{\tilde y}_{q,\mathbf{4}}\\
        \underline{\tilde y}_{q,\mathbf{6}}
    \end{pmatrix}\,,  & &\begin{array}{l}
         \underline{\tilde y}_{q,\mathbf{4}}\equiv \tilde y_q-\sqrt{\frac{3}{5}}\frac{t}{\Lambda}\tilde y'_{q,\mathbf{4}}  \\
          \underline{\tilde y}_{q,\mathbf{6}}\equiv \sqrt{\frac{2}{5}}\frac{t}{\Lambda}\tilde y'_{q,\mathbf{6}}
    \end{array}\,,
 \\    m_X&=\sqrt{\frac{2}{3}}\frac{y_X t^2}{\Lambda}\,. &&
\end{align}
Please note we chose matrix notation above to show the de-correlation between masses as evidenced by linearly independent rows in the matrices but there is no mixing, in fact is not allowed since they each have different electric charges. At the level of mass terms the 7 free couplings collapse into 5 yet these are enough to produce 5 uncorrelated masses for all coloured fermions. Fitting to the right spectrum then requires a rather precise cancellation between $y_{q,\mathbf{4}}$ and $y_{q,\mathbf{6}}$, but more relevant than fine tuning is the feature of masses of the new fermions being bounded from above. Indeed $t/\Lambda \leq1$ is demanded by EFT consistency and perturbative unitarity sets $y\leq 4\pi$ so the new fermions have at most a few TeV mass.

The lepton case requires now generating masses for all SM and BSM leptons, 8 of them. The operators needed for fully-decorrelated masses now reach to dimension 6, explicitly
\begin{align}
   \mathcal{H} \bar{\psi}_{4,L}  \psi_{7,R} & : \mathbf4 \otimes \mathbf4 \otimes \mathbf7 = \mathbf4\otimes(\mathbf4\oplus\mathbf{6}\oplus \mathbf{8}\oplus\mathbf{10})  & 1 &\textrm{\,Op\,}, 
   \\
    \widetilde{\mathcal{H}} \bar{\psi}_{4,L}  \psi_{7,R}  &: \mathbf4 \otimes \mathbf4 \otimes \mathbf7 = \mathbf4\otimes(\mathbf4\oplus\mathbf{6}\oplus \mathbf{8}\oplus\mathbf{10})  & 1 &\textrm{\,Op\,},
    \\
    \mathcal{T} \mathcal{H} \bar{\psi}_{4,L} \psi_{7,R} & : \mathbf4 \otimes \mathbf3 \otimes \mathbf4 \otimes \mathbf7= (\mathbf2 \oplus \mathbf4 \oplus \mathbf6) \otimes(\mathbf4\oplus\mathbf{6}\oplus \mathbf{8}\oplus\mathbf{10}) & 2& \textrm{\,Ops\,}, 
     \\
     \mathcal{T} \widetilde{\mathcal{H}}  \bar{\psi}_{4,L} \psi_{7,R} &: \mathbf4 \otimes \mathbf3 \otimes \mathbf4 \otimes \mathbf7= (\mathbf2 \oplus \mathbf4 \oplus \mathbf6) \otimes(\mathbf4\oplus\mathbf{6}\oplus \mathbf{8}\oplus\mathbf{10}) & 2& \textrm{\,Ops\,}, 
     \\
    \mathcal{T}^2 \mathcal{H}  \bar{\psi}_{4,L} \psi_{7,R} & : \mathbf{3}^2\otimes\mathbf{4}  \otimes \mathbf{4} \otimes \mathbf{7} =(\mathbf{4}\oplus \mathbf{2}\oplus\mathbf{4}\oplus\mathbf{6}\oplus \mathbf{8})\otimes(\mathbf4\oplus\mathbf{6}\oplus \mathbf{8}\oplus\mathbf{10})& 4& \textrm{\,Ops\,}, 
    \\
   \mathcal{T}^2 \widetilde{\mathcal{H}}  \bar{\psi}_{4,L} \psi_{7,R} & : \mathbf{3}^2\otimes\mathbf{4}  \otimes \mathbf{4} \otimes \mathbf{7} =(\mathbf{4}\oplus \mathbf{2}\oplus\mathbf{4}\oplus\mathbf{6}\oplus \mathbf{8})\otimes(\mathbf4\oplus\mathbf{6}\oplus \mathbf{8}\oplus\mathbf{10})& 4& \textrm{\,Ops\,}, 
\end{align}
whereas for the other fermion bilinear
\begin{align}
 \mathcal H  \bar \psi_{4,L} \nu_R &: \mathbf{4} \otimes \mathbf{4} \otimes \mathbf{1} &1 &\textrm{\,Op\,}, 
 \\
  \widetilde{\mathcal H}  \bar \psi_{4,L} \nu_R &: \mathbf{4} \otimes \mathbf{4} \otimes \mathbf{1} &1 &\textrm{\,Op\,}, 
  \\
  \mathcal T\mathcal H  \bar \psi_{4,L} \nu_R &: \mathbf{3} \otimes\mathbf{4} \otimes \mathbf{4} \otimes \mathbf{1}=(\mathbf2 \oplus \mathbf4 \oplus \mathbf6) \otimes\mathbf4 & 1& \textrm{\,Op\,}, 
 \\
  \mathcal T\widetilde{\mathcal H}  \bar \psi_{4,L} \nu_R &: \mathbf{3} \otimes\mathbf{4} \otimes \mathbf{4} \otimes \mathbf{1}=(\mathbf2 \oplus \mathbf4 \oplus \mathbf6) \otimes \mathbf4 &1 &\textrm{\,Op\,}, 
  \\
  \mathcal T^2\mathcal H  \bar \psi_{4,L} \nu_R &: \mathbf{3}^2 \otimes\mathbf{4} \otimes \mathbf{4} \otimes \mathbf{1} =(\mathbf{4}\oplus \mathbf{2}\oplus\mathbf{4}\oplus\mathbf{6}\oplus \mathbf{8})\otimes\mathbf{4}&2 &\textrm{\,Ops\,}, 
 \\
  \mathcal T^2\widetilde{\mathcal H}  \bar \psi_{4,L} \nu_R &: \mathbf{3}^2 \otimes\mathbf{4} \otimes \mathbf{4} \otimes \mathbf{1}=(\mathbf{4}\oplus \mathbf{2}\oplus\mathbf{4}\oplus\mathbf{6}\oplus \mathbf{8})\otimes\mathbf{4} &2&\textrm{\,Ops\,}. 
  \end{align}
These amount to more than 8 operators yet our study of quarks tells us that all combinations of scalars $\mathcal{T}^n\mathcal{H}$ that result in the same representation will become degenerate when it comes to generating mass terms. In order to agree with observation however we need not have all masses de-correlated, only for the SM quarks to be lighter than the rest; this in fact can be achieved with operators only up to dimension 5 as in the quark case. Let us show how next. 

For the lepton bilinear let us give the relevant elements after breaking:\\
i) for the $\mathbf{4}$ combination and $m=3/2$
\begin{align}
  -\sqrt{\frac{4}{7}} \underbrace{ (\bar{\psi}_{7,R})_3({\psi}_{4,L})_{-\frac{3}{2}} }_{\bar{e}_R\ell_L} +  \sqrt{\frac{2}{7}}\underbrace{ (\bar{\psi}_{7,R})_2 ({\psi}_{4,L})_{-\frac{1}{2}} }_{-\bar{E}'_R L''_L} - \sqrt{\frac{4}{35}} \underbrace{ (\bar{\psi}_{7,R})_1 ({\psi}_{4,L})_{\frac{1}{2}}}_{\bar{E}''_R L''_R} + \sqrt{\frac{1}{35}} \underbrace{(\bar{\psi}_{4,L})_{\frac{3}{2}} ({\psi}_{7,R})_0 }_{-\bar{\nu}'_R L'''_L}, \end{align}
  while the $m=-3/2$
 \begin{align}
 \sqrt{\frac{4}{7}} \underbrace{ (\bar{\psi}_{7,R})_{-3} ({\psi}_{4,L})_{\frac{3}{2}}}_{\bar{N}'''_R L'''_L} - \sqrt{\frac{2}{7}} \underbrace{(\bar{\psi}_{7,R})_{-2}({\psi}_{4,L})_{\frac{1}{2}}  }_{-\bar{N}''_R L''_L}+ \sqrt{\frac{4}{35}} \underbrace{(\bar{\psi}_{4,L})_{-\frac{1}{2}} ({\psi}_{7,R})_{-1} }_{\bar{N}'_R L'_L}  -\sqrt{\frac{1}{35}} \underbrace{ (\bar{\psi}_{7,R})_0({\psi}_{4,L})_{-\frac{3}{2}} }_{-\bar{\nu}'_R \ell_L},
\end{align}
ii) for the $\mathbf{6}$ combination and $m=3/2$
\begin{align}
    \frac{3}{2\sqrt{7}} \underbrace{ (\bar{\psi}_{7,R})_3({\psi}_{4,L})_{-\frac{3}{2}} }_{\bar{e}_R\ell_L} +  \sqrt{\frac{1}{14}}\underbrace{ (\bar{\psi}_{7,R})_2 ({\psi}_{4,L})_{-\frac{1}{2}} }_{-\bar{E}'_R L''_L} - \sqrt{\frac{7}{20}} \underbrace{ (\bar{\psi}_{7,R})_1 ({\psi}_{4,L})_{\frac{1}{2}}}_{\bar{E}''_R L''_R} + \frac{3}{\sqrt{35}} \underbrace{(\bar{\psi}_{4,L})_{\frac{3}{2}} ({\psi}_{7,R})_0 }_{-\bar{\nu}'_R L'''_L},\end{align}
  while the $m=-3/2$
 \begin{align}
 \frac{3}{2\sqrt{7}} \underbrace{ (\bar{\psi}_{7,R})_{-3} ({\psi}_{4,L})_{\frac{3}{2}}}_{\bar{N}'''_R L'''_L} + \sqrt{\frac{1}{14}} \underbrace{(\bar{\psi}_{7,R})_{-2}({\psi}_{4,L})_{\frac{1}{2}}  }_{-\bar{N}''_R L''_L} - \sqrt{\frac{7}{20}} \underbrace{(\bar{\psi}_{4,L})_{-\frac{1}{2}} ({\psi}_{7,R})_{-1} }_{\bar{N}'_R L'_L}  +\frac{3}{\sqrt{35}} \underbrace{ (\bar{\psi}_{7,R})_0({\psi}_{4,L})_{-\frac{3}{2}} }_{-\bar{\nu}'_R \ell_L} ,
\end{align}
with a single operator correlating the mass of 4 different fermions. 

The following Lagrangian contains all Yukawa terms up to and including dimension 5
\begin{align}\nonumber
    -\mathcal{L}_{\textrm{Yuk},L} =& y_\ell \bar{\psi}_{4,L} \mathcal{H} \psi_{7,R}+y_{\nu}^0\bar\psi_{4,L} \mathcal{H}\nu_R +\frac{y_{\nu}'}{\Lambda}\bar\psi_{4,L} \mathcal{T H}\nu_R+ \sum_{n=4,6}\frac{y'_{\ell,n}}{\Lambda}\bar{\psi}_{4,L} (\mathcal{T} \mathcal{H})_{n} \psi_{7,R}+
    h.c.\\ \nonumber
     &+\tilde y_\ell \bar{\psi}_{4,L} \mathcal{H} \psi_{7,R} +\tilde y_{\nu}^0\bar\psi_{4,L} \widetilde{\mathcal{H}}\nu_R+\frac{\tilde y_{\nu}'}{\Lambda}\bar\psi_{4,L} \mathcal T\widetilde{\mathcal{H}}\nu_R+ \sum_{n=4,6}\frac{\tilde y'_{\ell,n}}{\Lambda}\bar{\psi}_{4,L} (\mathcal{T} \mathcal{H})_{n} \psi_{7,R}
   +h.c.,
\end{align}
and the masses for the lepton sector are
\begin{align}
   & \begin{pmatrix}
        m_{N'}\\
        m_{N''}\\
        m_{N'''}
    \end{pmatrix}=\frac{v}{\sqrt{2}}\frac{1}{\sqrt{7}}\begin{pmatrix}
        \frac{2}{\sqrt{5}}&-\frac{7\sqrt{5}}{10}\\
        \sqrt{2}&-\frac{1}{\sqrt{2}}\\
        2&\frac{3}{2}
    \end{pmatrix}\begin{pmatrix}
        \underline{\tilde y}_{\ell,\mathbf{4}}\\
        \underline{\tilde y}_{\ell,\mathbf{6}}\\
    \end{pmatrix}, & 
 &\begin{array}{l}
         \underline{\tilde y}_{\ell,\mathbf{4}}\equiv \tilde y_\ell-\sqrt{\frac{3}{5}}\frac{t}{\Lambda}\tilde y'_{\ell,\mathbf{4}}  \\
          \underline{\tilde y}_{\ell,\mathbf{6}}\equiv \sqrt{\frac{2}{5}}\frac{t}{\Lambda}\tilde y'_{\ell,\mathbf{6}}
    \end{array},
    \\
    &\begin{pmatrix}\nonumber
        m_{e}\\
        m_{E'}\\
        m_{E''}
    \end{pmatrix}=\frac{v}{\sqrt{2}}\frac{1}{\sqrt{7}}\begin{pmatrix}
        -2&\frac{3}{2}\\
        -\sqrt{2}&-\frac{1}{\sqrt{2}}\\
        -\frac{2}{\sqrt{5}}&-\frac{7\sqrt{5}}{10}
    \end{pmatrix}\begin{pmatrix}
        \underline{ y}_{\ell,\mathbf{4}}\\
        \underline{ y}_{\ell,\mathbf{6}}
    \end{pmatrix},&&
    \begin{array}{l}
         \underline{y}_{\ell,\mathbf{4}}\equiv y_\ell+\sqrt{\frac{3}{5}}\frac{t}{\Lambda}y'_{\ell,\mathbf{4}}  \\
          \underline{y}_{\ell,\mathbf{6}}\equiv \sqrt{\frac{2}{5}}\frac{t}{\Lambda}y'_{\ell,\mathbf{6}}
    \end{array},
\end{align}
whereas for neutral leptons we have
\begin{align}
    \begin{pmatrix}
          \nu_L\\
        \nu'_L
    \end{pmatrix}^\dagger M_{\nu\nu'}
    \begin{pmatrix}
          \nu_R\\
        \nu'_R
    \end{pmatrix},& &M_{\nu\nu'}&=\frac{v}{\sqrt{2}}\begin{pmatrix}
        \tilde y_{\nu}&\underline{\tilde y}_{\nu'}\\
        y_{\nu}&\underline{y}_{\nu'}
    \end{pmatrix}, & &\begin{array}{l}
          \tilde{y}_{\nu'}=\frac{1}{\sqrt{35}}\underline{\tilde y}_{\ell,\mathbf{4}}-\frac{3}{\sqrt{35}}\underline{\tilde y}_{\ell,\mathbf{6}}\\
          y_{\nu'}=-\frac{1}{\sqrt{35}}\underline{y}_{\ell,\mathbf{4}}-\frac{3}{\sqrt{35}}\underline{y}_{\ell,\mathbf{6}}\\
          \tilde y_\nu=\tilde y_\nu^0-\sqrt{\frac{3}{5}}\frac{t}{\Lambda}\tilde y_\nu'\\
           y_\nu=y_\nu^0+\sqrt{\frac{3}{5}}\frac{t}{\Lambda} y_\nu'.
    \end{array}
\end{align}

A spectrum compatible with observations and reproducing the SM masses can be achieved as 
\begin{align}
    y_\nu&=0, & \tilde y_{\nu'}&=0  \Rightarrow 3\underline{\tilde y}_{\ell,\mathbf{6}}=\underline{\tilde y}_{\ell,\mathbf{4}}, & m_\nu&=\frac{v\tilde y_{\nu}}{\sqrt{2}},
    \\ m_{N'}&=\frac{v}{\sqrt{14}}\frac{\sqrt{5}}{6}\underline{ y}_{\ell,\mathbf{4}},&\frac{m_{N''}}{m_{N'}}&=\sqrt{10}, & \frac{m_{N'''}}{m_{N''}}&=\frac{3}{\sqrt{2}}.\label{eq:ratioN}
\end{align}
This choice cancels $\nu-\nu'$ mixing and gives active neutrinos a Dirac mass. Finally,
\begin{align}
   \frac{3}{2}\underline{y}_{\ell,\mathbf{6}}=&\frac{\sqrt{14}m_e}{v}+2\underline{y}_{\ell,\mathbf{4}},&
    m_{E'}&=\frac{v}{\sqrt{14}}\frac{5\sqrt{2}}{3}\underline{ y}_{\ell,\mathbf{4}} +\mathcal O\left(\frac{m_e}{v\underline{ y}_{\ell,\mathbf{4}}}\right),
   \end{align}
   a choice that predicts the following rations assuming a hierarchy of SM vs BSM masses $m_e/m_E\sim m_e/(v y_\ell)\ll1$
   \begin{align}
    \frac{m_{E''}}{m_{E'}}&=\frac{2\sqrt{2}}{\sqrt{5}}+\mathcal O\left(\frac{m_e}{v\underline{ y}_{\ell,\mathbf{4}}}\right),&\frac{m_{\nu'}}{m_{E''}}&=\frac{3}{4}+\mathcal O\left(\frac{m_e}{v\underline{ y}_{\ell,\mathbf{4}}}\right).\label{eq:ratioE}
   \end{align}

As is the case for coloured particles there is a need for non-renormalisable terms to make this a phenomenologically viable model; mass terms are of the order $yv$ or $yv (t/\Lambda)^p$ and for the EFT to apply we have $t/\Lambda<1$ so perturbative unitarity sets an upper bound on the masses of the new fermions of $\sim 4\pi v$.  A renormalisable alternative will be discussed momentarily.

This Lagrangian has been built with the assumption of Lepton and Baryon number conservation, the consequences of not assuming this are discussed below and include a naively vector-like fermion lepton sector in disagreement with observations. On the flavour sector, largely omitted here since it is not necessary to discuss gauge embeddings, let us mention that the same discussion as above follows with every Yukawa $y$ promoted to a complex $3\times 3$ matrix.

In summary, the Lagrangian for this model is
\begin{align}
    \mathcal L= &-\frac{1}{4}\sum_i F^i_{\mu\nu}F^{i,\mu\nu} + \textrm{Tr}\left(D_\mu \mathcal H^\dagger D^\mu \mathcal H\right)+ \frac12 D_\mu \mathcal T^\dagger D^\mu \mathcal T+i\sum_n\bar\psi_{n}\gamma^\mu D_\mu \psi_n \notag \\ &+\mathcal L_{\textrm{Yuk},Q}+\mathcal L_{\textrm{Yuk},L}-V(\mathcal H,\mathcal T).
\end{align}
For the new states one has new gauge bosons and a $T$ Higgs as well as the coloured fermion $X$ with masses around $t$ whereas the rest should have masses below $4\pi v$. 

\medskip

 \noindent$\bullet$~{\bf A renormalisable SU2Y model}. Let us propose here a modification that produces a UV complete model. It requires adding a new scalar representation $\Phi$ and replacing the scalar $\mathcal T$ for a new scalar $S$ as shown in Table ~\ref{tab:BosonUV} so that the total scalar field content is
 \begin{align}
     \mathcal H\,,&&\Phi&=\left(\phi_{-3}\,,\,\phi_{-2}\,,\,\phi_{-1}\,,\,\phi_{1}\,,\,\phi\,,\,\phi_2\right)\,, & 
     S=&\left(S_{-4},S_{-2},S_0,S_{2},S_{4}\right)\,,
 \end{align}
 this allows for the renormalisable Lagrangian
\begin{align}
    \mathcal L= &-\frac{1}{4}\sum_i F^i_{\mu\nu}F^{i,\mu\nu} + \textrm{Tr}\left(D_\mu \mathcal H^\dagger D^\mu \mathcal H\right)+\textrm{Tr}\left(D_\mu \Phi^\dagger D^\mu  \Phi\right)+\frac12 D_\mu \mathcal S^\dagger D^\mu \mathcal S\notag \\ &+i\sum_n\bar\psi_{n}\gamma^\mu D_\mu \psi_n +\mathcal L_{Q}+\mathcal L_{L}-V(\mathcal H,\Phi,S),
\end{align}
with
\begin{align}\nonumber
    \mathcal L_{Q}=&y_{q,\mathbf{4}} \bar{\psi}_{2,L} \mathcal{H} \psi_{5,R} + y_{q,\mathbf{6}}\bar{\psi}_{2,L} \Phi \psi_{5,R}+ y_X\bar X_L S \psi_{5,R}\\
    &+\tilde y_{q,\mathbf{6}} \bar{\psi}_{2,L}\widetilde{\mathcal{H}} \psi_{5,R} +\tilde y_{q,\mathbf{6}}\bar{\psi}_{2,L} \Phi \psi_{5,R}+h.c.\,,\\\nonumber
    \mathcal L_{L}=&
    \tilde y_{\ell,\mathbf{4}} \bar{\psi}_{4,L} \mathcal{H} \psi_{7,R}+ \tilde y_{\ell,\mathbf{6}}\bar{\psi}_{4,L} \Phi \psi_{7,R} +\tilde y_{\nu}\bar\psi_{4,L} \widetilde{\mathcal{H}}\nu_R\\
     &
      +y_{\ell,\mathbf{4}} \bar{\psi}_{4,L} \mathcal{H} \psi_{7,R} + y'_{\ell,\mathbf{6}}\bar{\psi}_{4,L} \Phi \psi_{7,R}+
    h.c.\,.
\end{align}
 The field $S$ supplants $\mathcal T$ and plays the same role of breaking $SU(2)_Y\to U(1)_Y$ while $\Phi$ does get a vev too so that the vevs read
 \begin{align}
         \langle  S\rangle \equiv (0,0,t,0,0), \qquad\langle \mathcal H\rangle=\begin{pmatrix}
        0&0&0&0\\
        0&0&0&v_1/\sqrt{2}
    \end{pmatrix}, \qquad \langle\Phi\rangle=\begin{pmatrix}
        0&0&0&0&0&0\\
        0&0&0&0&v_2/\sqrt{2}&0
    \end{pmatrix},
 \end{align}
 where $v_1^2+v_2^2=v^2$ and we have that the low energy description is a two Higgs doublet model with $H,\phi$.
\begin{table}[h]
    \centering
    \begin{tabular}{c||c||c}
                $SU(3)_c\times SU(2)_L \times SU(2)_Y$  &  $(\mathbf{1},\mathbf{1},\mathbf{5})$  &  $(\mathbf{1},\mathbf{2},\mathbf{6})$ \\ \hline
        \hline
        BSM fields  &  $S^{-4},S^{-2},S^0,S^{2},S^{4}$  & 
       $\phi_{-5}\,,\,\phi_{-3}\,,\,\phi_{-1}\,,\,\phi_{1}\,,\,\phi\,,\,\phi_5$  \\
         $n_6$  &  
         2,\,\,\,4,\,\,\,0,\,\,\,2,\,\,\,4 & 4,\,\,\, 0, \,\,\,2,\,\,\,\,\,\, 4,\,\,\,\,\,\,0,\,\,\,\,\,\,2
         \\ 
    \end{tabular}
    \caption{Scalar field content (in addition to $\mathcal{H}$) in the UV complete SU2Y model.}
    \label{tab:BosonUV}
\end{table}

The mass terms for quarks after the breaking are
\begin{align}
    \begin{pmatrix}
        m_d\\
        m_{U}
    \end{pmatrix}&=\frac{1}{\sqrt{2}}\frac{1}{\sqrt{5}}
    \begin{pmatrix}
        -1& -2  \\
        -2& 1
    \end{pmatrix}
    \begin{pmatrix}
        v_1{y}_{q,\mathbf{4}}\\
        v_2{y}_{q,\mathbf{6}}
    \end{pmatrix}, &
        \begin{pmatrix}
        m_u\\
        m_{D}
    \end{pmatrix}&=\frac{1}{\sqrt{2}}\frac{1}{\sqrt{5}}
    \begin{pmatrix}
        2& 1  \\
        1& -2
    \end{pmatrix}
    \begin{pmatrix}
        v_1{\tilde y}_{q,\mathbf{4}}\\
        v_2{\tilde y}_{q,\mathbf{6}}
    \end{pmatrix},
 \\    m_X&=y_Xt, &&
\end{align}
whereas for leptons we choose $3v_2{\tilde y}_{\ell,\mathbf{6}}={\tilde y}_{\ell,\mathbf{4}}v_1$
and
 \begin{align}
    m_\nu&=\frac{v_1\tilde y_{\nu}}{\sqrt{2}},
    & m_{N'}&=\frac{v_1}{\sqrt{14}}\frac{\sqrt{5}}{6}{ y}_{\ell,\mathbf{4}},&\frac{m_{N''}}{m_{N'}}&=\sqrt{10}, & \frac{m_{N'''}}{m_{N''}}&=\frac{3}{\sqrt{2}},
\end{align}
again with Dirac-mass for neutrinos, and finally
\begin{align}
   \frac{3}{2}v_2{y}_{\ell,\mathbf{6}}=&\frac{\sqrt{14}m_e}{v}+2v_1{y}_{\ell,\mathbf{4}}, & 
    m_{E'}&=\frac{v_1}{\sqrt{14}}\frac{5\sqrt{2}}{3}{ y}_{\ell,\mathbf{4}} +\mathcal O\left(\frac{m_e}{v{ y}_{\ell,\mathbf{4}}}\right),
   \end{align}
   a choice that produces the rations (assuming $m_e/m_E\sim m_e/(v y_\ell)\ll1$)
   \begin{align} \frac{m_{E''}}{m_{E'}}&=\frac{2\sqrt{2}}{\sqrt{5}}+\mathcal O\left(\frac{m_e}{v{ y}_{\ell,\mathbf{4}}}\right),&\frac{m_{\nu'}}{m_{E''}}&=\frac{3}{4}+\mathcal O\left(\frac{m_e}{v{ y}_{\ell,\mathbf{4}}}\right).
   \end{align}
   
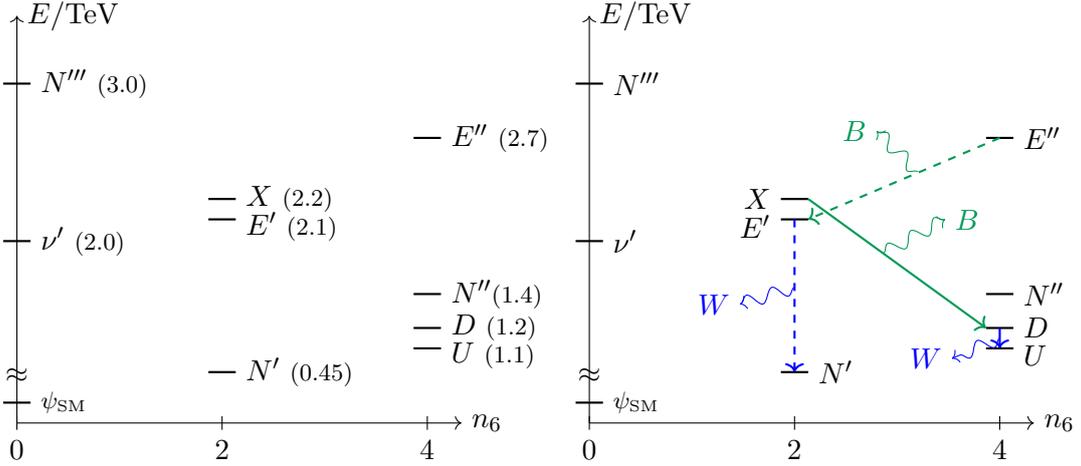
\begin{figure}[h!]
    \centering 
\begin{tikzpicture}[scale=0.9]

\draw[->] (0,0) -- (0,6) node[right] {$E/\text{TeV}$};
\draw[->] (0,0) -- (6.5,0) node[right] {$n_6$};

\foreach \x \label in {0/0, 3/2, 6/4} {
    \draw (\x,0.1) -- (\x,-0.1) node[below] {\label};
}

\draw[thick] (5.8,4.2) -- (6.2,4.2);
\node[right] at (6.2,4.2) {$E''$ {\footnotesize (2.7)}};
\draw[thick] (5.8,1.9) -- (6.2,1.9);
\node[right] at (6.2,1.9) {$N''${\footnotesize (1.4)}};
\draw[thick] (5.8,1.1) -- (6.2,1.1);
\node[right] at (6.2,1.0) {$U$ {\footnotesize (1.1)}};
\draw[thick] (5.8,1.4) -- (6.2,1.4);
\node[right] at (6.2,1.4) { $D$ {\footnotesize (1.2)}};

\draw[thick] (2.8,0.75) -- (3.2,0.75);
\node[right] at (3.2,0.75) {$N'$ {\footnotesize (0.45)}};
\draw[thick] (2.8,3.3) -- (3.2,3.3);
\node[right] at (3.2,3.3) {$X$ {\footnotesize (2.2)}};
\draw[thick] (2.8,3) -- (3.2,3);
\node[right] at (3.2,2.9) {$E'$ {\footnotesize (2.1)}} ;

\draw[thick] (-0.2,5.0) -- (0.2,5.0);
\node[right] at (0.2,5.0) {$N'''$  {\footnotesize (3.0)}};

\draw[thick] (-0.2,2.68) -- (0.2,2.68);
\node[right] at (0.2,2.68) {$\nu'$  {\footnotesize (2.0)}};

\draw[thick] (-0.2,0.3) -- (0.2,0.3);
\node[right] at (0.2,0.3) {\footnotesize$\psi_{\text{SM}}$};

\node at (0,0.7) {$\approx$};
\end{tikzpicture}
\begin{tikzpicture}[scale=0.9]

\draw[->] (0,0) -- (0,6) node[right] {$E/\text{TeV}$};
\draw[->] (0,0) -- (6.5,0) node[right] {$n_6$};

\foreach \x \label in {0/0, 3/2, 6/4} {
    \draw (\x,0.1) -- (\x,-0.1) node[below] {\label};
}

\draw[thick] (5.8,4.2) -- (6.2,4.2);
\node[right] at (6.2,4.2) {$E''$ };
\draw[thick] (5.8,1.9) -- (6.2,1.9);
\node[right] at (6.2,1.9) {$N''$};
\draw[thick] (5.8,1.1) -- (6.2,1.1);
\node[right] at (6.2,1.0) {$U$};
\draw[thick] (5.8,1.4) -- (6.2,1.4);
\node[right] at (6.2,1.4) {$D$};

\draw[thick] (2.8,0.75) -- (3.2,0.75);
\node[right] at (3.2,0.75) {$N'$ };
\draw[thick] (2.8,3.3) -- (3.2,3.3);
\node[left] at (2.8,3.3) {$X$};
\draw[thick] (2.8,3) -- (3.2,3);
\node[left] at (2.8,2.9) {$E'$} ;

\draw[thick,->,ForestGreen,dashed]  (6,4.2) -- (3.2,3);
\draw [decorate,decoration=snake,->,ForestGreen]  (4.8,3.7) -- (4.2,4.3) node[left] {$B$};
\draw[thick,->,ForestGreen]  (3.2,3.3) -- (5.8,1.4);
\draw [decorate,decoration=snake,->,ForestGreen]  (4.3,2.5) -- (5.2,3) node[right] {$B$};

\draw[thick] (-0.2,5.0) -- (0.2,5.0);
\node[right] at (0.2,5.0) {$N'''$ };

\draw[thick] (-0.2,2.68) -- (0.2,2.68);
\node[right] at (0.2,2.68) {$\nu'$  };

\draw[thick] (-0.2,0.3) -- (0.2,0.3);
\node[right] at (0.2,0.3) {\footnotesize$\psi_{\text{SM}}$};

\node at (0,0.7) {$\approx$};
\draw[thick,->,blue,dashed] (3,3) -- (3,0.75);
\draw [decorate,decoration=snake,->,blue]  (3,2.0) -- (2.2,1.75)
node[left] {$W$};
\draw[thick,->,blue] (6,1.4) -- (6,1.1);
\draw [decorate,decoration=snake,->,blue]  (6,1.25) -- (5.3,0.95) 
node[left] {$W$};
\end{tikzpicture}
\caption{Benchmark mass scales of the BSM particles arising in the SU2Y model plotted against the $n_6$ values. The numbers in the brackets are the predicted mass of the particles in units of TeV and this instance has that $N'$ and $U$ are stable given they are the lightest $n_6\neq 0$ particles with $L$ and $B$ number respectively. The RHS in addition illustates a possible decay chain for quarks (leptons) with the solid (dashed) arrows.}
    \label{322mass}
\end{figure}

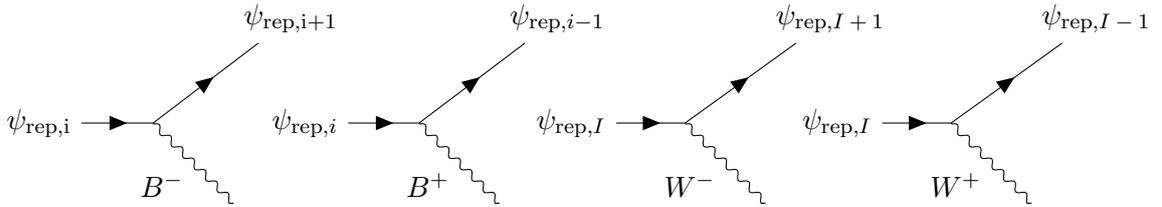
\begin{figure}[h!]
\begin{tikzpicture}
  \begin{feynman}
    \vertex (a) {\(\psi_{\text{rep,i}}\)};
    \vertex [right=of a] (b);
    \vertex [above right=of b] (f1) {\(\psi_{\text{rep,i}+1}\)};
    \vertex [below right=of b] (c);

    \diagram* {
      (a) -- [fermion] (b) -- [fermion] (f1),
      (b) -- [boson, edge label'=\(B^{-}\)] (c),
    };
  \end{feynman}
  \hspace{3.5cm}
    \begin{feynman}
    \vertex (a) {\(\psi_{\text{rep,$i$}}\)};
    \vertex [right=of a] (b);
    \vertex [above right=of b] (f1) {\(\psi_{\text{rep,$i$}-1}\)};
    \vertex [below right=of b] (c);
    \diagram* {
      (a) -- [fermion] (b) -- [fermion] (f1),
      (b) -- [boson, edge label'=\(B^{+}\)] (c),
    };
  \end{feynman}
    \hspace{3.5cm}
    \begin{feynman}
    \vertex (a) {\(\psi_{\text{rep,$I$}}\)};
    \vertex [right=of a] (b);
    \vertex [above right=of b] (f1) {\(\psi_{\text{rep,$I+1$}}\)};
    \vertex [below right=of b] (c);
    \diagram* {
      (a) -- [fermion] (b) -- [fermion] (f1),
      (b) -- [boson, edge label'=\(W^{-}\)] (c),
    };
  \end{feynman}
      \hspace{3.5cm}
    \begin{feynman}
    \vertex (a) {\(\psi_{\text{rep,$I$}}\)};
    \vertex [right=of a] (b);
    \vertex [above right=of b] (f1) {\(\psi_{\text{rep,$I-1$}}\)};
    \vertex [below right=of b] (c);
    \diagram* {
      (a) -- [fermion] (b) -- [fermion] (f1),
      (b) -- [boson, edge label'=\(W^{+}\)] (c),
    };
  \end{feynman}
\end{tikzpicture}
\caption{Baryon and lepton number conserving interactions in the $SU(3)\times SU(2)^2$ model, with $i$ being the $SU(2)_Y$ index and $I$ the $SU(2)_L$ one. The interactions must conserve $n_6$ which accidentally enforces $B$ and $L$ to be conserved.}
\label{feyndiagram}
    \end{figure} 

 \noindent$\bullet$~{\textbf{Phenomenology.}}
Our analysis of the conservation of $n_6$ guarantees that the lightest of the new particles with $n_6\neq 0$ is stable. The distinctive signal at colliders of this model is therefore fractionally charged leptons, baryons, scalars and vector bosons. The conservation of $B$ and $L$ further implies that the lightest leptonic and baryonic new state will be stable (even if for the latter it has a choice of channels to hadronise into).  Searches at the LHC exist that constrain fractionally charged particles~\cite{ATLAS:2019gqq,CMS:2024eyx} yet these are mostly based in supersymmetric models. These searches can be extended beyond supersymmetry and into fractionally charged particles as done in~\cite{Koren:2024xof} yet ideally dedicated experimental analyses will be produced in the future taking into account factors beyond the reach of theorists. On the other hand as remarked in the introduction cosmological constraints do imply in general that a low scale of inflation is needed for TeV scale particles.

It is not the aim of this paper to elaborate on the phenomenology of this model but rather to point out some features on top of studies such as~\cite{Koren:2024xof}. For this purpose we chose a benchmark scenario which provides a spectrum as shown in Fig.~\ref{322mass}. Details of the spectrum would change with other numerical choices but let us highlight the generic features of the spectrum. The mass of the new gauge bosons $B^\pm$ as well as that of the neutral quark $X$ are of order $t$ and hence potentially heavier than the rest. For the rest of the fermion spectrum no mass can surpass $4\pi v$ as imposed by unitarity and in addition the ratio of masses in the leptonic sector is as shown in eqs.~\eqref{eq:ratioN},\eqref{eq:ratioE} so that there are only two free mass parameters.

In the following for simplicity we restrict to gauge interactions. As for the $SU(2)_Y$ sector it is useful to remark that the $T_{Y}^\pm$ operators that accompany the gauge bosons $B^\pm$ in the covariant derivatives are ladder operators and as such have only non-vanishing entries in the $(T_Y)^+_{i,i+1}$ $(T_Y)^-_{i,i-1}$ elements and hence in three-point vertexes connect fermions to their nearest neighbours, as shown in Fig.~\ref{feyndiagram} side by side with the $W$ couplings. These two are the type of couplings that would mediate the decay of new particles.

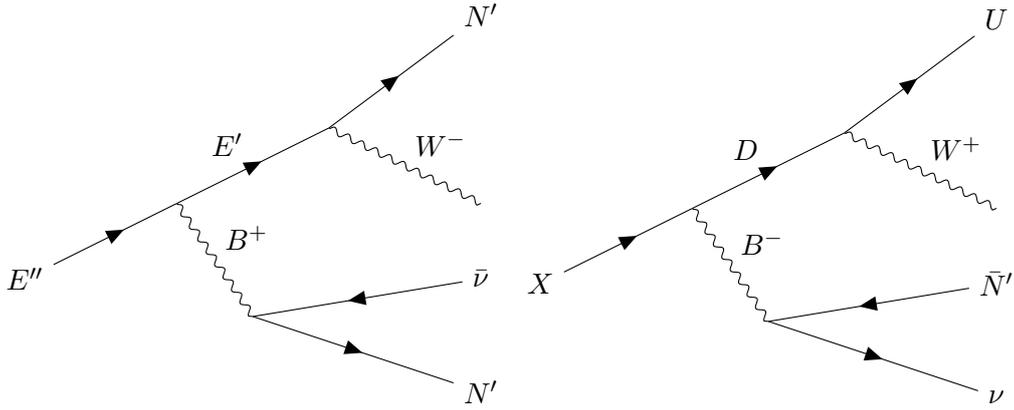
\begin{figure}[h!]
    \centering
    \begin{tikzpicture}
  \begin{feynman}
    \vertex (b) at (0,0) {\(E''\)};
    \vertex (i1) at (2,1);
    \vertex (mid) at (3,-0.5);
    \vertex (i2) at (4,2);
    \vertex (i3) at (6,1);
    \vertex (c) at (6,3.5) {\(N'\)};
    \vertex (d) at (6,0) {\(\bar \nu\)};
    \vertex (e) at (6,-1.5) {\(N'\)};
    \diagram* {
      (b) -- [fermion] (i1) -- [fermion, edge label=\(E'\)] (i2) -- [fermion] (c),
      (i1) -- [boson, edge label=\(B^+\)] (mid),
      (i2) -- [boson, edge label=\(W^-\)] (i3),
      (d) -- [fermion] (mid) -- [fermion] (e),
    };
  \end{feynman}
\end{tikzpicture}
    \begin{tikzpicture}
  \begin{feynman}
    \vertex (b) at (0,0) {\(X\)};
    \vertex (i1) at (2,1);
    \vertex (mid) at (3,-0.5);
    \vertex (i2) at (4,2);
    \vertex (i3) at (6,1);
    \vertex (c) at (6,3.5) {\(U\)};
    \vertex (d) at (6,0) {\(\bar N'\)};
    \vertex (e) at (6,-1.5) {\(\nu\)};
    \diagram* {
      (b) -- [fermion] (i1) -- [fermion, edge label=\(D\)] (i2) -- [fermion] (c),
      (i1) -- [boson, edge label=\(B^-\)] (mid),
      (i2) -- [boson, edge label=\(W^+\)] (i3),
      (d) -- [fermion] (mid) -- [fermion] (e),
    };
  \end{feynman}
\end{tikzpicture}
    \caption{New SU2Y particles cannot solely decay to SM without producing other BSM ones.}
    \label{BSMtreelevel}
\end{figure}

One of the features that this model makes explicit is that the spectrum of new particles would extend well beyond a single new fractionally charged particle. The particle pair produced at say the LHC would then need not be stable itself and promptly decay to other BSM and SM states. We can illustrate one such instance for both a quark and a lepton with non-trivial $n_6$ as 
\begin{align}
    X&\to \underbrace{B^-}_{\to \bar N'  \nu} (D\to U W^+), & E''&\to \underbrace{B^+}_{\to \bar\nu  N'} (E'\to W^- N'),
\end{align}
these decays are indicated in the RHS of Fig.~\ref{322mass} and their Feynman diagrams are shown in Fig.~\ref{BSMtreelevel} where we restricted ourselves to gauge interactions.

\begin{figure}[h!]
    \centering
    \begin{tikzpicture}
  \begin{feynman}
    \vertex (v1); 
    \vertex [right=2cm of v1] (v2);
    \vertex [right=2cm of v2] (v3); 
    \vertex [right=2cm of v3] (v4); 
    \vertex [right=2cm of v4] (v5);
    \vertex [below=2.5cm of v2] (b1);
    \vertex [below=2cm of v3] (b2);
    \vertex [below=2.5cm of v4] (b3);
    \node (a) [above right=0.05cm of v1] {\(\nu'\)};
    \node (b) [above right=0.05cm of v2] {\(E''\)};
    \node (c) [above right=0.05cm of v3] {\(E'\)};
    \node (d) [above right=0.05cm of v4] {\(e\)};
    \node at ($(b1) + (-0.5cm,0.8cm)$) {\(B\)};
    \node at ($(b2) + (-0.5cm,0.3cm)$) {\(B\)};
    \node at ($(b3) + (-0.5cm,0.8cm)$) {\(B\)};
    \diagram* {
      (v1) --[fermion] (v2) --[fermion] (v3) --[fermion] (v4) --[fermion] (v5),
      (v2) --[boson] (b1),
      (v3) --[boson] (b2),
      (v4) --[boson] (b3)
    };
    \path let \p1 = (b2) in
      coordinate (loopcenter) at (\x1, {\y1 - 1.2cm});
    \draw[thick] (loopcenter) ellipse [x radius=2.5cm, y radius=1.2cm];
    \coordinate (ellipright) at ($(loopcenter) + (2.5cm, 0)$);
    \coordinate (Wend) at ($(ellipright) + (1.5cm, 0)$);
    \draw[boson] (ellipright) -- (Wend);
    \node at ($(Wend) + (0.5cm, 0)$) {\(W^+\)};
  \end{feynman}
\end{tikzpicture}
    \caption{Loop level proc``ess for $\nu' \rightarrow E'' + B \rightarrow E' + 2B \rightarrow e + 3B$. The $B$ bosons combine to make a $W$ boson which can decay further to SM particles.}
    \label{fig:loopn60}
\end{figure}
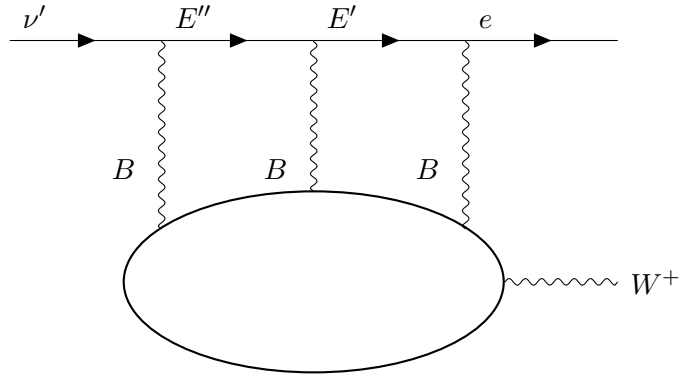

One last remark is that in the case of a new particle with $n_6=0$ where the decay into SM particles is allowed, the decay itself might be suppressed due to the structure of the gauge couplings. Take $\nu'$ as an example, via gauge interactions it can change to an $E''$ and a $B$ yet these individually have $n_6$ numbers so cannot decay to SM forbidding a tree level diagram for $\nu's$ decay; we could instead have $\nu'\to N''' W$ but $N'''$ is itself a BSM state which would face the very same problem $\nu'$ did, it can transtion to $B +N''$ but these have $n_6\neq 0$ and again forbid a tree-level decay. One possible diagram for $\nu'$ to decay into SM particles is shown in Fig.~\ref{fig:loopn60}.
    
\medskip

 \noindent$\bullet$~{\textbf{Baryon number, Lepton number and $B-L$ symmetry.}}
All colour triplet quarks have baryon number $B=1/3$ and leptons have $L=1$ in this model. The quarks and leptons are not in the same multiplets, and as as opposed to say $SU(5)$, one can conserve separately $B$ and $L$ at the perturvatibe level. This model, however, has a complete singlet in $\nu$ and just like in the Seesaw it could have a Majorana mass. We choose here to \emph{impose} $L$ conservation, since once one opens the door to Majorana mass terms, all lepton representations in fact could have bare Majorana mass terms, in particular, making explicit only the $SU(2)_Y$ structure
\begin{align}
    \nu\nu,&&\psi_{4,L}\varepsilon {\psi}_{4,L}^T,& & &\psi_{7,L} \varepsilon \psi_{7,L}^T.
\end{align}
In fact the reason one does not have a $\ell_L\tilde\ell_L^c$ term in the SM is hypercharge, but now hypercharge has been promoted to $SU(2)_Y$ and is a pseudo-real group itself with the tilde in the equation above acting as well in $SU(2)_Y$ space. In practice this means that every representation contains a given $\tilde T_{Y}$ eigenvalue particle and its opposite sign, so the Majorana mass term is in fact pairing these up to form a Dirac fermion (as otherwise cannot be any other way if they have electric charge). The exception to this are $\nu_R$ and $\nu_R'$ with $\tilde T_{Y}=0$ which are QED singlets and get a bona-fide Majorana mass term. This scenario is not phenomenologically viable, both because it would give all multiplet entries the same mass but even if degeneracies are broken, the pairing of same-representation particles would lead to a vector like theory and hence e.g. $W$ would couple to the $V$ and not a $V-A$ current.

    Both baryon and lepton numbers are therefore conserved at the classical level in this model just like in the SM, but one has that in the SM $B$ and $L$ are not quantum mechanically conserved due to anomalies~\cite{tHooft:1976rip} which are sourced by electroweak (EW) instantons~\cite{Belavin:1975fg,tHooft:1976snw}. These anomalous processes are unsuppressed at high temperature in the early universe and violate $B+L$~\cite{Kuzmin:1985mm} but conserve $B-L$ symmetry, which is is an anomaly-free global symmetry of the SM. It is then necessary to revisit the possible anomalies of $B$ and $L$ in this theory, which now has two non-Abelian and chiral sectors in $SU(2)_L$ and $SU(2)_Y$
\begin{align}
    U(1)_B& \times SU(2)_L^2  & & \frac{1}{3}N_c\times 2(\psi_{2,L})=2,\\
        U(1)_L& \times SU(2)_L^2  & & 4(\psi_{4,L})=4,\\
    U(1)_B& \times SU(2)_Y^2  & & \frac{1}{3}N_c\left[2C_\mathbf{2}(\psi_{2,L})-C_\mathbf{5}(\psi_{5,R})\right]=-36,\\ 
    U(1)_L& \times SU(2)_Y^2  & & 2C_{\mathbf{4}}(\psi_{4,L})-C_\mathbf{7}(\psi_{7,R})=-72,
\end{align}
where  $C_\mathbf{d}=$Tr$(\tilde T_Y^2)=\mathbf{d}(\mathbf{d}^2-1)/3$. Certain combination of $B$ and $L$ is guaranteed to be conserved in the SM since there is only one chiral non-Abelian group, not the case any more in this model where no linear combination could have been preserved; it is remarkable that the combination $2B-L$ is conserved simultaneously by the two $SU(2)$'s at the quantum level.
 \medskip
 
 \noindent$\bullet$ {\bf Quotient of the embedding group}. Although we started this section with the aim of finding a $p=1$ embedding, one has that this model admits a quotient. This fact can be spotted looking at all $n_6$ values; they are all even when they could have been in the whole $0$ to $5$ range.  Should one add another representation with $n_6=1$,
 as say $(\mathbf{1},\mathbf{1},\mathbf{2})$ for a scalar field or a pair of such representations for fermions to cancel global $SU(2)$ anomalies, no quotient should be possible.

As stressed in the introduction however if one is to talk about the quotients of the SM group the quotients of the BSM group should be specified and the different choices are different models so let us define
\begin{align}\label{eq:SU2Yquot}
    \frac{SU(3)_L\times SU(2)_L\times SU(2)_Y}{Z_2} \quad \textrm{with}\quad Z_2=\left\{1,e^{i\pi (\tilde T_{Y}+\tilde T_{3L})}\right\} \quad \Rightarrow \, G_2,\\
    SU(3)_L\times SU(2)_L\times SU(2)_Y \quad \textrm{e.g. SU2Y model \,+\,} (\mathbf{1},\mathbf{1},\mathbf{2})\, \textrm{reps} \quad  \Rightarrow \,G_1.
\end{align}

\subsection{Embedding hypercharge and weak isopin}
\label{sec:3.2}

One would expect this case to tie together the titular sectors of the Standard Model group and naturally lead to $G_2$, as discussed in sec.~\ref{sec:2} in fact it leads to $p\geq2$. 

\begin{figure}[h!]
    \centering    
\begin{tikzpicture}
    \draw [thick] (-1,0) circle (3pt);
    \draw [thick,dashed] (0,0) circle (3pt);
    \draw [thick,blue,<-] (0.5,0)-- (2,0);
    \draw [thick] (2.5,0.5) circle (3pt);
    \draw [thick] (3.5,0.5) circle (3pt);
    \draw [thick] (2.65,0.5) -- (3.35,0.5);
    \draw [thick] (2.75,-0.25) circle (3pt);
    \draw [thick] (2.75,-0.75) circle (3pt);
    \draw (-0.5,-1) node {$SU(2)_L\times U(1)_Y$};
    \draw (4.5,0.5) node {$SU(3)$};
    \draw (4.95,-.5) node {$SO(4)= SU(2)^2/Z_2$};
\end{tikzpicture}
    \caption{Embeddings in the smallest simple group (RHS) of the SM group $SU(2)\times U(1)$ (LHS), the hypercharge Cartan element is represented dashed to highlight there's no SU(2) associated to it, but it does originate from some element in the embedding group's Cartan subalgebra}
    \label{DynkinWY}
\end{figure}
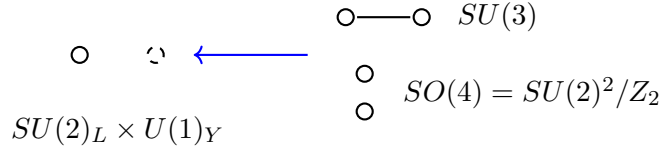

The Dynkin diagram-like completion is shown in Fig.~\ref{DynkinWY}. Since $SU(2)\times U(1)$ has a dimension-2 Cartan subalgebra, we need a group with at least such dimension, either $SU(3)$ or $SU(2)^2/Z_p$. The latter is in fact discussed in sec.~\ref{sec:3.1} with the two choices of $Z_p=Z_1,Z_2$ and it can lead to either $p=1$ or $p=2$. Hence we concentrate here on an $SU(3)$ embedding.

The following embedding would realise the simplest encompassing group
\begin{align}
\label{eq:su3sq}
     SU(3)_c&\times(SU(3)_{LY}\to SU(2)_L \times U(1)_Y/Z_p), & 6Q_Y&=\tilde\lambda_{8LY}.
\end{align}

For the value of $p$ one finds the built-in relation between hypercharge and isospin characteristic of $p=2$, i.e.
\begin{align}\label{eq:su3p2}
    \xi^3=\exp\left(\pi i(2\tilde \lambda_8+3\tilde T_{3L}+6Q_Y) \right)=\exp\left(\pi i(\tilde \lambda_{3L}+\tilde\lambda_{8LY}) \right)=\exp\left(\pi i\left(\begin{array}{ccc}
        2 &  &  \\
         & 0 &  \\
         &  & -2
    \end{array}\right)\right)=1,
\end{align}
and indeed $p=2$. With the given group and hypercharge identification finding the matter content is straight-forward, e.g. $q_L\in(\mathbf{3},\mathbf{3})$ and it would require another LH BSM hypercharge $-1/3$ fermion, yet as in the SU2Y model this means a proliferation of new states. Instead of pursuing this model however we turn to a modification of the group that allows for more minimal `packing' of matter yet staying with $p\geq 2$.\medskip 

\noindent\textbf{Trinification model.}
The minimal pattern is instead that of the $SU(3)_c\times SU(3)_L \times SU(3)_R$ Trinification model~\cite{Glashow:1984gc,Babu:1985gi}
first considered by
de~R\'ujula, Georgi and Glashow 
\begin{align}
\label{eq:trinQY}
     SU(3)_c&\times(SU(3)^2\to SU(2)_L \times U(1)_Y/Z_p) & -6Q_Y&=\tilde\lambda_{8L}+(3\tilde\lambda_{3R}+\tilde\lambda_{8R}),\\
     & &-6Q_Y&=\tilde\lambda_{8L}-2\tilde \lambda_{8R'},
\end{align}
where RH generator $\tilde \lambda_{8R'}=$Diag$(-2,1,1)$, is a permuted version of $\tilde\lambda_8$ whose addition nonetheless maintains the periodicity of $6Q_Y$ at $2\pi$.
 \medskip

     \noindent$\bullet$ \textbf{Value of } $p$.
From the hypercharge definition, we have now two more components in the new $SU(3)_R$ group, one can check if they spoil or not the relation in eq.~\eqref{eq:su3p2},
\begin{align}
    \xi^3=&\exp\left(\pi i(2\tilde \lambda_8+3\tilde T_{3L}+6Q_Y) \right)=\exp\left(\pi i(3\tilde \lambda_{3L}+\tilde\lambda_{8L}-(3\lambda_{3R}+\lambda_{8R}))\right)
\\
    =&\exp\left(\pi i\left(\begin{array}{ccc}
        4 &  &  \\
         & -2 &  \\
         &  & -2
    \end{array}\right)\right)\otimes\exp\left(\pi i\left(\begin{array}{ccc}
        -4 &  &  \\
         & 2 &  \\
         &  & 2
    \end{array}\right)\right),
\end{align}
they do not and once more we conclude $p=2$,
\begin{align}
  G_2 \,\,\, : \,\,\, e^{6\pi i Q_Y}\,e^{3\pi i n_L} = 1.
\end{align}
The possibility of even higher $p$ is dependent on the spectrum and as is norm postponed to the end of this subsection.

It is useful to present the generators for the hypercharge and for the electric charge, $Q_Y$ and $Q_{em}$, explicitly for fundamental representation of both groups $SU(3)_L$ and $SU(3)_R$,
\begin{eqnarray}
\label{eq:QYF}
Q_Y^{\,F}\,&=\, -\frac{1}{6} \tilde{\lambda}_{8L} - \frac{1}{6}(3 \tilde{\lambda}_{3R} + \tilde{\lambda}_{8R})
\,&=\, \begin{pmatrix} -\frac{1}{6}& &\\ &-\frac{1}{6}&\\ & &\frac{1}{3}\,\end{pmatrix}_L
\oplus
\begin{pmatrix} -\frac{2}{3}& &\\ &\frac{1}{3}&\\ & &\frac{1}{3}\,\end{pmatrix}_R,
\\
\label{eq:QemF}
Q_{em}^{\,F}\,&=\, Q_Y^{\,F} + 
\begin{pmatrix} \frac{1}{2}& &\\ &-\frac{1}{2}&\\ & &0\,\end{pmatrix}_L
\,&=\, \begin{pmatrix} \frac{1}{3}& &\\ &-\frac{2}{3}&\\ & &\frac{1}{3}\,\end{pmatrix}_L
\oplus
\begin{pmatrix} -\frac{2}{3}& &\\ &\frac{1}{3}&\\ & &\frac{1}{3}\,\end{pmatrix}_R.
\end{eqnarray}

 \medskip
 
 \noindent$\bullet$ \textbf{Fermionic matter content} is explained briefly below and compiled in Table~\ref{tab:GGRmatter}. 
For the LH and RH quarks we have,
\begin{align}
\label{eq:QLtri}
    Q_L : (\bold 3,\bar{\bold 3}, \bold1) = 
    \begin{pmatrix} (\tilde q_L^\dagger)^1& D^1\\
    (\tilde q_L^\dagger)^2 & D^2\\ (\tilde q_L^\dagger)^3 & D^3
    \end{pmatrix}_L, \\ 
\label{eq:QRtri}    
    Q_R : (\bold 3, \bold1,\bar{\bold 3}) = 
    \begin{pmatrix} u^1& d^1& D^1\\
    u^2& d^2 & D^2\\ u^3 &d^3 & D^3
    \end{pmatrix}_R,
 \end{align}   
 where $\tilde q_L^\dagger=(\epsilon q_L^*)^\dagger=(d_L,-u_L)$, while the leptons are assembled into
 \begin{equation}
    L : (\bold1, \bold 3, \bar{\bold 3}) = \begin{pmatrix}
        L_1 & L^- & \nu_L \\
        L^+ & L_2 & e_L \\
        e_R^\dagger & \nu_R^\dagger & L_3
    \end{pmatrix}.
\label{eq:Leptri}    
\end{equation}

LH quark matrix $Q_L$ is acted by $SU(3)_c$ from the left and $SU(3)_L$ from the right, and the RH quarks $Q_R$ are acted by $SU(3)_R$ from the right, as e.g. for $Q_R$
\begin{align}
    Q_R\to U_c Q_R U^\dagger_R.
\end{align}
Since the LH quarks $Q_L$ transform in 
$\bar{\bold 3}$ of $SU(3)_L$ their electric charges are given by $(-1)$ times the first matrix on the RHS of eq.~\eqref{eq:QemF}. Similarly the electric charges for the RH quarks in $Q_R$ (which transform in 
$\bar{\bold 3}$ of $SU(3)_R$) are given by by $(-1)$ times the second matrix on the RHS of eq.~\eqref{eq:QemF}. This gives correct values for
$Q_{em}$ for $u$ and $d$ quarks, and  
 $D$ is a new quark, singlet under $SU(2)_L$ with electromagnetic charge $-1/3$. 
 
 The lepton matrix $L$ in eq.~\eqref{eq:Leptri} transforms under $SU(3)_L$ from the left and under $SU(3)_R$ from the right (and is a singlet under colour). It is easy to verify using eqs.~\eqref{eq:QYF}-\eqref{eq:QemF} that $e_{L/R}$ and $\nu_{L/R}$ in eq.~\eqref{eq:Leptri} have correct values of the SM hypercharge and the electric charge.
 New BSM states in eq.~\eqref{eq:Leptri} are $L_1, L_2, L_3$ and $L^\pm$ where
 $L_1, L_2, L_3$ are neutral leptonic states (independent of each other) and $L^\pm$ are charged leptons.
\begin{table}[h]
    \centering
    \begin{tabular}{c||c||c||c}
        $\mathcal G=SU(3)_c\times SU(3)_L\times SU(3)_R$ & $ (\mathbf{3}, \bar{\mathbf{3}}, \bold 1)$ & $ (\mathbf{3}, \bold 1,\bar{\mathbf{3}})$ & $ (\bold1,\mathbf{3}, \bar{\mathbf{3}})$ \\ \hline
         SM fields  & $u_L, d_L$ & $u_R,d_R$ & $\nu_L, e_L,e_R^\dagger$\\
         $n_6$ & 0 & 0 & 0 \\
         \hline
          BSM fields  & $D_L$ & $D_R$ & $L_1, L_2, L_3, L^+, L^-, \nu_R^\dagger$\\
         $n_6$ & 0 & 0 & 0
    \end{tabular}
    \caption{Fermion fields in $SU(3)_c\times SU(3)_L\times SU(3)_R$ Trinification theory following eqs.~\eqref{eq:QLtri}-\eqref{eq:Leptri} and their $n_6$ values. }
    \label{tab:GGRmatter}
\end{table}

These can decompose into SM representations, with notation as $(R_{SU(3)_c}, R_{SU(2)_L})_{Q_Y}$, as follows
\begin{align}
    (\bold{3}, \Bar{\bold 3}, \bold 1)& \rightarrow (\bold 3, \bold 2)_{1/6} \oplus (\bold 3, \bold 1)_{-1/3}, \\
    (\bold 3, \bold1, \Bar{\bold 3}) &\rightarrow (\bold 3, \bold1)_{2/3} \oplus 2\times (\bold 3, \bold1)_{-1/3}, \\
    (\bold 1, \bold 3, \Bar{\bold 3}) &\rightarrow 2\times (\bold1, \bold 2)_{-1/2} \oplus (\bold1, \bold 2)_{1/2} \oplus 2\times (\bold1, \bold1)_0 \oplus (\bold1, \bold1)_1.
\end{align}

\textbf{Particle count:}  this model adds 8 chiral fermions, two of which are coloured and form a massive Dirac fermion, another 3 get Majorana mass terms, two new charged leptons get a Dirac mass and finally $\nu_R$ provides a Dirac mass for active neutrinos. The mass terms, for leptons in particular, necessitate a scalar field.

The values of $n_6$ for each fermion are given in Table~\ref{tab:GGRmatter} where one finds they are all $0$ even though the $p=2$ case only constrains them to be even, note especially that the color triplet $D$ has the opposite electric charge as D in the SU2Y model and is in this sense a heavier copy of the SM $d$. This is a sign of the possibility of taking a quotient of the group as discussed momentarily.
 \medskip

 \noindent$\bullet$ \textbf{Bosonic matter content and symmetry breaking.} To break the Trinification group to the SM,  one requires at least two flavours $f$ of bi-triplet Higgs fields $\mathcal{H}_f$~\cite{Babu:1985gi,Hetzel:2015bla},
where each bi-triplet flavour is of the form
 \begin{equation}
    \mathcal{H}_f: (\bold1, \bold 3, \bar{\bold 3}) = \begin{pmatrix}
        H_1 & H_2 & H_3 \\
        S_1^+ & S_2 & S_3 
    \end{pmatrix}_f \,, \qquad f=1,2.
\label{eq:Htri}    
\end{equation} 
Here $H_1$,  $H_2$ and  $H_3$ are $SU(2)_L$ doublets and one of them can be identified with the SM Higgs, for example we can take
$H_3:=\tilde H =\epsilon H^* = \begin{pmatrix} (h^0)^* \\ -(h^+)^* \end{pmatrix}$, where $H$ is the SM Higgs. The scalar fields $S_2$, $S_3$ and $S_1^+$ are the neutral and charge-one complex singlets.
The hypercharges and electric charges of the individual Higgs field components in eq.~\eqref{eq:Htri}  are the same as those of the leptons in eq.~\eqref{eq:Leptri}  since both of them transform as $(\bold1, \bold 3, \bar{\bold 3})$. 

The Higgs fields $\mathcal{H}_{f=1,2}$ (and their components that encompass $H$) can realise the required breaking pattern from
  the Trinification group to the Standard Model
\begin{align}
  \nonumber  & SU(3)_c \times SU(3)_L \times SU(3)_R \xrightarrow{(\bold1,\bold 3,\bar{\bold 3})} SU(3)_c \times SU(2)_L \times SU(2)_R \times  U(1)_{B-L} \\
  \nonumber  &\xrightarrow{(\bold1,\bold 2,\bold 2)}\widetilde{G}_{SM}/Z_p
\end{align}
with $p\geq3$.
\begin{table}[h]
    \centering
    \begin{tabular}{c||c||c}
        $\mathcal G=SU(3)_c\times SU(3)_L\times SU(3)_R$  & $ (\bold1,\mathbf{3}, \bar{\mathbf{3}})$ & $(\bold1,\mathbf{3},\bar{\mathbf{3}})$  \\
        \hline
         Scalar fields  & $\mathcal H_1$ &  $\mathcal H_2$\\
         $n_6$ & 0 &0 
    \end{tabular}
    \caption{Scalar fields in the $SU(3)_c\times SU(3)_L\times SU(3)_R$ Trinification model. Note that in $\mathcal{H}_{1,2}$ there are six doublets under $SU(2)_L$, and the SM Higgs can be identified with a linear combination of these, other linear combinations could potentially be light too in a two(or more) Higgs doublet model.}
    \label{tab:SU3boson}
\end{table}

\textbf{Particle count:}  
Twelve (8+8-4) more massive gauge bosons and (36-12-4)=20 real Higgs-like scalar degrees of freedom obtained after subtracting longitudinal broken gauge boson and $H$ d.o.f.

\medskip

Yukawa interactions generate Dirac masses for quarks and Majorana-type masses for leptons, schematically as
\begin{equation}
-\, \mathcal{L}_{\rm yuk} \,=\, y_{Q\,{ijf}}\, \bar Q_{Ri} Q_{Lj} \mathcal{H}_f \,+\, 
 y_{L\,{ijf}}\, L_{Li} L_{Lj} \mathcal{H}_f \,+\, h.c.\,,
\end{equation}
where $i,j$ label the three generations of quarks and leptons.
\medskip

 \noindent$\bullet$ \textbf{Quotients of the embedding group.}
The fact that $n_6=0$ for all representations is the sign of a possible quotient. The maximal centre of the group for a matter-free theory is $\mathcal{Z}(SU(3)^3) = Z_{3c} \times Z_{3L} \times Z_{3R}$. Adding fundamental matter will necessarily not respect this but inspecting the different representations one sees they are always of the form $\mathbf{3}$, $\bar{\mathbf{3}}$ which means the diagonal subgroup $Z_{3}$ is preserved, which reads
\begin{align}
    Z_3 = \{1,\zeta,\zeta^2\}=\{\mathbb{I}, \,\, e^{\frac{2\pi i}{3}(n_c+n_{3L}+n_{3R})}, \,\, e^{\frac{4\pi i}{3}(n_c+n_{3L}+n_{3R})}\},
\end{align}
with $n_c, n_{3L}, n_{3R}$ the triality for each group taking values of $\{0,1,2\}$, $\bold1$ being the fundamental $\textbf{3}$ and $-1\,\text{mod}\,3=2$ for the anti-fundamental $\bar{\textbf{3}}$.
Indeed
\begin{align}
    &(\bold3,\bar{\bold3}, \bold1) : \zeta Q_L = e^{2\pi i/3} e^{-2\pi i/3} e^0 Q_L = Q_L,\\
    &(\bold3, \bold1, \bar{\bold3}) : \zeta Q_R = e^{2\pi i/3} e^0 e^{-2\pi i/3} Q_R = Q_R,\\
    &(\bold1,\bold3, \bar{\bold3}) : \zeta L =  e^0  e^{2\pi i/3} e^{-2\pi i/3} L = L.
\end{align}

Imposing the quotient means any other rep $R$  should be invariant under $\zeta$
\begin{align}
    \zeta R = R\,\,\, ; \,\,\,
    \zeta = e^{2\pi i n_{c}/3} e^{2\pi in_{3L}/3} e^{2\pi in_{3R}/3}.
    \label{zeta}
\end{align}
A complete singlet of this GUT $(\bold1,\bold1,\bold1)$, as well as representations $(\ydiagram{1},\ydiagram{1},\ydiagram{1})$ and $(\ydiagram{2},\ydiagram{2},\ydiagram{2})$, with $\ydiagram{1} = \bold{3}$ denoting the fundamental representation of $SU(3)$, also obey this invariance. However, there are representations that pick up a phase under the action of $\zeta$, meaning they are not invariant under the quotient and thus are disallowed under $SU(3)^3/Z_3$. Alternatively if these are found one concludes the group is $SU(3)^3$ with no quotient. Such representations are shown in Table~\ref{tab:GGRQuotientMatter} with their corresponding $n_6$ values.
\begin{table}[h]
    \centering
    \begin{tabular}{c|c}
        $n_6$ & Disallowed $\frac{SU(3)^3}{Z_3}$ states \\ \hline
        0 & ($\bold1,\bold1,\bold1$), ($\ydiagram{1}$,$\ydiagram{1}$,$\bold1$), ($\ydiagram{1}$,$\bold1$,$\ydiagram{1}$) \\
        & ($\ydiagram{1}$,$\ydiagram{2}$,$\ydiagram{2}$),($\ydiagram{2}$,$\ydiagram{1}$,$\ydiagram{1}$) \\ \hline
        2 & ($\bold1$,$\ydiagram{1}$,$\bold1$), ($\bold1$,$\bold1$,$\ydiagram{1}$), ($\ydiagram{1}$,$\ydiagram{1}$,$\ydiagram{1}$) \\
        & ($\ydiagram{2}$, $\ydiagram{2}$,$\ydiagram{1}$), ($\ydiagram{2}$,$\ydiagram{1}$,$\ydiagram{2}$) \\
        & ($\ydiagram{2}$,$\bold1$,$\bold1$) \\ \hline
        4 & ($\bold1$,$\ydiagram{1}$,$\ydiagram{1}$),($\ydiagram{1}$,$\bold1$,$\bold1$) \\
        & ($\bold1$,$\bold1$,$\ydiagram{2}$),($\bold1$,$\ydiagram{2}$,$\bold1$) \\
        & ($\ydiagram{1}$,$\ydiagram{1}$,$\ydiagram{2}$), ($\ydiagram{1}$,$\ydiagram{2}$,$\ydiagram{1}$) \\
        & ($\ydiagram{2}$,$\ydiagram{2}$,$\ydiagram{2}$)
    \end{tabular}
    \caption{$SU(3)^3$ representations that pick up a phase when taking the $Z_3$ quotient and are not present in the spectrum of $SU(3)^3/Z_3$, finding these in Nature therefore would lead to invalidating $SU(3)^3/Z_3$ in favour of $SU(3)^3$. Other representations with the same properties as the above can be obtained if $SU(3)$ reps are subbed by others with as many young tableaux mod 3.}
    \label{tab:GGRQuotientMatter}
\end{table}

As far as the group breaking into the SM the expectation is
\begin{align}
    SU(3)^3 &\rightarrow \frac{\widetilde{G}_{SM}}{Z_2}, & \frac{SU(3)^3}{Z_3} &\rightarrow \frac{\widetilde{G}_{SM}}{Z_2 \otimes Z_3} = \frac{\widetilde{G}_{SM}}{Z_6}.
\end{align}
One could show that $Z_2\times Z_3=Z_6$ by making explicit the action of $Z_3$ and comparing with $Z_6$, yet the most straightforward approach is to show that $\zeta=\xi$. Let us start from 
\begin{align}
    \xi=&\exp 2\pi i \left(\frac{n_c}{3}+\frac12 \tilde T_{3L}+Q_Y \right)=\exp 2\pi i \left(\frac{n_c}{3}+\frac12 \tilde \lambda_{3L}-\frac{\lambda_{8L}}{6}+\frac{\tilde \lambda_{8R'}}{3} \right)\\
     =&e^{2\pi i n_c/3}e^{2\pi i n_{3R}/3}\exp\frac{2\pi i}{3}\begin{pmatrix}
         1 & &\\
          & -2&\\
          & &1
     \end{pmatrix}=\zeta,
\end{align}
where we have used that $\tilde \lambda_{8R'}$ and in fact any other permutation of $\tilde\lambda_{8}$, such as the one appearing in the LH sector, generates triality when exponentiated times $2\pi i/3$.
 \medskip

 \noindent$\bullet$ {\textbf{Baryon number, Lepton number and $B-L$ symmetry.}} In Trinification, $B$ and $L$ are not fundamental symmetries of the theory, and $B-L$ can emerge as a global symmetry in the minimal model. Extended versions can gauge $B-L$ symmetry by embedding it via adding a new $U(1)$. Proton decay is model-dependent (on the Higgs content and breaking pattern) and can be suppressed if accidental symmetries protect baryon number.

\subsection{Embedding hypercharge and colour}
\label{sec:3.3}

\begin{figure}[h!]
    \centering    
\begin{tikzpicture}
    \draw [thick] (-0.9,0) circle (3pt);
    \draw [thick] (-1,0) -- (-1.6,0);
    \draw [thick] (-1.7,0) circle (3pt);
    \draw [thick,dashed] (0,0) circle (3pt);
    \draw [thick,blue,<-] (0.5,0)-- (2,0);
    \draw [thick] (2.7,0.5) circle (3pt);
    \draw [thick] (3.5,0.5) circle (3pt);
    \draw [thick] (2.8,0.5) -- (3.4,0.5);
    \draw [thick] (3.6,0.5) -- (4.2,0.5);
    \draw [thick] (4.3,0.5) circle (3pt);
    \draw [thick] (3.15,-0.5) circle (3pt);
    \draw [thick] (3.25,-0.55) -- (3.65,-0.75);
    \draw [thick] (3.25,-0.45) -- (3.65,-0.25);
    \draw [thick] (3.75,-0.75) circle (3pt);
    \draw [thick] (3.75,-0.25) circle (3pt);
    \draw (-0.5,-1) node {$SU(3)_c\times U(1)_Y$};
    \draw (5.25,0.5) node {$SU(4)$};
    \draw (5.25,-.5) node {$SO(6)$}; 
\end{tikzpicture}
    \caption{The embedding of $SU(3)_c \times U(1)_{B-L}$  in either $SU(4)_c$ or $SO(6)$ is group theoretically identical by their isomorphic Lie algebras.}
    \label{DynkincPS}
\end{figure}
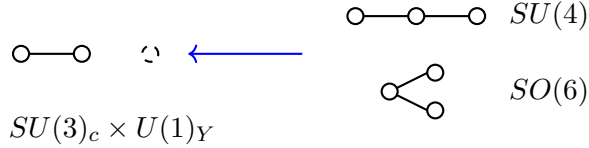

The subgroup of $\widetilde{G}_{SM}$ here considered has Cartan subalgebra of dimension 3 now, the Dynkin diagram for its embedding in Fig.~\ref{DynkincPS} shows that minimal groups that cover it are $SU(4)$ or $SO(6)$. One has that $SU(4)$ is the double cover of $SO(6)$, i.e. $SO(6)=SU(4)/Z_2$, which is itself pointing at the possibility of taking a quotient if, in this case, the matter conent is compatible with $SO(6)$. As usual however we make explicit first the matter content to later address this question.

Let us take the universal cover $SU(4)$ and assume a breaking as
\begin{align}
    SU(2)&\times(SU(4)\to SU(3) \times U(1)), &  6Q_Y&= \tilde T_Y,
\end{align}
with $\tilde T_Y=\tilde\lambda_{15}=$Diag$(1,1,1,-3)$. This suffices to write
\begin{align}\label{eq:PatSalp}
    \xi^2=\exp(2\pi i(\frac{2\tilde \lambda_{8c}}{3}+2Q_Y))=\exp\frac{2\pi i}{3} 
    \left(\begin{pmatrix}
        2&&&\\
        &2&&\\
        &&-4&\\
        &&&0\\
    \end{pmatrix}+\begin{pmatrix}
        1&&&\\
        &1&&\\
        &&1&\\
        &&&-3
    \end{pmatrix}\right)=1,
\end{align}
which confirms $p\geq 3$. The matter content then could be placed as
$q_L,\ell_L\in \mathbf{4}$, $ d_R,e_R\in \mathbf{10}$  $u_R \in \mathbf{15}$ but here again we will not pursue this model but rather the more matter-packing efficient version.
 \medskip

\noindent\textbf{Pati--Salam.} In this model, in addition to the $SU(4)_c$ group, hypercharge also has a component along a new $SU(2)_R$ group. The breaking pattern then could have an intermediate stage as
\begin{align}
    (SU (4)_c \rightarrow SU(3)_c \times U(1)_{B-L}) \times SU(2)_L \times SU(2)_R.
\end{align}
 Here, $SU(2)_L$ is the standard weak gauge group and $SU(3)_c$ is embedded in $SU(4)_c$ as the upper $3 \times 3$ block. The new colour group $SU(4)_c \rightarrow SU(3)_c \times U(1)_{B-L}$ with lepton number as the `fourth colour' i.e., $\nu$, $e$ being the $u$, $d$ quarks of `lepton colour'. 
Hence the SM hypercharge arises in the breaking
$U(1)_{B-L} \times SU(2)_R \rightarrow U(1)_Y$ and the operator reads
\begin{align}
   6Q_Y= \tilde T_{Y}=\tilde \lambda_{15}+3\tilde T_{3R}=3T_{B-L}+3\tilde T_{3R}.
\end{align}

 \noindent$\bullet$ {\textbf{The value of} $p$. The change from eq.~\eqref{eq:PatSalp} is simply the addition of $3\tilde T_{3R}$ so one has 
 \begin{align}
     \xi^2=1\times\exp 2\pi i \tilde T_{3R}=1,
 \end{align}
so that $p\geq 3$,
\begin{align}
  G_3 \,\,\, : \,\,\, e^{4\pi i Q_Y}\,e^{4\pi i n_c/3} = 1.
\end{align}

 \noindent$\bullet$ {\textbf{The matter content.}} This model puts together leptons and quarks such that only two representations are needed for fermions as
\begin{align}
    \psi_L = (q_L, \ell_L) = \begin{pmatrix}
        u_r & u_b & u_g & \nu \\
        d_r & d_b & d_g & e
    \end{pmatrix}_L \,\,\, ; \,\,\, \psi_R = 
    \begin{pmatrix}
       u_r & u_b & u_g & \nu \\  d_r & d_b & d_g & e
    \end{pmatrix}_R.
\end{align}

\begin{table}[h]
    \centering
    \begin{tabular}{c||c||c}
        $\mathcal G=SU(4)_c\times SU(2)_L\times SU(2)_R$   & $ (\mathbf{4},\mathbf{2}, \bold1)$ & $ (\mathbf{4},\bold1,\mathbf{2})$ \\ \hline
         SM fields  & $u_L, d_L, \nu_L, e_L$ & $u_R, d_R, e_R$ \\
         BSM fields  &  & $\nu_R$ \\
         $n_6$ & 0 &0 \\
    \end{tabular}
    \caption{Fermionic matter content in the Pati--Salam model.}
    \label{tab:PSmatter}
\end{table}

Where the representations each break down after the high energy breaking as
\begin{align}
    (\bold4,\bold2, \bold1) \rightarrow (\bold3,\bold2)_{1/6} \oplus (\bold1,\bold2)_{-1/2}, \\
    (\bold4, \bold1, \bold2) \rightarrow (\bold3, \bold1)_{-1/3} \oplus (\bold3, \bold1)_{2/3} \oplus (\bold1, \bold1)_{-1} \oplus (\bold1, \bold1)_0.
\end{align}

\textbf{Particle count:} 1 more fermion only, the RH neutrino. The fermions are collected in Table~\ref{tab:PSmatter} and all have $n_6=0$.


 \medskip

 \noindent$\bullet$~{\textbf{Bosonic content and symmetry breaking.}}
The are two different breaking patterns from Pati--Salam to SM, \\
Pathway A:
\begin{align}
  \nonumber & SU(4)_c \times SU(2)_L \times SU(2)_R \xrightarrow{(\bold{10},\bold1,\bold3)\,\text{or}\,(\bold4,\bold2,\bold1)} \widetilde{G}_{SM}/Z_3, 
  \end{align}
or Pathway B:
  \begin{align}
  \nonumber  & SU(4)_c \times SU(2)_L \times SU(2)_R \xrightarrow{(\bold{15},\bold1,\bold1)} SU(3)_c \times U(1)_{B-L} \times SU(2)_L \times SU(2)_R \xrightarrow{(\bold1,\bold1,\bold3)} \\ & \nonumber \widetilde{G}_{SM}/Z_3 .
\end{align}
To find the $n_6$ values the Higgses will take, we have to decompose them into SM reps
\begin{align}
    & (\bold{15},\bold1,\bold1) \rightarrow (\bold8,\bold1)_0 \oplus (\bold3,\bold1)_{2/3} \oplus (\bar{\bold3},\bold1)_{-2/3} \oplus (\bold1,\bold1)_0, \\ 
    & (\bold1,\bold1,\bold3) \rightarrow (\bold1,\bold1)_{+1} \oplus (\bold1,\bold1)_0 \oplus (\bold1,\bold1)_{-1}, \\
    & (\bold{10},\bold1,\bold3) \rightarrow (\bold6_{{1}/3},\bold1,\bold3) \oplus (\bold3_{-{1}/3},\bold1,\bold3) \oplus (\bold1_{-{1}},\bold1,\bold3),
\end{align}
where the subscript outside the parenthesis denotes hypercharge while the subindex inside the parenthesis for colour representations gives  $(B-L)/2$.

\begin{table}[h]
    \centering
    \begin{tabular}{c||c||c||c||c||c}
        $\mathcal G=SU(4)_c\times SU(2)_L\times SU(2)_R$ & $(\mathbf{4},\mathbf{2}, \bold1)$ & $ (\mathbf{15},\bold1,\bold1)$ & $(\bold1,\mathbf{2},\mathbf{2})$ &$ (\mathbf{10},\bold1,\mathbf{3})$ & $(\bold1,\bold1,\mathbf{3})$ \\
        \hline
         SM fields  &  & & $H$& \\
         $n_6$ & 0 &0 & 0 &0 & 0
    \end{tabular}
    \caption{Possible scalar fields in the Pati--Salam model, a given pathway for the breaking will select some of the fields above and the SM-scale BSM model is a two Higgs doublet model.}
    \label{tab:PSboson}
\end{table}

\textbf{Particle counting:} (15+3+3-12) = 9 more gauge bosons. As for the new Higgses,

\begin{itemize}
    \item Pathway A : $(\bold{10},\bold1,\bold3)$ has 60 real d.o.f. or for a more economic breaking the choice of $(\bold4,\bold2,\bold1)$ has 16 real d.o.f. In each case, $9$ of these will be swallowed in the heavy gauge bosons, with the remaining being heavy Higgses.
    \item Pathway B : $(\bold{15},\bold1,\bold1)$ has 15 real d.o.f. and $(\bold1,\bold1,\bold3)$ 6 real d.o.f so 21 in total, out of which 9 are the longitudinal components of the heavy gauge bosons.
\end{itemize}
Lastly, the Higgs doublet of the SM brings along another doublet in $(\mathbf{1},\mathbf{2},\mathbf{2})$ and hence 4 massive degrees of freedom.
 \medskip

\noindent$\bullet$~{\textbf{Quotients of the embedding group.}} 
The centre of the Pati--Salam group $\mathcal{Z}(G_{PS}) = Z_4 \times Z_2 \times Z_2$ is generated by $\omega_4=e^{\pi in_4/2}$, $\omega_{2L}=e^{\pi n_L}$ and $\omega_{2R}=e^{\pi n_R}$ with the n-alities $n_N$ defined as positive integers mod $N$ for each of the $SU(N)$ groups; $n_4 = \{0, 1, 2, 3\}, \, n_L = \{0, 1\}$ and $ n_R = \{0, 1\}$. The matter content is not invariant under the full centre but it is under the subgroup
\begin{equation}
    Z_2=\{1,\tilde{\omega}\}, \qquad  \tilde \omega= \omega_4^2\times\omega_{2L}\times\omega_{2R}) = e^{\pi i n_4} \, e^{\pi i n_L} \, e^{\pi i n_R},
\end{equation}
which indeed acting on $\psi_L$
\begin{equation}
    \Tilde{\omega} \, \psi_L = (e^{\pi i} \, e^{\pi i} \, e^0) \psi_L = \psi_L,
\end{equation}
as $\psi_L$ belongs in the $(\bold4, \bold2,\bold1)$ representation for which $n_4 = 1, \, n_L = 1, \, n_R =0$. Similarly, for $\psi_R \in (\bold4, \bold1, \bold2)$ with $n_4 = 1, \, n_L = 0, \, n_R =1$ we get the same invariance.

Finding representations in the Pati--Salam model that do not obey this invariance under the action of the generating element would mean that we can disregard the $Z_2$ quotient and consider $G_{PS}$ as the proper group. We have listed such representations for which $\Tilde{\omega} \, \psi = -\psi$ in Table ~\ref{tab:PSQuotientMatter}.

\begin{table}[h!]
    \centering
    \begin{tabular}{c|c}
        $n_6$ & Disallowed $\frac{G_{PS}}{Z_2}$ states \\ \hline
        0 & ($\bold1$,$\bold1$,$\bold1$),($\ydiagram{2}$,$\bold1$,$\bold1$) \\ \hline
        3 & ($\bold1$,$\bold2$,$\bold1$), ($\bold1$,$\bold1$,$\bold2$), ($\ydiagram{1}$,$\bold1$,$\bold1$), ($\ydiagram{1}$,$\bold2$,$\bold2$) \\
        & ($\ydiagram{2}$,$\bold2$,$\bold1$), ($\ydiagram{2}$,$\bold1$,$\bold2$) \\
        & ($\ydiagram{3}$,$\bold1$,$\bold1$), ($\ydiagram{3}$,$\bold2$,$\bold2$) 
    \end{tabular}
    \caption{Sample of Pati--Salam representations that pick up a phase when taking the $Z_2$ quotient, with $\ydiagram{1} = \bold{4}$ being the fundamental representation of $SU(4)$.}
    \label{tab:PSQuotientMatter}
\end{table}

To verify that the $Z_2$ quotient of Pati--Salam gives the $Z_6$ quotient of the Standard Model group, let us first write the expected result.
\begin{align}
    G_{PS} &\rightarrow \frac{\widetilde{G}_{SM}}{Z_3} \,\,\, ,& \,\,\, \frac{G_{PS}}{Z_2}& \rightarrow \frac{\widetilde{G}_{SM}}{Z_3 \otimes Z_2} = \frac{\widetilde{G}_{SM}}{Z_6},
\end{align}
where $Z_2^{PS} = \{1, \Tilde{\omega}\}$ and $Z_3^{SM} = \{1, \xi^2, \xi^4\}$. As in the case of Trinification let us rewrite the generator, in particular, in place of $n_4$ we can substitute $\tilde \lambda_{15}$
\begin{align}
    \tilde \omega=\exp 2\pi i\left(\frac{\tilde \lambda_{15}+\tilde T_{3L}+\tilde T_{3R}}{2}\right)=\exp 2\pi i\left(\frac{\tilde \lambda_{15}+\tilde T_{3L}+3\tilde T_{3R}}{2}\right)=e^{2\pi i(3 Q_Y+\frac12 \tilde T_{3L})},
\end{align}
whereas one has
\begin{align}
    \xi^3=\exp 3\times2\pi i\left(\frac{n_c}{3}+\frac{\tilde T_{3L}}{2}+Q_Y\right)=e^{2\pi i(3 Q_Y+\frac12 \tilde T_{3L})},
\end{align}
and we have indeed $Z_2\times Z_3=Z_6$ and $p=6$.
 \medskip

 \noindent$\bullet$~{\textbf{Baryon number, Lepton number and $B-L$ symmetry.}} In Pati--Salam, $B$ and $L$ are not individually conserved, as quarks and leptons are grouped in the same multiplet, but $B-L$ is conserved and in fact it is a gauge symmetry embedded in the structure of $SU(4)$ (it breaks at a lower scale when $SU(4)_c \rightarrow SU(3)_c \times U(1)_{B-L}$). Proton decay is not automatic in minimal PS; it depends on the Higgs content and the symmetry breaking pattern. Majorana masses for $\nu_R$ are naturally generated in PS via the seesaw mechanism and imply lepton number violation through processes like neutrinoless double beta decay. 

\subsection{Embedding all of the SM group in a simple group}
\label{sec:3.4}

When all three SM interactions are embedded into a single simple group we have a Grand Unified Theory (GUT). 
In this case the smallest simple group is also the one that leads to the smallest matter content, $SU(5)$. This is realised by the
\textbf{Georgi--Glashow $\mathbf{SU(5)}$} model~\cite{Georgi:1974sy} which combines quarks and leptons into single
irreducible representations. The group structure follows from the Dynkin diagram of Fig.~\ref{DynkincSU(5)} and reads
\begin{align}
    SU(5) \rightarrow SU(3)_c \times SU(2)_L \times U(1)_Y,
\end{align}
with 
\begin{align}
    6Q_Y=&\tilde T_Y, &T_Y^{(F)}&=\textrm{Diag} (-2,-2,-2,3,3),\label{eq:QYSU5}
\end{align}
where we note that $\tilde T_Y$ is given in the fundamental which is in fact not the representation with the smallest $\tilde T_{Y}$ eigenvalue, it is rather the anti-symmetric $\mathbf{10}$.
\begin{figure}[h!]
    \centering    
\begin{tikzpicture}
    \draw [thick] (-2,0) circle (3pt);
    \draw [thick] (-2.1,0) -- (-2.7,0);
    \draw [thick] (-2.8,0) circle (3pt);
    \draw [thick] (-1,0) circle (3pt);
    \draw [thick,dashed] (0,0) circle (3pt);
    \draw [thick,blue,<-] (0.5,0)-- (2,0);
    \draw [thick] (2.7,0) circle (3pt);
    \draw [thick] (2.8,0) -- (3.4,0);
    \draw [thick] (3.5,0) circle (3pt);
    \draw [thick] (3.6,0) -- (4.2,0);
    \draw [thick] (4.3,0) circle (3pt);
    \draw [thick] (4.4,0) -- (5,0);    
    \draw [thick] (5.1,0) circle (3pt);
    \draw (-1,-1) node {$SU(3)_c\times SU(2)_L\times U(1)_Y$};
    \draw (6,0) node {$SU(5)$};
\end{tikzpicture}
    \caption{Embedding of the SM groups in the SU(5)}
    \label{DynkincSU(5)}
\end{figure}
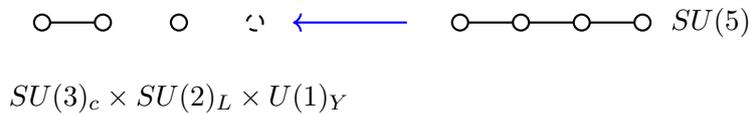

The hypercharge generator in this case is structured as $$\begin{pmatrix}
    SU(3)_c \\& SU(2)_L 
\end{pmatrix} \in SU(5).$$

 \noindent$\bullet$~{\textbf{Value of} $p$.}
Using the realisation of hypercharge $Q_Y$ in the $SU(5)$ GUT of eq.~\eqref{eq:QYSU5}, one has
\begin{align}
    \xi = \exp{\left(\frac{2\pi}{6} i( 2\tilde{\lambda}_8\ + 3\tilde T_{3L} + \tilde T_Y) \right)} 
    = \exp\left(2\pi i\left(\begin{array}{ccccc}
        0 &  & & \\
         & 0 & & \\
         &  & -1 & \\
         & & & 1 & \\
         &&&&0
    \end{array}\right)\right)
\end{align}
so this case leads to $p=6$,
\begin{align}
  G_6 \,\,\, : \,\,\, e^{2\pi i Q_Y}\,e^{2\pi i n_c/3}\, e^{\pi i n_L} = 1.
\end{align}
In this case there is no higher $p$ possible to make this choice variable.

 \noindent$\bullet$~{\textbf{Matter content.}} Representations for the LH and RH fermions are combined and fully fill up new representations as 
\begin{align}
  \Bar{\textbf{5}}  = \begin{pmatrix}
    d^\dagger_r \\ d^\dagger_b \\ d^\dagger_g \\ e \\ -\nu_e
\end{pmatrix} \,\,\, ; \,\,\,
 \textbf{10} = \begin{pmatrix}
    0 & u^\dagger_g & -u^\dagger_b & u_r & d_r \\ 
    - u^\dagger_g & 0 & u^\dagger_r & u_b & d_b \\
    u^\dagger_b & -u^\dagger_r & 0 & u_g & d_g \\ 
    -u_r & -u_b & -u_g & 0 & e^\dagger \\
    -d_r & -d_b & -d_g & -e^\dagger & 0
\end{pmatrix},
\end{align}
and are collected in Table ~\ref{tab:GGmatter}.
\begin{table}[h]
    \centering
    \begin{tabular}{c||c||c||c}
        $\mathcal G=SU(5)$ reps  & $ \bar{\bold5}$ & $\bold{10}$ & $\bold1$ \\ \hline
         SM fields  & $d^{\dagger}_R, \ell_L$ & $q_L, u^{\dagger}_R, e^{\dagger}_R$ & $ \nu^{\dagger}_R$\\
         $n_6$ & 0 &0 &0
    \end{tabular}
    \caption{$SU(5)_{\rm GUT}$ fermions.}
    \label{tab:GGmatter}
\end{table}


After the breaking of $SU(5)$ one picks up the pieces in terms of SM reps as
\begin{align}
  & \Bar{\textbf{5}} \rightarrow (\bar{\bold3}, \bold1)_{1/3} \oplus (\bold1, \bold2)_{-1/2} : d^{\dagger}, \ell,\\
 & \textbf{10} \rightarrow (\bold3, \bold2)_{1/6} \oplus (\bar{\bold3}, \bold1)_{-2/3} \oplus (\bold1, \bold1)_1 : q, u^{\dagger}, e^{\dagger},\\
  & \textbf{1} \rightarrow (\bold1, \bold1)_0 : \nu.
\end{align} 

\textbf{Particle count:}  No new fermions, this being the most economical model from the fermion count side. Also note that necessarily in this case all the spectrum has $n_6=0$. 
 \medskip

 \noindent$\bullet$~{\textbf{Symmetry breaking.}}
An adjoint Higgs field $\Sigma$ in the \textbf{24} representation of $SU(5)$ breaks $SU(5)\rightarrow \widetilde{G}_{SM}/Z_6$ with vev $\langle \Sigma \rangle = v_{GUT} T_{24}$ and gives mass to heavy bosons X,Y, where $T_{24}$ is the diagonal generator of $SU(5)$.
Then, a Higgs field $H$ in the fundamental representation \textbf{5} with $\langle H\rangle = (0,0,0,v,0)$ breaks electroweak symmetry and gives mass to SM fermions and EW bosons. The breaking symmetry pattern is 
\begin{equation}
SU(5) \xrightarrow[v_{GUT}]{\bold{24}} \widetilde{G}_{SM}/Z_6. 
\end{equation} 

To figure out the electromagnetic charge $Q_{em}$ and new quantum number $n_6$ of the Higgses, we have to look at the generators of the representations they belong in,

\begin{align}
    \bold 5 \rightarrow (\bold3,\bold1)_{-1/3} \oplus (\bold1,\bold2)_{+1/2}.
\end{align}
 The hypercharge generator for the fundamental as we saw previously is defined as $Q_Y=\text{diag}(-1/3, -1/3, -1/3, 1/2, 1/2)$, while the $SU(2)_L$ generator is $T_{3L}=\text{diag}(0, 0, 0, 1/2, -1/2)$.

As for the GUT Higgs field, looking at the decomposition of the 24 representation into SM ones will help us determine its hexality,
\begin{equation}
    \bold{24} \rightarrow (\bold8,\bold1)_0 \oplus (\bold1,\bold3)_0 \oplus (\bold1,\bold1)_0 \oplus (\bold3,\bold2)_{-5/6} \oplus (\bar{\bold3},\bold2)_{5/6}.
\end{equation}

\textbf{Particle count:} 12 more gauge bosons, a new colour triplet Higgs coming from the $\bold5$ (which also includes the SM Higgs doublet) and 12 massive Higgses.

 \medskip

 \noindent$\bullet$~{\textbf{Quotients of the embedding group.}} The centre of the Georgi--Glashow group $\mathcal{Z}(SU(5))=Z_5$ yet the matter content does not allow one to take the quotient and $Z_5$ has no subgroups.
$SU(5)$ representations like the anti-fundamental $\boldsymbol{\bar{5}}$ and the antisymmetric $\bold{10}$, which make up a generation of quarks and leptons, pick up a phase $\propto e^{2\pi i/5}$ and are thus not invariant under the action of the quotient. Should one insist on taking the quotient fermions would be forced to higher $n_5=0$ representations like $\bold{24}$, with a dramatic increase in the number of new particles needed.
 \medskip

 \noindent$\bullet$~{\textbf{Baryon number, Lepton number and $B-L$ symmetry.}} Quarks and leptons are combined in irreducible representations in the $SU(5)$ GUT, so the $B$ and $L$ numbers are explicitly broken. $SU(5)$ predicts $B$ and $L$ number violating processes (like proton decay via 6-dimensional operators) but $B-L$ remains conserved.



\section{The particle spectrum with compositeness degree}\label{sec:PSpeck}

We have seen instances of group embeddings that lead to each $p=1,2,3,6$ with $k=0$. For each of the groups considered the $p$ value is built into the theory, a relation that the generators satisfy regardless of the matter content. One can then keep the same $p$ by staying with the same $\mathcal G$ group but change $k$ by placing SM particles in a different, higher dimensional, representations. It all can be reduced to the placement of $q_L$, given all other SM representations have hypercharge that is a multiple of $q_L$'s. Let us outline next for each group the first few instances of $k\neq 0$ as follow from the representaion $q_L$ is put in, with the convention in the following for simplicity that the subindex is $\tilde T_Y$ rather than $Q_Y$.   
\begin{itemize}
    \item $SU(3)_c\times SU(2)_L\times SU(2)_Y$. Here the decorrelation among colour, weak-iosspin and hyperhcarge  means that one can put the LH quark in the representation immediately above the $\mathbf{2}$ of $SU(2)_Y$, i.e.
    \begin{align}\label{eq:k1SU2Y}
        q_L\in \left(\mathbf{3},\mathbf{2},\mathbf{3}\right),
    \end{align}
    this requires revisiting the connection of hypercharge
    \begin{align}\label{eq:k1SU2Y2}
        \tilde T_{Y}q_L\,=\,2q_L\,, \qquad 12Q_Y=\tilde T_{Y}.
    \end{align}
    The group itself satisfies the very same relation defining $p$, $e^{2\pi i \tilde T_{Y}}=1$ which however reads differently in terms of the newly branded hypercharge
    \begin{align}\label{eq:k1SU2Y3}
          e^{2\pi i \tilde T_{Y}}&=e^{2\pi i 12 Q_Y}=e^{2\pi 6(1+k)Q_Y},
    \end{align}
    from whence it follows that $k=1$.
    \item $(SU(3)_c\times SU(2)_L\times SU(2)_Y)/Z_2$. This instance restricts the choices; in particular, the choice in eq.~\eqref{eq:k1SU2Y} is no longer allowed since $Z_2$ as given in eq.~\eqref{eq:SU2Yquot} requires the $SU(2)_L$ and $SU(2)_Y$ dualities to add up to an even number $n_{L}+n_{Y}=0\mod 2$. Instead the next representation available is
    \begin{align}
        q_L=(\mathbf{4},\mathbf{2})_{3}\in \left(\mathbf{3},\mathbf{2},\mathbf{4}\right)\,, \qquad  \tilde T_Y q_L&=3q_L,
    \end{align}
    where $(\mathbf{3},\mathbf{2})_{3}$ is the representation under $(SU(3)_c,SU(2)_L)_{\tilde T_Y}$,
    and the connection of hypercharge and $T_Y$ is now a factor of  $18$ which when equated to $6(1+2k)$ returns $k=1$.
    \item $SU(3)_c\times SU(3)_L\times SU(3)_R$. The case of $p=2$ has more structure to it, in particular, if we are to put $q_L$ in a representation of $SU(3)_L$ with  $\tilde\lambda_{8L}$ eigenvalue different from $1$, it should still be a doublet under $SU(2)_L$, i.e. have $\tilde T_{3L}$ eigenvalues $\pm1$. The possible eigenvalues of $\tilde T_{3L}, \tilde T_{8L}$ are however correlated, as one can see inspecting the weight diagrams of the $SU(3)$ Lie algebra, one of them the renowned eight-fold way. Let us proceed by inspection of the first few $SU(3)_L$ representations recalling $\tilde T_{Y}=(-\tilde\lambda_{8L})\oplus(-3\tilde\lambda_{3R}-\tilde \lambda_{8R})$,
    \begin{align}\label{eq:kneq0Tri}
        \bar{\mathbf{3}}\to \mathbf{2}_1\oplus \mathbf{1}_{-2} \,, \quad  \mathbf{3}\to\mathbf{2}_{-1}\oplus \mathbf{1}_{2}\,, \quad 
         \mathbf{8}\to \mathbf{3}_0\oplus\mathbf{2}_{3}\oplus \mathbf{2}_{-3}\oplus\mathbf{1}_0.
    \end{align}
    Each of these new possibilities will require rescaling the relation between $\tilde T_{Y}$ and $Q_Y$ as follows
    \begin{align}
        q_L\in\,\mathbf{3} \Rightarrow (1+2k)=-1\,, \qquad 
        q_L\in\,\mathbf{8} \Rightarrow (1+2k)=\pm 3,
    \end{align}
    which would lead respectively to $k=-1,1,-2$. Let us highlight that the relative change in hypercharge definition is now not any integer but has to conform to $1+2k$ and the fact that it does is guaranteed by $p$ being set to 2 in our group $\mathcal G$ choice.
    \item $(SU(3)_c\times SU(3)_L\times SU(3)_R)/Z_3.$ This case sets extra constraints since representations should be invariant under $e^{2\pi i(n_c+n_{3L}+n_{3R})/3}$. Given $q_L$ has $n_c=1$ one needs $(n_{3L},n_{3R})= (2,0),(0,2)$ or $(1,1)$. Revisiting the choices in eq.~\eqref{eq:kneq0Tri} one has that for $q_L \in \mathbf{3}$ of $SU(3)_L$, we need $n_{3R}=1$ while for $q_L \in \mathbf{8}$ of $SU(3)_L$ $n_{3R}=2$, two possibilities would be
    \begin{align}
    q_L \in & (\mathbf{3},\mathbf{3},\mathbf{3})\to( \mathbf{3},(\mathbf{2}_{-1}\oplus \mathbf{1}_{2})\otimes (\mathbf{1}_{-4}\oplus\mathbf{2}_2))=(\mathbf{3},\mathbf{2})_{-5}\oplus \dots ,\\
     q_L \in &  \left(\mathbf{3},\mathbf{8},\bar{\mathbf{3}}\right)\to (\mathbf{3},(\mathbf{3}_0\oplus\mathbf{2}_{3}\oplus \mathbf{2}_{-3}\oplus\mathbf{1}_0)\otimes (\mathbf{1}_{4}\oplus\mathbf{2}_{-2}))=(\mathbf{3},\mathbf{2})_{7}\oplus(\mathbf{3},\mathbf{2})_{1} \oplus \dots ,
    \end{align}
    and we get that the first few ratios of hypercharge are
    \begin{align}
        (1+6k)= 1,-5,7,
    \end{align}
    in accordance with $p=6$.
    
    \item $SU(4)\times SU(2)_L\times SU(2)_R$. For Pati--Salam, the next representations break down as (with $\mathbf{d}_{\tilde T_Y}$ and $\mathbf{d}$ an $SU(3)_c$ irrep)
    \begin{align}
        \mathbf{6}\to \,\bar{\mathbf{3}}_2\oplus \mathbf{3}_{-2} \,, \quad  \mathbf{10}\to \mathbf{6}_2\oplus\mathbf{3}_{-2}\oplus \mathbf{1}_{-6} 
        \,, \quad \mathbf{15}\to \mathbf{8}_0\oplus\mathbf{3}_4\oplus\bar{\mathbf{3}}_{-4}\oplus\mathbf{1}_0,
    \end{align}
    again we can see the pattern and obtain the first two non-trivial $k$ values
    \begin{align}
        q_L\in\,\mathbf{6}\textrm{\,\,or\,\,}\mathbf{10} \Rightarrow (1+3k)=-2 \,, \qquad 
        q_L\in\,\mathbf{15} \Rightarrow (1+3k)=4.
    \end{align}
    \item $(SU(4)\times SU(2)_L\times SU(2)_R)/Z_2$. In this case, it is $(n_4+n_L+n_R)$ that should add up to an even number. Both our examples above have $n_4$ even, yet for a $SU(2)_L$ doublet we need $n_L=1$ which forces $n_R=1$ and our solutions are upgraded to
    \begin{align}
        q_L\in(\mathbf{6},\mathbf{2},\mathbf{2})\to &((\bar{\mathbf{3}}_2\oplus \mathbf{3}_{-2})\otimes(\mathbf{1}_{3}\oplus \mathbf{1}_{-3}), \mathbf 2)\\&=(\mathbf{3},\mathbf{2})_{1}\oplus (\mathbf{3},\mathbf{2})_{-5}\oplus \dots ,\\
        q_L\in(\mathbf{15},\mathbf{2},\mathbf{2})\to &((\mathbf{8}_0\oplus\mathbf{3}_4\oplus\bar{\mathbf{3}}_{-4}\oplus\mathbf{1}_0)\otimes(\mathbf{1}_{3}\oplus \mathbf{1}_{-3}),\mathbf 2)\\&=(\mathbf{3},\mathbf{2})_{1}\oplus (\mathbf{3},\mathbf{2})_{7}\oplus \dots ,
    \end{align}
    while the subindexes in $(\mathbf{3},\mathbf{2})_{\tilde T_Y}$ are all of the form $(1+6k)$, compatible with $p=6$ as expected.
    \item In $SU(5)$, we have that the pattern is more spread out in hypercharge; the first two options can be found as follows
    \begin{align}
        \mathbf{10} &\to (\mathbf{3},\mathbf{2})_{1}\oplus\dots , &\mathbf{24}&\to (\mathbf{3},\mathbf{2})_{-5}\oplus\dots , & \ydiagram{2,2,1,1}=\mathbf{45}&\to (\bar{\mathbf{3}},\mathbf{2})_{-7}\oplus\dots
    \end{align}
    The theory would lead to a much larger matter content, but here the emphasis is on how $p=6$ is built into the possible hypercharges and one has indeed that
    \begin{align}
       q_L\in &\,\,\mathbf{24} \quad (1+6k)=-5,& q_L\in \,\,&\mathbf{45} \quad (1+6k)=7.
    \end{align}
\end{itemize}

Let us now turn to a model that was considered in the days of GUTs as an illustrative case of fractional charge, yet it is based on a simple embedding group, SU(7) \cite{Li:1981un,Farhi:1979zx,Umemura:1981bw}.
The model partially uses the matter assignment of $SU(5)$ but crucially extends hypercharge out of the $SU(5)$ Lie algebra and into $SU(7)$ as
\begin{align}
    Q_Y=\textrm{Diag}(1/3,1/3,1/3,-1/2,-1/2,q,-q)\equiv \frac{1}{6}\tilde T_{24}+q\,\tilde T_{3},
\end{align}
where, given $SU(3)_c\times SU(2)_L$ are embedded as in $SU(5)$, the electric charge of the two new states is $q$, $-q$. We will leave $q$ free but it is worth considering which values it could take. A full analysis would look at the roots and weights of $\tilde T_{24}$ and $\tilde T_{3}$ including possible correlations between the two, here we simply note that with our definition of tilded operators the smallest weight is $1$ and any other representation will have hence integer weights under each, dubbed $w_{24}$ and $w_3$. Let $|w,R\rangle$ be an element of a representation $R$ which develops a vev and breaks the symmetry, it should still leave $U(1)_Y$ unbroken, i.e.
\begin{align}
    \left(\frac{1}6 \tilde T_{24}+q\tilde T_3\right)|w,R\rangle=
    \left(\frac{1}6 w_{24}+q w_3\right)|w,R\rangle=0,
\end{align}
we conclude $q=-w_{24}/(6w_3)$ with $w_{3},w_{24}\in \mathbb{Z}$. This simple analysis then tells us that $q$ could be a rational number but never irrational. 

It is not the purpose of this case study however to map which representations realise a breaking consistent with a given $q$ (even `simple' values of $q$ as $1/3$ need scalar representations such as the $\mathbf{21}$ and $\mathbf{840}$~\cite{Li:1981un}) but instead to show the predictive power of our group analysis of the Standard Model.

Given $q$ is rational we can always write it as $q=N/D$ with $N,D\in \mathbb{Z}$ and $N,D$ co-primes, i.e. they have no common divisor other than one. For coprimes a useful result that follows from B\'ezout's identity is that $N$ has a multiplicative inverse modulo $D$, which we can use to find the smallest possible (colour neutral) electric charge  as follows. The mathematical result reads
\begin{align}
    q=\frac{N}{D}\,\, \textrm{with}\,\, N,D\,\,\textrm{coprimes} \Rightarrow \exists\,\, y \in\mathbb{Z} \,\,\textrm{such that} \quad y N= 1 \mod D,
\end{align}
which we can rewrite as
\begin{align}
    x,y\in\mathbb{Z} \quad \textrm{such that} \quad y\frac{N}{D}-x=y q-x=\frac{1}{D},
\end{align}
and read this result as a physicist: combining $y$ particles of charge $q$ and $x$ charge $-1$ particles (take electrons if concreteness helps) we find the smallest quantum of charge to be $1/D$. On the other hand in the deep infrared of QED we derived the smallest quantum of charge as a function of $p,k$ so we simply equate
\begin{align}
    D=|6/p+6k|.\label{eq:Dqpk}
\end{align}
As advertised, the RHS covers the whole of the positive integers in a one to one correspondence with $(p,k)$ so solving the equation above uniquely determines $G_p$ and $k$. The values so found can be checked against the defining property of $G_p$,
\begin{align}\xi^{6/p}=e^{2\pi i\frac{Q_6(k)}{p}}&=\left(\begin{array}{ccc}
        \mathbb{I}_{3\times 3}e^{2\pi i(1+pk-1)2/p} &  &\\
         &\mathbb{I}_{2\times 2}e^{2\pi i(1+pk-1)3/p} &\\
         &&e^{2\pi i ((6/p+6k)q)\tilde T_3}
    \end{array}\right)\\
    &=\left(\begin{array}{ccc}
        \mathbb{I}_{3\times 3}e^{2\pi i2k} &  &\\
         &\mathbb{I}_{2\times 2}e^{2\pi i3k} &\\
         &&e^{\pm 2\pi i N\tilde T_3}
    \end{array}\right)=1.
\end{align}

In summary, our approach tells us that e.g. the choice of $q=1/3$ which is taken in~\cite{Li:1981un} will lead to $p=2$ regardless of the details of the breaking. Further, were one to set $q=1/7$, the SM group would be $G_6$ and compositeness degree $k=1$.

\section{The monopole spectrum}
\label{sec:mono}

The aim of this section is to construct the spectrum of magnetically charged states in the SM that is consistent with its global gauge group structure. These states are magnetic monopoles that carry conserved magnetic charges of the QED sector of the SM, as well as the non-Abelian magnetic charges associated with the strong and weak sectors $SU(3)_c$ and $SU(2)_L$. 

The central question is:
which magnetically charged states can be allowed in the SM with the gauge group $G_p$? We shall address it in two parts, first by asking which magnetically charged probes can be consistently added to the to the theory. Then we can discuss whether these probe charges can be realised by dynamical magnetic monopoles, i.e. non-singular stable field configurations of finite energy.

\subsection{Spectrum of `t~Hooft lines in $\mathbf{G_p}$}

Inclusion of magnetic probes (i.e. non-dynamical magnetic degrees of freedom) is tantamount to adding 't~Hooft line operators to the theory. These 't~Hooft lines are the world-lines of Dirac monopoles; they are analogous to the Wilson lines that describe world-lines of the more familiar electric probes of the theory, e.g. heavy non-dynamical quarks or leptons charged under $G_p$. The spectrum of Dirac monopoles or 't~Hooft lines in a given theory is determined operationally by solving constraints imposed by the Dirac quantisation conditions.  

QED provides the simplest example to illustrate this approach. 
A Dirac monopole of magnetic charge $q_m$ in the $U(1)_{\textrm{em}}$ theory
is described by the gauge field configuration,
\begin{equation}
{A}_\mu
\,=\, \frac{q_m}{2}\, (1-\cos \theta) \, \partial_\mu \phi\,,
\label{eq:ADirqm}
\end{equation}
where $\theta$ and $\phi$ denote the conventional polar and azimuthal angles of the spherical polar coordinates and the monopole is at the origin, i.e. $r=0$. This gauge potential is singular on the ray at $\theta=\pi$ which corresponds to the Dirac string that originates from the monopole centre at the origin and stretches along the negative $z$-axis.\footnote{The alternative but physically equivalent 
approach~\cite{Wu:1975es} which does not involve the Dirac string is to define the monopole by two distinct gauge potentials on the upper and the lower hemisphere, and request that they are related by a gauge transformation on the equator eqs.~\eqref{eq:WuYangdef}-\eqref{eq:Uwuyang}. The existence of this gauge map  is equivalent to the requirement below in eq.~\eqref{eq:Wloop1}.}
For a matter field $\psi_{q_e}$ with an electric charge $q_e$, the Dirac quantisation condition for the pair $(q_e,q_m)$ is derived by considering a parallel transport 
of $\psi_{q_e}(x)$ along a closed contour winding around the monopole's Dirac string. 
The un-observability of the Dirac string implies that the Wilson loop corresponding to
 $\psi_{q_e}$ along the closed contour, is trivial,
\begin{equation}
e^{i q_e \oint_C {A}_\mu dx^\mu}\, =\, 1\,.
\label{eq:Wloop1}
\end{equation}
Evaluating the exponent on the monopole configuration and imposing the constraint eq.~\eqref{eq:Wloop1} results in\footnote{Note that Stokes's theorem allows to turn this line integral into the surface integral of the magnetic field~\cite{Corrigan:1976wk}, i.e. the magnetic flux $2\pi q_m$ or conserved magnitude of the 1-form magnetic symmetry discussed in sec.~\ref{sec:1.2}.}
\begin{equation}
 q_e \oint_C {A}_\mu{|}_{\theta=\pi} \, dx^\mu\, =\, q_e q_m \int_0^{2\pi} d\phi \,\,=\, 2\pi n,
\label{eq:Wloop11}
\end{equation}
implies that 
\begin{equation}
 U(1) \,\,{\rm theory}\,:\qquad  q_e q_m 
\,\in\,  \mathbb{Z}\,.
\label{eq:Wloop2}
\end{equation}
In particular, if the minimal electric charge in the $U(1)$ model is $|q_e|=1$, as is the case for electrons/positrons in QED, the minimally charged Dirac monopole has $q_m=1$, and for the anti-monopole, $q_m=-1$. Higher integer-valued $q_m$ correspond to multi-monopole Dirac configurations. In summary, the spectrum of magnetically charged states in QED is given by Dirac monopoles or 't~Hooft lines with $U(1)$ magnetic charges $q_m = (1/{q_e^{\,\rm min}})  \,\mathbb{Z}$, where $q_e^{\,\rm min}$ is the 
minimal electric charge of the matter fields present in the theory, and both electric and magnetic charges are quantised so that the Dirac quantisation conditions eq.~\eqref{eq:Wloop2} are satisfied for all  $(q_e,q_m)$.

\medskip

't~Hooft lines in a general non-Abelian gauge theory $G$ are obtained by embedding the gauge configuration of the Dirac monopole probe
eq.~\eqref{eq:ADirqm} into the Cartan subalgebra of the Lie group. The monopole gauge configuration (on the Northern hemisphere) 
takes the form,
\begin{equation}
{A}_\mu
\,=\, \frac{\vec{g}_m\cdot \vec{T}}{2}\, (1-\cos \theta) \, \partial_\mu \phi\,,
\label{eq:ADirG}
\end{equation}
where $T^1,\ldots,T^r$ are the generators of the Cartan subalgebra of $G$ and $g_m^1,\ldots,g_m^r$ are constants that have the meaning of the monopole's magnetic charges for each of the Abelian subgroups generated by $T^i$s.
The Dirac quantisation condition eq.~\eqref{eq:Wloop1} now involves the sum of the contributions of each Cartan factor in the exponent,
\begin{equation}
e^{2\pi i \, \sum_{i=1}^r q_e^i g_m^i  T^i }\, =\, 1\,.
\label{eq:WloopSM}
\end{equation}
Here $q_e^i$ denote the `electric' charges of the matter fields under each Cartan generator of the SM gauge group. 
In the non-Abelian sectors of the SM,  $q_e^i =1$, otherwise the corresponding $q_e$ is given by the hypercharge $Q_Y$ of the matter field,
\begin{align}
	q_e^i = \left\{\begin{aligned}
1 \qquad 		& \text{for } \, SU(3)_c \,\,  \text{and} \,\, SU(2)_L\\
{\,Q_Y} \quad \,				& \text{for } \, U(1)_Y 
\end{aligned}
 \right. .
\end{align}
Having specified the values of electric charges $q_e^i$, we can now write down the general solution of the Dirac quantisation conditions eq.~\eqref{eq:WloopSM} and thus determine the spectrum of magnetic probe charges allowed in the SM with the general global gauge group $G_p$. These are given by the gauge field potential eq.~\eqref{eq:ADirG} with the magnetic flux matrix which we write in the form:
\begin{equation}
\vec{g}_m\cdot \vec{T}
\,\,=\,\, \begin{pmatrix} n_1-\frac{N}{3}& &\\ & n_2 -\frac{N}{3}&\\ & & -\frac{N}{3}
\end{pmatrix} 
\oplus
\begin{pmatrix} \frac{m}{2} & \\ &-\frac{m}{2}
\end{pmatrix} 
\oplus \, q_m\,,
\label{eq:FmgSM1}
\end{equation}
where $N\equiv n_1+n_2$ so that $(n_1,n_2,m,q_m)$ parametrise an arbitrary element of the Cartan subalgebra and we have chosen the fundamental representation (both of $SU(3)_c$ and $SU(2)_L$) for an explicit form of $\vec T$. Having parametrised $\vec g_m$ we can solve for the constraint in eq.~(\ref{eq:WloopSM}); a prerequisite for it to be satisfied is each of the different sectors producing an overall colour- and isospin-independent phase. This is automatic for $q_m$ but for the non-Abelian sector it requires $N$, $n_1$, $m$ being integer-valued, so that
\begin{equation}
e^{2\pi i \, \sum_{i=1}^r q_e^i g_m^i  T^i }=e^{2\pi i \,\left( -n_c \frac{N}{3} \,+\,n_L  \frac{m}{2} \, +\, Q_Y q_m \right) }\, =\, 1\,,
\label{eq:DQC2}
\end{equation} 
Finally, one can input the constraint on the spectrum that correlates different sectors across themselves, eq.~\eqref{eq:1.6}, here reproduced for convenience
\begin{equation}
 6Q_Y+2n_c+3n_L=p\, \mathbb{Z}\,,
\end{equation}
to find
\begin{equation}
e^{2\pi i \,\left( -n_c \frac{q_m+N}{3} \,+\,n_L  \frac{m-q_m}{2} \, +\, \frac{p}{6}q_m \mathbb Z  \right) }\, =\, 1\,.
\label{eq:DQC3}
\end{equation} 
This substitution effectively takes into account the restriction on the possible hypercharges, but the equation above should still be true for any $n_c$ and $n_L$, hence:
\begin{align}\label{eq:MplSol}
    q_m&=\frac{6}{p}\mathbb{Z}  \,\,, \qquad
    N=-q_m+3\mathbb Z  \,\,,\qquad
    m=q_m+2\mathbb Z    \,.     
\end{align}
These three equations capture all $p$ cases, from $p=1$ in which $q_m=6\mathbb{Z}$ and $N$ and $m$ decorrelate since $-6\mathbb{Z}+3\mathbb{Z}=3\mathbb{Z}$, $6\mathbb{Z}+2\mathbb{Z}=2\mathbb{Z}$, to the fully correlated $p=6$ case. Specifically, for all values of $p$, we have:
 \begin{eqnarray}
 \label{eq:Nmq1}
 p=1: \qquad &N=3\mathbb{Z}, \quad m=2\mathbb{Z}, \quad &q_m =6\mathbb{Z},\\
  p=2: \qquad &N=3\mathbb{Z}, \quad m=\mathbb{Z}, \quad &q_m =3m +6\mathbb{Z},\\ 
  \label{eq:Nmq3}
  p=3: \qquad &N=\mathbb{Z}, \quad m=2\mathbb{Z}, \quad &q_m =2N +6\mathbb{Z},\\ 
  \label{eq:Nmq6}
  p=6: \qquad &N=\mathbb{Z}, \quad m=\mathbb{Z}, \quad &q_m =2N+3m +6\mathbb{Z}.
\end{eqnarray}
Equations~\eqref{eq:Nmq1}-\eqref{eq:Nmq6} specify magnetic fluxes for all `t~Hooft lines allowed in the Standard Model with the gauge group $G_p$ for all values of $p$. The richest spectrum on `t~Hooft lines occurs in the minimal Standard Model with the gauge group $G_6$ and their magnetic flux numbers in eq.~\eqref{eq:Nmq6} are in agreement with the results of Ref.~\cite{vanBeest:2023mbs}. 

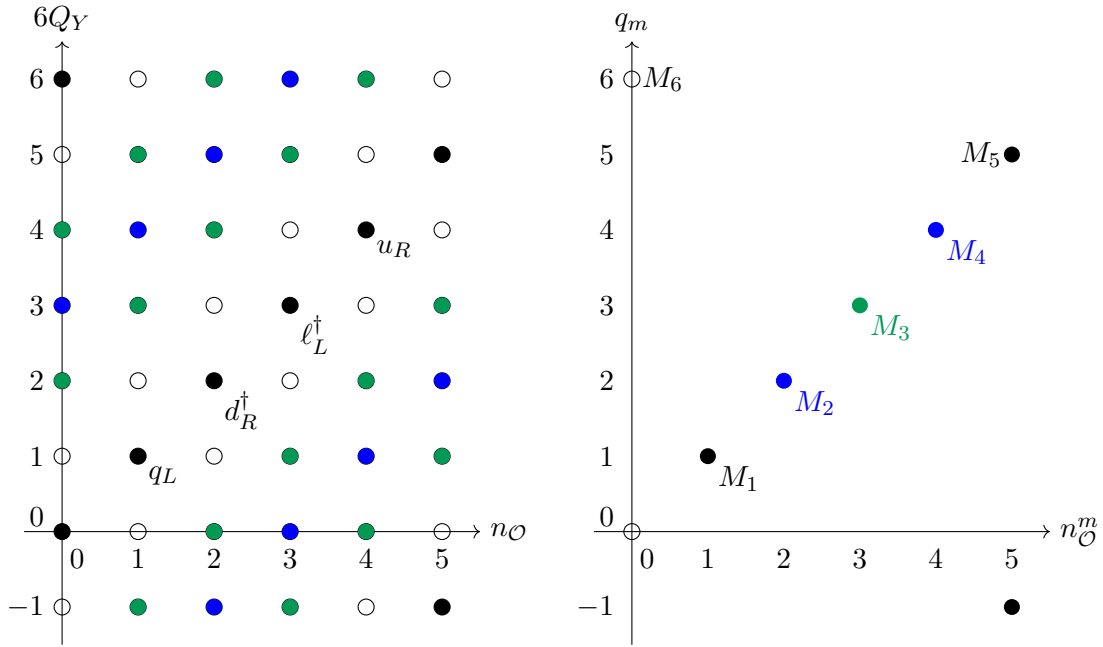
\begin{figure}
    \centering
    \begin{tikzpicture}[scale=1]
  \draw[->] (-0.5,0) -- (5.5,0) node[right] {$n_{\mathcal{O}}$};
  \draw[->] (0,-1.5) -- (0,6.5) node[above] {$6Q_Y$};
    \node[left] at (-0.1,-1) {$-1$};
  \node[left] at (-0.1,0.2) {$0$};
  \node[left] at (-0.1,1) {$1$};
  \node[left] at (-0.1,2) {$2$};
  \node[left] at (-0.1,3) {$3$};  
  \node[left] at (-0.1,4) {$4$};
  \node[left] at (-0.1,5) {$5$};
    \node[left] at (-0.1,6) {$6$};
  \foreach \x in {1,...,5}
    \node[below] at (\x,-0.1) {\x};
  \node[below] at (0.2,-0.1) {0};  
\foreach \x in {0,...,5} {
    \foreach \y in {-1,...,6} {
      \draw (\x,\y) circle (3pt);
    }
  }
  \fill (0,0) circle (3pt);
  \fill (5,-1) circle (3pt);
  \fill (0,6) circle (3pt);
  \fill (1,1)  node [anchor=north west] {$q_L$} circle (3pt);
  \fill (2,2)  node [anchor=north west] {$d_R^\dagger$}circle (3pt);
  \fill (3,3)  node [anchor=north west] {$\ell_L^\dagger$} circle (3pt);
  \fill (4,4)  node [anchor=north west] {$u_R$} circle (3pt);
  \fill (5,5) circle (3pt);
  \fill [blue](0,3) circle (3pt);
  \fill [blue](4,1) circle (3pt);
  \fill [blue](5,2) circle (3pt);
  \fill [blue](3,0) circle (3pt); 
  \fill [blue](1,4) circle (3pt);
  \fill [blue](2,5) circle (3pt);
  \fill [blue](3,6) circle (3pt);
  \fill [blue](2,-1) circle (3pt);
  \fill [ForestGreen] (0,2) circle (3pt);
  \fill [ForestGreen] (1,3) circle (3pt);
  \fill [ForestGreen] (0,4) circle (3pt);
  \fill [ForestGreen] (1,5) circle (3pt);
  \fill [ForestGreen] (2,0) circle (3pt);
  \fill [ForestGreen] (3,1) circle (3pt);
  \fill [ForestGreen] (2,4) circle (3pt);
  \fill [ForestGreen] (3,5) circle (3pt);
  \fill [ForestGreen] (4,0) circle (3pt);
  \fill [ForestGreen] (5,1) circle (3pt);
  \fill [ForestGreen] (4,2) circle (3pt);
  \fill [ForestGreen] (5,3) circle (3pt);
  \fill [ForestGreen] (4,6) circle (3pt);
  \fill [ForestGreen] (2,6) circle (3pt);
  \fill [ForestGreen] (1,-1) circle (3pt);
  \fill [ForestGreen] (3,-1) circle (3pt);
\end{tikzpicture}\quad\begin{tikzpicture}[scale=1]
  \draw[->] (-0.5,0) -- (5.5,0) node[right] {$n_{\mathcal{O}}^m$};
  \draw[->] (0,-1.5) -- (0,6.5) node[above] {$q_m$};
  \node[left] at (-0.1,-1) {$-1$};
  \node[left] at (-0.1,0.2) {$0$};
  \node[left] at (-0.1,1) {$1$};
  \node[left] at (-0.1,2) {$2$};
  \node[left] at (-0.1,3) {$3$};  
  \node[left] at (-0.1,4) {$4$};
  \node[left] at (-0.1,5) {$5$};
  \node[left] at (-0.1,6) {$6$};
  \foreach \x in {1,...,5}
    \node[below] at (\x,-0.1) {\x};
  \node[below] at (0.2,-0.1) {0};
  \draw (0,0) circle (3pt);
  \draw (0,6) node[right] {$M_6$} circle (3pt);
    \fill (5,-1) circle (3pt);
  \fill (1,1)  node [anchor=north west] {$M_1$} circle (3pt);
  \fill [blue] (2,2)  node [anchor=north west] {$M_2$}circle (3pt);
  \fill [ForestGreen] (3,3)  node [anchor=north west] {$M_3$} circle (3pt);
  \fill [blue](4,4)  node [anchor=north west] {$M_4$} circle (3pt);
  \fill (5,5) node[left] {$M_5$} circle (3pt);
\end{tikzpicture}
    \caption{Stairway structure for electric charges on the LHS (see Fig.~\ref{fig:novsQY}) and for monopoles on the RHS. RHS: empty nodes are possible for all $p$ and are the only ones possible for $p=1$. For $p=2$ ($p=3$) in addition to empty, green (blue) nodes are possible, while $p=6$ allows for all the above and black nodes.}
    \label{fig:MplStairway}
\end{figure}

In general there is an infinite 4-dimensional lattice of 't~Hooft lines available for each SM group $G_p$, 
characterised by the set of four integers 
$(n_1,n_2,m,q_m)$ subject to the constraints in eq.~\eqref{eq:MplSol} or, equivalently, in eqs.~\eqref{eq:Nmq1}-\eqref{eq:Nmq6}. One can further remove the residual discrete gauge degeneracy between some of these monopole states 
by requiring for example that $n_1\ge n_2\ge 0$ and $m\ge 0$. This condition fixes the ordering of the diagonal elements  
of the magnetic flux matrix eq.~\eqref{eq:FmgSM1}.  To visualise this 4-dimensional lattice of magnetic probes in the $G_p$ theory, one can select three arbitrary integers $(n_1,n_2,m)$ and compute $q_m$ using eqs. \eqref{eq:Nmq1}-\eqref{eq:Nmq6} as
\begin{equation}
 q_m\,=\, n_\mathcal{O}^m +6w\,, \quad \textrm{where} \quad
 n_\mathcal{O}^m \,:=\,  2N+3m \mod 6\,, \quad \textrm{and} \quad
 w=\mathbb{Z}.
\end{equation}
As the result, there is a magnetic probe or a `t~Hooft line for each choice
of the four integers ($n_1,n_2,m,w$). Let us note that for $G_p$ theories with $p<6$ the monopole spectrum is a subset of $p=6$ in the inverse situation of the electric spectrum where $p=6$ has the smallest elecctric spectrum.

One can plot the spectrum of allowed monopoles in the $n_\mathcal{O}^m$ vs $q_m$ plane as shown in Fig.~\ref{fig:MplStairway}, side by side with the analogous plot for electric charges. 
The benefit of displaying the monopole spectrum on the $(n_\mathcal{O}^m\,,\,q_m)$ plane is that each node specified by a fixed value of $n_\mathcal{O}^m$ coordinate is a representative of a class of monopoles
which can decay into each other by emitting massless non-Abelian gauge bosons (see the next subsection for more details). But nodes with different values of $n_\mathcal{O}^m$ cannot be connected to each other by emitting such non-Abelian gauge boson radiation. Indeed $n_\mathcal{O}^m=2N+3m \mod 6$ is insensitive to $N\to N+3\mathbb{Z}$ and $m\to m+2\mathbb{Z}$ i.e. non-Abelian gauge boson emission, whereas for each different value of $q_m$, topology provides a conservation law. 

\medskip

 We can now verify the validity of the 
 equation~\eqref{eq:n6mdefExp} that we quoted in sec~\ref{sec:1.2} for all `t~Hooft lines in eqs.~\eqref{eq:Nmq1}-\eqref{eq:Nmq6}
  and determine the values of $n_6^m$.
  Note that in the notation used in eq.~\eqref{eq:n6mdefExp}, the generators in the linear combination $\vec g_m \cdot\vec T$
  were normalised in such a way as to include the electric charges of the SM matter field representations. 
  In the normalisation we use now in eq.~\eqref{eq:FmgSM1}, we need to multiply the generators by their electric charges $q_e^i$ and consider the exponent in the form of eq.~\eqref{eq:WloopSM} on the LHS of eq.~\eqref{eq:n6mdefExp}.
  Let us substitute the solution for $N,m$ in eq.~\eqref{eq:MplSol} back in eq.~\eqref{eq:DQC2}
 \begin{align}
\label{eq:n6mdefExp11}
e^{2\pi iq_m\left(Q_Y+\frac{n_L}{2}+\frac{n_c}{3}\right)}=e^{2\pi iq_m\frac{Q_6}{6}}
   \,=\, \exp\left[2\pi i\,\frac{n_6^\textrm{m} Q_6}{6}\right],
\end{align}
where the last line reproduces eq.~\eqref{eq:n6mdefExp}. The conclusion is therefore that the linear combination of triality, duality and hypercharge does align along the $Q_6$ direction, as was assumed in sec.~\ref{sec:1.2}, and further the connection between the hypercharge flux and $n_6^m$ (recalling it is defined from 0 to 5) is simply
\begin{align}
    n_6^m=q_m\mod6.
\end{align}
This equation provides the explicit realisation of the statement that 1-form symmetries provide a finer grading of the topological classification, any stable monopole can be characterized by $q_m$, a topologically conserved $U(1)$ flux. However topology does not distinguish between monopoles that also have non-Abelian quantum numbers which is where the finer grading of 1-form symmetry comes in, i.e. $q_m=n_6^m+6\mathbb{Z}$ with the allowed values of $n_6^m$ given by $Z^{\textrm{mag},(1)}_{p}$.


The parallels and differences between the electric and magnetic spectrum are as follows. The monopole quantum numbers $n_1,n_2$ and $m$ label and specify non-Abelian properties, for electric non-Abelian charges one can also label the $SU(2)_L$ representation by an integer (the dimension) and for $SU(3)_c$ two integers to specify the highest weight or equivalently how many quarks and anti-quarks (boxes and two-box column in young tableaux diagrams) it takes to build the colour representation. Out of these one can define a magnetic triality and duality as 
\begin{align}
    n_c^m&=-N \mod 3,\, & n_L^m&=m\mod 2,\,
\end{align}
but where any $n_c$ and $n_L$ values are allowed for the electric spectrum for all $G_p$, allowed magnetic triality and duality depends on $p$. These can be worked out from $n_6^m\,=\,q_m \mod 6\,=\, n_\mathcal{O}^m\, =\, 2N+3m$ with takes values on $6 \mathbb{Z}/p$. The correlation $n_6^m,n_\mathcal{O},q_m$ is in fact a feature of all $p$ in the magnetic spectrum which always lie in a stairway in the $(n_\mathcal{O}^m \,,\, q_m)$ plane and yield a less populated lattice when compared to the electric one, as Fig.~\ref{fig:MplStairway} shows.


\subsection{Dynamical monopoles}

Having established the spectrum of `t~Hooft lines admissible in all $G_p$ SM theories, we can now ask which of them can be realised by dynamical magnetic monopoles. To establish this we first assume that there exist non-singular finite-mass `t~Hooft--Polyakov monopole solutions~\cite{tHooft:1974kcl,Polyakov:1974ek} in the appropriate UV-embedding of the SM which leads to the $G_p$ theory in the IR. This, in fact, is always the case in the setting considered in this paper where $U(1)_Y$ is embedded into a larger non-Abelian group. For the $G_1$ model considered in section~\ref{sec:3.1} the symmetry breaking is $SU(2)_Y \to U(1)_Y$, enabled by the adjoint Higgs field of $SU(2)_Y$. This field content ensures the existence of the `t~Hooft--Polyakov monopole solution in the $SU(2)_Y$ sector. Similarly the required `t~Hooft--Polyakov monopole solutions exist in the Trinification, Pati--Salam and 
Grand Unified models~\cite{Dokos:1979vu,Gardner:1983uu,Preskill:1984gd} of sections~\ref{sec:3.2}-\ref{sec:3.4} which realise the remaining $G_{p=6,3,2}$ embeddings of the Standard Model.

 \begin{table}[t]
\begin{equation} \nonumber
\begin{array}{lc|c|c|c}
 \text{Monopoles} :&(n_1,n_2,m;q_m)\,\,\,& \text{magnetic flux matrix} &\,G_p  &\quad n_6^m\quad
\\ & & & & \\ \hline  & & & &\\
\quad {M_1}  &(1,1,1;1)& \begin{pmatrix} \frac{1}{3}& &\\ & \frac{1}{3}&\\ & & -\frac{2}{3}\,
\end{pmatrix} 
\oplus
\begin{pmatrix} \frac{1}{2} & \\ &-\frac{1}{2}
\end{pmatrix} 
\oplus \, 1\quad
& G_6\, & \quad 1 \quad
\\ & & & &
\\ \hline  & & & &\\
\quad {M_2}  &(-1,-1,0;2)& \begin{pmatrix} -\frac{1}{3}& &\\ &-\frac{1}{3}&\\ & & \frac{2}{3}\,
\end{pmatrix} 
\oplus
\begin{pmatrix} 0 &  \\  & 0
\end{pmatrix} 
\oplus \, 2 \quad
& G_3\,, \, G_6 \,& \quad 2 \quad
\\ & & & & \\ \hline  & & & &\\
\quad {M_3}  &(0,0,1;3)&\begin{pmatrix} 0& &\\ &0&\\ & & 0 \,
\end{pmatrix} 
\oplus
\begin{pmatrix} \frac{1}{2} & \\ &-\frac{1}{2}
\end{pmatrix} 
\oplus \, 3 \quad
& G_2\,, \, G_6  \, & \quad 3\quad
\\ & & & & \\ \hline  & & &  &\\
\quad {M_4}  &(1,1,0;4)& \begin{pmatrix} \frac{1}{3}& &\\ &\frac{1}{3}&\\ & & -\frac{2}{3}\,
\end{pmatrix} 
\oplus
\begin{pmatrix}  0 & \\ & 0
\end{pmatrix} 
\oplus \, 4 \quad
& G_3\,, \, G_6 \,& \quad 4 \quad
\\ & & & &\\ \hline  & & & & \\
\quad {M_5}  &(-1,-1,1;5)& \,\, \begin{pmatrix} -\frac{1}{3}& &\\ &-\frac{1}{3}&\\ & & \frac{2}{3}\,
\end{pmatrix} 
\oplus
\begin{pmatrix} \frac{1}{2} & \\ &-\frac{1}{2}
\end{pmatrix} 
\oplus \, 5 \,\quad
& G_6 \, & \quad 5 \quad
\\ & & & &\\ \hline  & & & &\\
\quad {M_6}  &(0,0,0;6)&\quad \begin{pmatrix} 0& &\\ &0&\\ & &0\,
\end{pmatrix} 
\oplus
\begin{pmatrix} 0 &  \\  & 0
\end{pmatrix} 
\oplus \, 6 \quad
& \,\, G_1\,,\,G_2\,,\,G_3\,, \, G_6\,& \quad 0 \quad
\\ & & & &\\ \hline  
\end{array}
\end{equation}
\caption{Standard Model monopoles and their magnetic fluxes for different choices of the SM gauge group $G_p$. The Abelian magnetic flux number $q_m$ is the topological charge and plays the role of the conserved monopole number for each monopole $M_{q_m}$, and we also display $n_6^m =q_m \mod 6$. }
\label{Tab:MG_p}
\end{table}

At distances greater than their non-Abelian monopole cores, these `t~Hooft--Polyakov solutions in unitary gauge reduce to the Dirac monopole configuration eq.~\eqref{eq:ADirG} with the magnetic flux matrices parameterised by eq.~\eqref{eq:FmgSM1}. 
Their Abelian magnetic flux number $q_m$ is protected by topology~\cite{Lubkin:1963zz}. 
$q_m$ is the winding number of the closed contour $S^1$ around the equator of the Dirac monopole
 around the $U(1)_Y$ group manifold, $\pi_1(U(1)_Y) \in \mathbb{Z}$.  The non-Abelian fluxes $n_i$ and $m$ on the other hand
 can be reduced by the monopole emitting massless non-Abelian vector bosons. Specifically, the monopole stability requirements against decaying into massless gluons and weak vector boson radiation
 amounts to the following constraints on their relative flux numbers~\cite{Brandt:1979kk,Coleman:1982cx},
 \begin{equation}
 \{\pm m\,, \pm n_1\,, \pm n_2\,, \pm(n_1-n_2)\} \,=\, 0\,, \pm 1.
 \label{eq:Mstab}
 \end{equation}
We can label the monopoles by their topological charge $q_m$, and for each value of $q_m$ select the values of their non-Abelian fluxes that satisfy eqs.~\eqref{eq:Nmq1}-\eqref{eq:Nmq6} as well as the stability constraints eq.~\eqref{eq:Mstab}. 
It then follows, as was first shown in Ref.~\cite{vanBeest:2023mbs},  that for each value of $q_m$ the non-Abelian fluxes are 
completely fixed\footnote{Up to a discrete gauge transformation permuting diagonal elements in the flux matrices eq.~\eqref{eq:FmgSM1}.}
by the monopole stability conditions. Thus for $1\le q_m \le 6$ we can have 6 types of monopoles~\cite{vanBeest:2023mbs,Khoze:2024hlb} that carry SM magnetic fluxes and are perturbatively stable in the massless SM i.e. before the EWSB takes place.
They are listed in Tab.~\ref{Tab:MG_p} which also shows in which of the $G_p$ models each of these monopoles may be present.
 
 \medskip
 
The question of whether or not all of the monopoles in Tab.~\ref{Tab:MG_p} are truly elementary goes beyond the SM group itself and depends on the structure of the UV theory. In the minimal $SU(5)_{GUT}$ embedding of $G_6$ the lightest monopole is $M_1$, but nevertheless 
all of the monopoles in~Tab.~\ref{Tab:MG_p}, except $M_5$, are known~\cite{Gardner:1983uu} to be stable against decay into combinations of monopoles with lower charges $q_m$. 
On the other hand, in the $SO(10)_{\rm GUT}$ settings the monopole $M_2$ can be much lighter than $M_1$, depending on the symmetry-breaking hierarchy structure~\cite{Preskill:1984gd}.  Also remarkable, as we can see from Tab.~\ref{Tab:MG_p}, the $M_1$ monopole is only allowed in a $G_6$ theory, while in a $G_3$ the monopole with minimal magnetic charge is $M_2$; for $G_2$ it is $M_3$, and in the $G_1$ settings, only $M_6$ is allowed.

\medskip
 
The monopoles $M_{q_m}$ in Tab.~\ref{Tab:MG_p} form a 1-dimensional tower labelled by $q_m =\mathbb{Z} $ which is a 
1-dimensional reduction of the 4-dimensional lattice of  `t~Hooft lines generated by non-dynamical magnetic probes in the $G_p$ Standard Model as othewise depicted in Fig.~\ref{fig:MplStairway}.

\subsection{Magnetic spectrum after EWSB}
\label{sec:M-after-EWSB}

 \begin{table}[t]
\begin{equation} \nonumber
\begin{array}{lc|c|c|c}
 \text{Monopoles} :&(n_1,n_2;m=q_m)\,\,\,& \text{magnetic flux matrix} &\,G_p  \, & \,(n_6^m,n_c^m)
\\ \text{after EWSB} & & & & \\ \hline  & & & & \\
\quad {M_1'=M_1}  &(1,1;1)& \begin{pmatrix} \frac{1}{3}& &\\ & \frac{1}{3}&\\ & & -\frac{2}{3}\,
\end{pmatrix} 
\oplus
\begin{pmatrix} \frac{1}{2} & \\ &-\frac{1}{2}
\end{pmatrix} 
\oplus \, 1\quad
& G_6\,& (1,1)
\\ & & & &
\\ \hline  & & & &\\
\quad {M_2'}  &(-1,-1;2)& \begin{pmatrix} -\frac{1}{3}& &\\ &-\frac{1}{3}&\\ & & \frac{2}{3}\,
\end{pmatrix} 
\oplus
\begin{pmatrix} 1 &  \\  & -1
\end{pmatrix} 
\oplus \, 2 \quad
& G_3\,, \, G_6 \, &(2,2)
\\ & & & & \\ \hline  & & & & \\
\quad {M_3'}  &(0,0;3)&\begin{pmatrix} 0& &\\ &0&\\ & & 0 \,
\end{pmatrix} 
\oplus
\begin{pmatrix} \frac{3}{2} & \\ &-\frac{3}{2}
\end{pmatrix} 
\oplus \, 3 \quad
& G_2\,, \, G_6  \, & (3,0)
\\ & & & & \\ \hline  & & & & \\
\quad {M_4'}  &(1,1;4)& \begin{pmatrix} \frac{1}{3}& &\\ &\frac{1}{3}&\\ & & -\frac{2}{3}\,
\end{pmatrix} 
\oplus
\begin{pmatrix}  2 & \\ & -2
\end{pmatrix} 
\oplus \, 4 \quad
& G_3\,, \, G_6 \, & (4,1)
\\ & & & &\\ \hline  & & & &\\
\quad {M_5'}  &(-1,-1;5)& \,\, \begin{pmatrix} -\frac{1}{3}& &\\ &-\frac{1}{3}&\\ & & \frac{2}{3}\,
\end{pmatrix} 
\oplus
\begin{pmatrix} \frac{5}{2} & \\ &-\frac{5}{2}
\end{pmatrix} 
\oplus \, 5 \,\quad
& G_6 \, & (5,2)
\\ & & & & \\ \hline  & & & &\\
\quad {M_6'}  &(0,0;6)&\quad \begin{pmatrix} 0& &\\ &0&\\ & &0\,
\end{pmatrix} 
\oplus
\begin{pmatrix} 3 &  \\  & -3
\end{pmatrix} 
\oplus \, 6 \quad
& \,\, G_1\,,\,G_2\,,\,G_3\,, \, G_6\, & (0,0)
\\ & & & &\\ \hline  
\end{array}
\end{equation}
\caption{Monopoles $M_{q_m}'$ in the Standard Model with different global $G_p$ structures after EWSB. We also display their $n_6^m =q_m\mod 6$ and $n_c^m=-N$ magnetic numbers.
}
\label{Tab:Mpcpr}
\end{table}

As was recently emphasised in~\cite{Khoze:2024hlb}, 
the onset of the electroweak symmetry breaking necessarily modifies the long-range fields of SM monopoles 
by screening their non-Abelian magnetic fluxes associated with massive $Z^0$ bosons. At distances 
exceeding the inverse $Z$-boson mass, the monopole field in the electroweak sector should be aligned with the photon (in our conventions with the generator  $Q_{\textrm{em}}=Q_Y+\tilde T_{3L}/2$
and vanish along the orthogonal $Z^0$ direction, $\tilde T_{3L}/2 -Q_Y$. 
This requires that we set $m=q_m$ in the monopole magnetic flux matrix eq.~\eqref{eq:FmgSM1} and rewrite the `t~Hooft lines 
spectrum eqs.~\eqref{eq:Nmq1}-\eqref{eq:Nmq6}
as follows
\begin{eqnarray}
 \label{eq:Nmq1H}
 p=1: \qquad &N=3\mathbb{Z},  \quad &q_m =6\mathbb{Z},\\
  p=2: \qquad &N=3\mathbb{Z},  \quad &q_m =3\mathbb{Z},\\ 
  \label{eq:Nmq3H}
  p=3: \qquad &N=\mathbb{Z},  \quad &q_m =2N +6\mathbb{Z},\\ 
  \label{eq:Nmq6H}
  p=6: \qquad &N=\mathbb{Z}, \quad &q_m =-N +3\mathbb{Z}.
\end{eqnarray}
The magnetic lattice is  now 3-dimensional, parameterised by 3 integers $(n_1,n_2, q_m=m)$.
Monopole stability conditions are also modified since the $SU(2)_L$-sector gauge fields are no longer massless 
\begin{equation}
 \{ \pm n_1\,, \pm n_2\,, \pm(n_1-n_2)\} \,=\, 0\,, \pm 1,
 \label{eq:MstabH}
 \end{equation}
and the monopoles with any $m=q_m$ are stable in the  $SU(2)_L$  sector post EWSB. 
The resulting stable dynamical magnetic monopole spectrum in the Standard Model for different choices of the global group $G_p$
is presented in~Tab.~\ref{Tab:Mpcpr} (generalising the $G_6$ construction in~\cite{Khoze:2024hlb} to general $p$) and their disposition in the $n_c^m\equiv -N \mod 3$ vs $q_m$ plane shown in Fig.~\ref{fig:MplstarEWSB}.
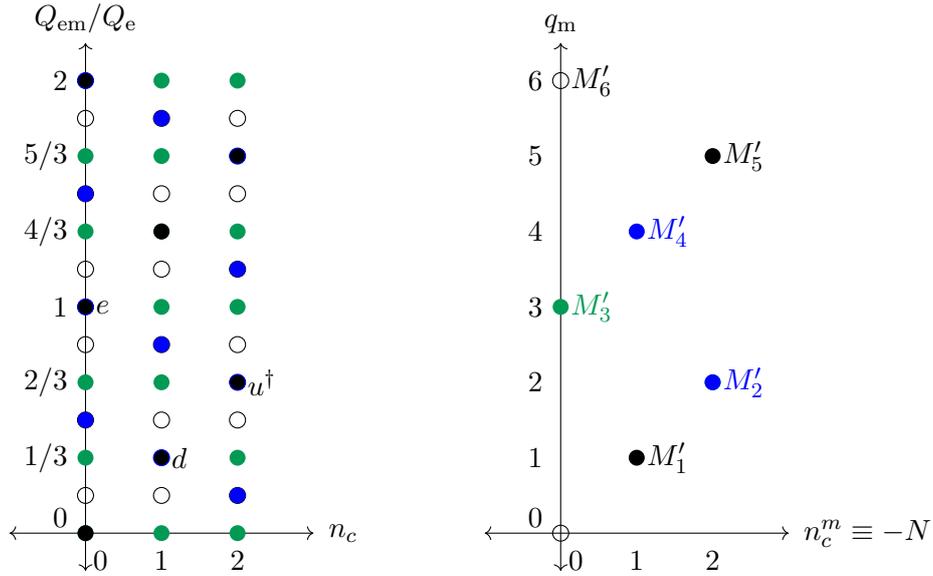
\begin{figure}
    \centering\begin{tikzpicture}[scale=1]
  \draw[<->] (-1,-3) -- (3,-3) node[right] {$\,n_{c}$};
  \draw[<->] (0,-3.5) -- (0,3.5) node[above] {$Q_{\textrm{em}}/Q_{\textrm{e}}$};
    \foreach \x in {1,...,2}
    \node[below] at (\x,-3.1) {\x};
  \node[below] at (0.2,-3.1) {0};  
\foreach \x in {0,...,2} {
    \foreach \y in {-3,...,3} {
      \fill [ForestGreen] (\x,\y) circle (3pt);
    }
  }
  \foreach \x in {0,...,2} {
    \foreach \y in {-2.5,...,2.5} {
      \draw (\x,\y) circle (3pt);
    }
  }
  \node[left] at (-0.1,-2.8) {$0$};
  \node[left] at (-0.1,1) {$4/3$};
  \node[left] at (-0.1,2) {$5/3$};
  \node[left] at (-0.1,3) {$2$};
  \node[left] at (-0.1,-1) {$2/3$};
  \node[left] at (-0.1,-2) {$1/3$};
  \node[left] at (-0.1,0) {$1$};
  \fill  (2,-1)  node[right] {$u^\dagger$} circle (3pt);
  \fill  (1,-2)  node[right] {$d$} circle (3pt);
  \fill  (0,-3) circle (3pt);
  \fill  (0,0)  node[right] {$e$} circle (3pt);
  \fill  (1,1) circle (3pt);
  \fill  (2,2) circle (3pt);
  \fill  (0,3) circle (3pt);
    \fill [blue] (2,1/2) circle (3pt);
  \fill [blue](0,3/2) circle (3pt);
    \draw [blue](2,4/2) circle (3pt);
  \fill [blue](1,5/2) circle (3pt);
  \draw [blue](0,6/2) circle (3pt);
  \draw [blue](0,0) circle (3pt);
  \fill [blue](1,-1/2) circle (3pt);
  \draw [blue](2,-2/2) circle (3pt);
  \fill [blue](0,-3/2) circle (3pt);
  \draw[blue] (1,-4/2) circle (3pt);
  \fill [blue](2,-5/2) circle (3pt);
\end{tikzpicture}\qquad\qquad
\begin{tikzpicture}[scale=1]
  \draw[<->] (-1,-3) -- (3,-3) node[right] {$\,n_c^m \equiv -N$};
  \draw[<->] (0,-3.5) -- (0,3.5) node[above] {$q_{\textrm{m}}$};
    \foreach \x in {1,...,2}
  \node[below] at (\x,-3.1) {\x};
  \node[below] at (0.2,-3.1) {0};  
  \node[left] at (-0.1,0) {$3$};
  \node[left] at (-0.1,1) {$4$};
  \node[left] at (-0.1,2) {$5$};
  \node[left] at (-0.1,3) {$6$};
  \node[left] at (-0.1,-1) {$2$};
  \node[left] at (-0.1,-2) {$1$};
  \node[left] at (-0.1,-2.8) {$0$};
  \fill [ForestGreen] (2,-1) circle (3pt);
  \draw (0,-3) circle (3pt);
  \fill [blue] (1,1)  node[right] {$M_4'$} circle (3pt);
  \fill  (2,2)  node[right] {$M_5'$} circle (3pt);
  \draw (0,3) node[right] {$M_6'$} circle (3pt);
  \fill [ForestGreen] (0,0)  node[right] {$M_3'$} circle (3pt);
  \fill [blue](2,-2/2)  node[right] {$M_2'$}circle (3pt);
  \fill  (1,-4/2)  node[right] {$M_1'$} circle (3pt);
\end{tikzpicture}
    \caption{Stairway structure after EWSB for electric charges on the LHS ({\it cf.}~Fig.~\ref{fig:kne0SU3}) and monopoles on the RHS. The node colour scheme on the RHS: Empty nodes are possible for all $p$ and are the only ones possible for $p=1$. For $p=2$ ($p=3$) in addition to empty, green (blue) nodes are possible while $p=6$ allows for all the above and black nodes.}
    \label{fig:MplstarEWSB}
\end{figure}
\subsection{Monopole solutions in the UV theory} 

All magnetic monopoles that we have discussed up to now have been described in terms of their magnetic flux matrices or, equivalently, their magnetic flux numbers
$n_1, n_2, m, q_m$ as the relevant variables in the IR theory. To realise these monopoles explicitly as non-singular finite-mass `t~Hooft--Polyakov 
solutions~\cite{tHooft:1974kcl,Polyakov:1974ek}, one necessarily requires the knowledge of the non-Abelian embedding of the SM theory in the UV. The the goal of this section is to connect the two.

\subsubsection{Monopoles in the UV embedding of $\mathbf{G_1}$}

The monopole configuration with the minimal magnetic charge that can be consistent with $G_1$ is the monopole $M_6$.
This conclusion follows immediately from inspecting the list of monopoles in Tab. ~\ref{Tab:MG_p} 
which shows that only the $M_6$ monopole given by the last entry in the table can exist in $G_1$
\begin{equation}
M_6\,: \quad \begin{pmatrix} 0& &\\ &0&\\ & &0\,\end{pmatrix} 
\oplus \begin{pmatrix} 0 &  \\  & 0\end{pmatrix} 
\oplus \, 6 \,\, \in \,\,
  \frac{SU(3)_c \times SU(2)_L\times U(1)_Y }{Z_1}.
\end{equation}
This applies at energy scales above (or distances below) the scale of the electroweak symmetry breaking, and if we want instead to look for monopoles modified by the onset of EWSB, the
$M_6$ monopole becomes $M_6'$ given by the last entry in Tab.~\ref{Tab:Mpcpr}.

The minimal non-Abelian embedding of the $G_1$ group in our approach is provided  by the $SU(3)_c \times SU(2)_L \times SU(2)_Y$ model constructed in section~\ref{sec:3.1}.
Here $SU(2)_Y$ is broken by the adjoint Higgs field to $U(1)_Y$ of the SM. This implies that the natural (and the only) candidate for $M_6$ is the 
`t~Hooft--Polyakov monopole of the $SU(2)_Y$ gauge theory with the adjoint Higgs.
To determine the topological charge of this $SU(2)_Y$ `t~Hooft--Polyakov configuration, we recall the expression for the hypercharge in terms of the $SU(2)_Y$ generator $\tilde T_Y$
({\it cf.}~eq.~\eqref{Q_em for 322}),
\begin{align}
    Q_Y = \frac{\tilde{T}_Y}{6} \,\,\, ; \qquad \tilde{T}_Y^{(F)} = \text{diag} (1, -1).
    \label{eq:322more}
\end{align}
With this normalisation understood, the magnetic flux matrix ($\vec g_m\cdot \vec T$) of the $M_6$ monopole in the non-Abelian UV theory is given by
\begin{equation}\label{eq:M6SU2Y}
M_6\,: \quad \begin{pmatrix} 0& &\\ &0&\\ & &0\,\end{pmatrix} 
\oplus \begin{pmatrix} 0 &  \\  & 0\end{pmatrix} 
\oplus  \begin{pmatrix} 1 &  \\  & -1\end{pmatrix}
\,\, \in \,\,
SU(3)_c \times SU(2)_L \times SU(2)_Y.
 \end{equation}
We conclude that $M_6$ is realised as the \emph{minimal} `t~Hooft--Polyakov solution in the $SU(2)_Y$ sector.\footnote{Here by \emph{minimal} we mean the 
topological soliton with the minimal value = 1 of its topological charge
$\pi_2(SU(2)_Y/U(1)_Y )= \mathbb{Z}.$}

\medskip

\underline{After} the electroweak symmetry breaking, the 
$SU(2)_L$-flux of this monopole is appropriately modified, following earlier discussion in sec.~\ref{sec:M-after-EWSB}. This results in the monopole configuration $M_6'$ (see Tab.~\ref{Tab:Mpcpr}) which for our non-Abelian embedding  reads
\begin{equation}
M_6'\,: \quad \begin{pmatrix} 0& &\\ &0&\\ & &0\,\end{pmatrix} 
\oplus \begin{pmatrix} 3 &  \\  & -3\end{pmatrix} 
\oplus  \begin{pmatrix} 1 &  \\  & -1\end{pmatrix}
\,\, \in \,\,
SU(3)_c \times SU(2)_L \times SU(2)_Y.
 \end{equation}

\medskip

\subsubsection{Monopoles in the UV embedding of $\mathbf{G_2}$}

\medskip
The monopoles in the $G_2$ theory are $M_3$ and $M_6$, both of which have trivial chromomagnetic fluxes.
The minimally charged monopole is $M_3$ 
\begin{equation}
\label{eq:m3IR}
M_3\,: \quad \begin{pmatrix} 0& &\\ &0&\\ & &0\,\end{pmatrix}_c \oplus \begin{pmatrix} \frac{1}{2} & \\ &-\frac{1}{2}
\end{pmatrix}_L 
\oplus \, 3 \quad \in \quad
  \frac{SU(3)_c \times SU(2)_L\times U(1)_Y }{Z_2}.
\end{equation}

As we have seen in eq.~\eqref{eq:SU2Yquot}, the \textbf{SU2Y} group embedding admits a quotient, in which case the low energy group is $G_2$ and as such it should have another type of monopole in the spectrum (in addition to $M_6$ in eq.~\eqref{eq:M6SU2Y}). Indeed given $e^{i\pi(\tilde T_{3L}+\tilde T_{Y})}R=R$ as a condition on the spectrum we have
\begin{align}
    M_3\,: \quad &\vec g_m \vec T\,=\,\frac12 \tilde T_{3L}\oplus \frac{1}{2}\tilde T_{3Y}=\frac12 \tilde T_{3L}\oplus 3 Q_Y\\
    =&
    \begin{pmatrix} 0& &\\ &0&\\ & &0\,\end{pmatrix}_c \oplus 
\begin{pmatrix} 1/2 &  \\  & -1/2\end{pmatrix}_L 
\oplus  \begin{pmatrix} 1/2 &  \\  & -1/2\end{pmatrix}_R
\,\, \in \,\,
\frac{SU(3)_c \times SU(2)_L \times SU(2)_Y}{Z_2}.
\end{align}

Another UV embedding of the $G_2$ Standard Model group was considered in section~\ref{sec:3.2} and is provided by the \textbf{Trinification model} 
$SU(3)_c \times SU(3)_L \times SU(3)_R$. 
The SM hypercharge $Q_Y$ is realised in this model as the linear combination of the $SU(3)_L$ and $SU(3)_R$ generators given by~{\it cf.} eq.~\eqref{eq:trinQY}
\begin{equation}
Q_Y\,=\, -\frac{1}{6} \tilde{\lambda}_{8L} - \frac{1}{6}(3 \tilde{\lambda}_{3R} + \tilde{\lambda}_{8R}).
\end{equation}
The $U(1)_Y$ flux factor 
of the $M_3$ monopole is
\begin{equation}
3\,Q_Y\,=\, \begin{pmatrix} -\frac{1}{2}& &\\ &-\frac{1}{2}&\\ & &1\,\end{pmatrix}_L
\oplus
\begin{pmatrix} -2& &\\ &1&\\ & &1\,\end{pmatrix}_R,
\end{equation}
where the RHS is written assuming the fundamental representation for the generators of both groups.
The $SU(3)_L$ magnetic flux of the $M_3$ monopole follows immediately from eq.~\eqref{eq:m3IR} 
\begin{equation}
\frac{1}{2} \tilde T_{3L}\,=\,
\begin{pmatrix} \frac{1}{2}& &\\ &-\frac{1}{2}&\\ & &0\,\end{pmatrix}_L.
\end{equation}
Combining the two equations above we find the magnetic flux matrix of $M_3$ embedded in the Trinification model,
\begin{equation}
\label{eq:m3UV}
M_3\,: \quad 
\begin{pmatrix} 0& &\\ &0&\\ & &0\,\end{pmatrix}_c \oplus \begin{pmatrix} 0& &\\ &-1&\\ & &1\,\end{pmatrix}_L
\oplus
\begin{pmatrix} -2& &\\ &1&\\ & &1\,\end{pmatrix}_R
 \, \in \,
SU(3)_c \times SU(3)_L \times SU(3)_R.
\end{equation}
This is interpreted as minimal (soliton charge 1) `t~Hooft--Polyakov $SU(2)$ monopole configuration in the unitary gauge embedded into the lower $2 \times 2$ corner of the $SU(3)_L$ as well as into 
the  $SU(3)_R$ matrix in accordance with eq.~\eqref{eq:m3UV}.

The monopole $M_6$ embedding is
\begin{equation}
\label{eq:m6UV}
M_6\,: \quad 
\begin{pmatrix} 0& &\\ &0&\\ & &0\,\end{pmatrix}_c \oplus
 \begin{pmatrix} -1& &\\ &-1&\\ & &2\,\end{pmatrix}_L
\oplus
\begin{pmatrix} -4& &\\ &2&\\ & &2\,\end{pmatrix}_R 
 \, \in \,
SU(3)_c\times SU(3)_L \times SU(3)_R.
\end{equation}
Both $M_3$ and $M_6$ monopoles have no $SU(3)_c$ field components. 

\medskip

\underline{After} the EWSB, the monopoles $M_3$ and $M_6$ given in
eqs.\eqref{eq:m3UV}, \eqref{eq:m6UV} are modified by changing their $SU(2)_L$ sector magnetic fluxes in accordance with the data in the Tab.~\ref{Tab:Mpcpr},
\begin{equation}
M_3'\,:\, \frac{3}{2}\,\tilde T_{3L}\,\, , \qquad
M_6'\,:\, 3\,\tilde T_{3L},
\end{equation}
which result in the following final expressions for the magnetic fluxes of these monopoles in the Trinification model
\begin{align}
\label{eq:m3UVprime}
&M_3'\,: 
 &\begin{pmatrix} 1& &\\ &-2&\\ & &1\,\end{pmatrix}_L
\oplus
\begin{pmatrix} -2& &\\ &1&\\ & &1\,\end{pmatrix}_R
 \quad \in \quad
SU(3)_L \times SU(3)_R,
\\
\label{eq:m6UVprime}
&M_6'\,: 
 & \begin{pmatrix} 2& &\\ &-4&\\ & &2\,\end{pmatrix}_L
\oplus
\begin{pmatrix} -4& &\\ &2&\\ & &2\,\end{pmatrix}_R 
 \quad \in \quad
SU(3)_L \times SU(3)_R,
\end{align}
where for brevity we dropped the trivial $SU(3)_c$ magnetic fluxes.

\medskip

\subsubsection{Monopoles in the UV embedding of $\mathbf{G_3}$}

The relevant monopoles are $M_2$, $M_4$ and $M_6$ in Tab.~\ref{Tab:MG_p}. They all have vanishing $SU(2)_L$ magnetic fluxes.
The minimally charged monopole is $M_2$, which in the IR is described by
\begin{equation}
\label{eq:m2IR}
M_2\,: \quad 
\begin{pmatrix} -\frac{1}{3}& &\\ &-\frac{1}{3}&\\ & & \frac{2}{3}\,
\end{pmatrix}_c 
\oplus
\begin{pmatrix} 0& \\ & 0\,\end{pmatrix}_L
\oplus \, 2
\quad \in \,
  \frac{SU(3)_c\times SU(2)_L\times U(1)_Y }{Z_3}.
\end{equation}
The minimal embedding of the $G_2$ theory is the 
\textbf{Pati--Salam} $SU(4)_c \times SU(2)_L\times SU(2)_R$ model of section~\ref{sec:3.3}.
The hypercharge $Q_Y$ is defined in this case by
\begin{equation}
Q_Y\,=\, \begin{pmatrix} \frac{1}{6}& & &\\ &\frac{1}{6}& &\\ & &\frac{1}{6}& \\ & & & -\frac{1}{2}\,\end{pmatrix}_c
\oplus
\begin{pmatrix} \frac{1}{2}& \\ & -\frac{1}{2}\,\end{pmatrix}_R,
\end{equation}
again having adopted the generators in the fundamental representations of $SU(4)_c$ and $SU(2)_R$.
With this normalisation, the minimal monopole eq.~\eqref{eq:m2IR} embedded in the Trinification model has the magnetic flux
\begin{equation}
\label{eq:m2UV}
M_2\,: \quad 
\begin{pmatrix} 0& & &\\ &0& &\\ & &1& \\ & & & -1\,\end{pmatrix}_c
\oplus
\begin{pmatrix} 0& \\ & 0\,\end{pmatrix}_L
\oplus
\begin{pmatrix} 1& \\ & -1\,\end{pmatrix}_R
\quad \in \quad
 SU(4)_c \times SU(2)_L \times SU(2)_R.
\end{equation}
This flux is sourced by the minimal  `t~Hooft--Polyakov $SU(2)$ monopole solution where the $SU(2)$ is imbedded into the lower right $2\times2$ corner of the $SU(4)_c \times SU(4)_c$ and coincides with the $SU(2)_R$.

\medskip

\underline{After} the EWSB the minimal monopole is 
$M_2'$ in the Pati--Salam model with the magnetic flux 
\begin{equation}
\label{eq:m2UVprime}
M_2'\,: \quad 
\begin{pmatrix} 0& & &\\ &0& &\\ & &1& \\ & & & -1\,\end{pmatrix}_c
\oplus
\begin{pmatrix} 1& \\ & -1\,\end{pmatrix}_L
\oplus
\begin{pmatrix} 1& \\ & -1\,\end{pmatrix}_R
\quad \in \,
 SU(4)_c \times SU(2)_L \times SU(2)_R.
\end{equation}

\medskip

\subsubsection{Monopoles in the UV embedding of $\mathbf{G_6}$}
The monopole of smallest flux in the IR theory is
\begin{align}
\label{eq:M1form}
    M_1\,:\quad \vec{g}_m \cdot \vec T \,=\, \frac13 \tilde\lambda_8\oplus\frac{1}{2}\tilde T_{3L}\oplus Q_Y.
\end{align}
This can be obtained from theories with a simple non-Abelian group, such as the $SU(5)$ GUT model, but also from the direct product of non-Abelian groups that have a quotient, the two examples we have seen are Trinification and Pati--Salam.

\medskip

For \textbf{Trinification} the new monopole solutions have to do with the quotient by the $Z_3$ group generated by $\zeta=e^{2\pi i(n_c+n_{3L}+n_{3R})/3}$. New monopole solutions can now start at the identity for $\phi=0$ and end up at $\zeta$ for $\phi=2\pi$ yet these solutions would have to conform to the $M_1$ form in eq.~\eqref{eq:M1form}. One can connect both as follows,
\begin{align}
M_1\,:\quad \vec{g}_m \cdot \vec T \,=\,&\begin{pmatrix} 1/3& &\\ &1/3&\\ & &-2/3\,\end{pmatrix}_c
\oplus
\begin{pmatrix} 1/3& &\\ &-2/3&\\ & &2/3\,\end{pmatrix}_L
\oplus
\begin{pmatrix} -2/3& &\\ &1/3&\\ & &1/3\,\end{pmatrix}_R\\
\nonumber\\
& \in \,\,
 \frac{SU(3)_c \times SU(3)_L \times SU(3)_R}{Z_3},
\nonumber 
\end{align}
and indeed  $\exp(2\pi i \,\vec{g}_m \cdot \vec T)=\zeta$.

\medskip

The case of \textbf{Pati--Salam} model with $Z_2$ quotient would also allow
an $M_1$ monopole  in addition to the $M_2$ monopole in eq.~\eqref{eq:m2UV}. Here too one can use the known IR form of $M_1$ to find its embedding
\begin{align}
M_1\,:\quad \vec{g}_m \cdot \vec T \,=\,&
    \begin{pmatrix} 1/2& & &\\ &1/2& &\\ & &-1/2& \\ & & & -1/2\,\end{pmatrix}_c
\oplus
\begin{pmatrix} 1/2& \\ & -1/2\,\end{pmatrix}_L
\oplus
\begin{pmatrix} 1/2& \\ & -1/2\,\end{pmatrix}_R
\\
\nonumber\\
& \in \,\,
 \frac{SU(4)_c \times SU(2)_L \times SU(2)_R}{Z_2},
\end{align}
a solution which indeed has $\exp(2\pi i \,\vec{g}_m \cdot \vec T)=\exp(\pi i(n_4+n_L+n_R))$. 

Finally, we consider the simple group case. This can always be realised by a Grand Unified Theory with a simple group, as discussed in section~\ref{sec:3.4}. In this case, $G_6 \in \mathbf{SU(5)_{{GUT}}}$ and it is well-known that all six monopoles, $M_{1,\ldots,6}$, are realised by the `t~Hooft--Polyakov monopoles of the $SU(5)_{\textrm{GUT}}$ theory~\cite{Dokos:1979vu,Gardner:1983uu}, see also~\cite{Khoze:2024hlb} for the recent discussion and the overview of the $SU(5)_{\textrm{GUT}}$ magnetic fluxes for all six monopoles.

\section{The monopole spectrum with compositeness degree}
\label{sec:mono-comp}

Our discussion of the monopole spectrum can be generalised to the case where matter fields with non-trivial degree of compositeness $k$ are included. The relevant solutions of the Dirac quantisation conditions constraints are generalised by modifying the monopole magnetic charge $q_m$ by an overall multiplicative factor $(1+pk)$. The derivation is otherwise straightforward, the difference starting with the quantization condition 
\begin{align}
    &6(1+p k)Q_Y+2n_c+3n_L\,=\, p\,\mathbb{Z},
\end{align}
and its impact on possible magnetic fluxes is characterised by
\begin{align}
    &\exp\left[2\pi i\left(-\frac{n_c}{3}\left(N+\frac{q_m}{(1+p k)}\right)+\frac{n_c}{3}\left(m-\frac{q_m}{(1+p k)}\right)+\frac p6\frac{q_m\mathbb{Z}}{1+pk}\right)\right]
    \,=\,1
\end{align}
This condition is satisfied when
\begin{align}\label{eq:MplSol2}
    q_m&=\frac{6}{p}(1+pk)\mathbb{Z}  \,\,, \qquad
    N=-\frac{q_m}{1+pk}+3\mathbb Z  \,\,,\qquad
    m=\frac{q_m}{1+pk}+2\mathbb Z    \,.     
\end{align}
noting $q_m/(1+pk)=p\mathbb{Z}/6$ we remark that the factors that enter $N$ and $m$ are hence only $p$-dependent and on the $(n_{\mathcal{O}}^m\,,\,q_m)$ plane give rise to the stairway structure of Fig.~\ref{fig:MplStairway}; the difference being the vertical $q_m$ axis where the minimum magnetic hypercharge flux has increased by $(1+p k)$. This is made more evident if one instead writes the magnetic hypercharge in terms of the non-Abelian magnetic charges as
\begin{equation}
 q_m\,=\, (1+pk)(n_\mathcal{O}^m +6w)\,, \quad \textrm{where} \quad
 n_\mathcal{O}^m \,:=\,  2N+3m \mod 6\,, \quad \textrm{and} \quad
 w=\mathbb{Z},
\end{equation}
where the choice of $p$ determines the values of $n_\mathcal{O}^m=q_m=n_6^m$ as in Fig.~\ref{fig:MplStairway}. 
The $p$-explicit generalisation of eqs.\eqref{eq:Nmq1}-\eqref{eq:Nmq6} reads
 \begin{eqnarray}
 \label{eq:Nmq1k}
 p=1: \qquad &N=3\mathbb{Z}, \quad m=2\mathbb{Z}, \quad &q_m =(1+k)\,6\mathbb{Z},\\
  p=2: \qquad &N=3\mathbb{Z}, \quad m=\mathbb{Z}, \quad &q_m =(1+2k)\,(3m +6\mathbb{Z}),\\ 
  \label{eq:Nmq3k}
  p=3: \qquad &N=\mathbb{Z}, \quad m=2\mathbb{Z}, \quad &q_m =(1+3k)\,(2N +6\mathbb{Z}),\\ 
  \label{eq:Nmq6k}
  p=6: \qquad &N=\mathbb{Z}, \quad m=\mathbb{Z}, \quad &q_m =(1+6k)\,(2N+3m +6\mathbb{Z}).
\end{eqnarray}

One pertinent remark for the determination of the true group and compositeness degree is that $|6/p+6k|$ is a surjective function of $p,k$.
\begin{align}
    \boxed{\textrm{For each $p,k$ the minimum monopole charge is unique and equal to}\,\, |6/p+6k|.}
\end{align}

The case of non-zero $k$ for the models in sec.~\ref{sec:ThePSpec} as shown in sec.~\ref{sec:PSpeck} amounts to placing $q_L$ in higher reps and consequently redefining the hypercharge connection to $\tilde T_Y$. The implications for monopoles follow simply from this redefinition, in particular, monopole solutions are given by the group choice and are the exact same solutions in terms of magnetic fluxes of non-Abelian and $\tilde T_Y$ generators. The charge of monopoles in the SM however, is given by fluxes on the Lie algebra of $\widetilde{G}_{SM}$ and in particular, $q_m$ for $Q_Y$, so in order to obtain $q_m$ one needs the relation between $Q_Y$ and $\tilde T_Y$. 

Let us consider, as an example, the SU2Y model $SU(3)_c\times SU(2)_L\times SU(2)_Y$ where we decide to place the LH quark in the representation higher than the doublet of $SU(2)_Y$. For the choice of $\mathbf{3}$ of $SU(2)_Y$,
 worked out in eqs.~\eqref{eq:k1SU2Y}-\eqref{eq:k1SU2Y3}, we have
\begin{align}
 q_L \in \left(\mathbf{3},\mathbf{2},\mathbf{3}\right)\,, \quad 12 Q_Y&=\tilde T_{Y}\,, \quad k=1,
 \end{align}
 one has that the monopole flux is that of the $M_6$ monopole in eq.~\eqref{eq:M6SU2Y}, but with the different value of $q_m$ 
 due to the change of normalisation of $Q_Y$ 
 \begin{align}
 M_6 \textrm{-like}\,:&\quad \vec g_m\vec T= \begin{pmatrix} 0& &\\ &0&\\ & &0\,\end{pmatrix} 
\oplus \begin{pmatrix} 0 &  \\  & 0\end{pmatrix} 
\oplus  \begin{pmatrix} 1 &  \\  & -1\end{pmatrix}=\tilde{T}_Y=12 Q_Y\equiv q_m Q_Y.
\end{align}
So one obtains $q_m=12$ for the minimal flux which itself must equal to $|6(1+pk)|$ and one re-derives the known $(p,k)=(1,1)$ values. To avoid repetition rather than deriving each $q_m$ value for the rest of cases studied in sec.~\ref{sec:PSpeck}, we collect results in Tab. ~\ref{tab:qmkneq0}.

\begin{table}[h]
    \centering
    \begin{tabular}{c|c|c|c}
         $SU(3)_c\times SU(2)\times SU(2)$&  $(p,k)$& non-Abelian mg flux & monopole charge\\ \hline
         $q_L=(\mathbf{3},\mathbf{2})_{2} \in (\mathbf{3},\mathbf{2},\mathbf{3})$&$(1,1)$ & $M_6$-like & $q_m=12$  \\ 
         $q_L=(\mathbf{3},\mathbf{2})_{3} \in (\mathbf{3},\mathbf{2},\mathbf{4})$&$(1,2)$ & $M_6$-like & $q_m=24$  \\ \hline
         $SU(3)_c\times SU(2)\times SU(2)/Z_2$&  &  &\\ \hline
         $q_L=(\mathbf{3},\mathbf{2})_{3}\in (\mathbf{3},\mathbf{2},\mathbf{4})$&$(2,1)$ & $M_3$-like & $q_m=9$  \\ 
         $q_L=(\mathbf{3},\mathbf{2})_{5}\in (\mathbf{3},\mathbf{2},\mathbf{6})$&$(2,2)$ & $M_3$-like & $q_m=15$  \\ \hline
         $SU(3)_c\times SU(3)\times SU(3)$&  &  &\\ \hline
         $q_L\in (\mathbf{3},\mathbf{3},\mathbf{1})$&$(2,-1)$ & $M_3$-like & $q_m= -3$  \\ 
         $q\in (\mathbf{3},\mathbf{8},\mathbf{1})$&$(2,(1,-2))$ & $M_3$-like & $q_m= 9,-9$ \\\hline 
         $SU(4)\times SU(2)\times SU(2)$&  &  &\\ \hline
         $q_L\in (\mathbf{6},\mathbf{2},\mathbf{1})$&$(3,-1)$ & $M_2$-like & $q_m= -4$  \\
         $q_L\in (\mathbf{15},\mathbf{2},\mathbf{1})$&$(3,1)$ & $M_2$-like & $q_m= 8$  \\         \hline
         $SU(3)_c\times SU(3)\times SU(3)/Z_3$&  &  &\\ \hline
         $q_L\in (\mathbf{3},\mathbf{3},\mathbf{3})$&$(6,-1)$ & $M_1$-like & $q_m= -5$  \\ 
         $q_L=(\mathbf{3},\mathbf{2})_{7} \in (\mathbf{3},\mathbf{8},\bar{\mathbf{3}})$&$(6,1)$ & $M_1$-like & $q_m= 7$ \\
\hline 
         $SU(4)\times SU(2)\times SU(2)/Z_2$&  &  &\\ \hline
         $q_L=(\mathbf{3},\mathbf{2})_{-5}\in (\mathbf{6},\mathbf{2},\mathbf{2})$&$(6,-1)$ & $M_1$-like & $q_m= -5$  \\
         $q_L=(\mathbf{3},\mathbf{2})_{7}\in (\mathbf{15},\mathbf{2},\mathbf{2})$&$(6,1)$ & $M_1$-like & $q_m= 7$  \\\hline 
         $SU(5)$&  &  &\\ \hline
         $q_L\in \mathbf{24}$&$(6,-1)$ & $M_1$-like & $q_m= -5$ 
    \end{tabular}
    \caption{Magnetic monopoles for different examples of $(p,k)$ embeddings of the Standard Model. These monopoles in the UV theory are charcterised by their non-Abelian fluxes, as indicated, and also carry topologically conserved magnetic charges $q_m$ of QED.}
    \label{tab:qmkneq0}
\end{table}

\medskip 

Finally, we now revisit the other case study from sec.~\ref{sec:PSpeck}, namely the $SU(7)$ model. In eq.~\eqref{eq:Dqpk} we outlined how to obtain the values of $p,k$ from $q$, which itself uniquely determines the monopole spectrum. For complementarity, we will show here how to arrive at the same result by instead solving for the monopole fluxes. We first consider a $p=6$ model, so that a minimal $n_6^m=1$ monopole exists. One can try and find it with the ansatz for the magnetic fluxes in the form
\begin{align}
  M_1 \,:\quad  \vec g_m\cdot\vec T&=\frac{\tilde\lambda_{8}}{3}+\frac{\tilde T_{3L}}{2}+\alpha Q_Y
  \nonumber\\
    &=\textrm{Diag} \left(\frac{1-\alpha}{3},\frac{1-\alpha}3,\frac{-2-\alpha}3,\frac{1-\alpha}2,-\frac{\alpha+1}2,q\alpha,-q\alpha \right).
\end{align}
One would then need $e^{2\pi i\,\vec q_m\cdot\vec T}=1$ and hence 
\begin{align}
    \alpha&=q^{-1}\mathbb{Z}\,, & (1-\alpha)&=3\mathbb{Z}\,, &(1-\alpha)&=2\mathbb{Z},
\end{align}
from the last two equations one needs a multiple of 6 plus one, i.e. $\alpha=1+6k$ with $k$ an integer, and the first equation gives  $q^{-1}=(1+6k)\mathbb{Z}$. This relation between $q$ and $k$ was derived in the $p=6$ case (as assumed at the outset).
For $p\neq 6$, no $M_1$ monopole exists, and one moves to $q^{-1}=(1+pk)\mathbb{Z}$ and solves for $p$. 

The flux for our $M_1$ monopole is therefore
\begin{align}
M_1 \,:\quad  \vec g_m\cdot\vec T&\,=\,\frac{\tilde\lambda_{8}}{3}+\frac{\tilde T_{3L}}{2}+(1+6k) Q_Y\,=\,\frac{Q_6(k,p=6)}{6}.
\end{align}

\section{Conclusions}\label{sec:Concl}
The Standard Model gauge group is not determined in full, the remaining unknown being its `periodicity' or the times the universal cover wraps around the true group. This fundamental question is inextricably connected to charge quantisation both for electric and magnetic particles while recent developments in theory allowed for the problem to be formulated in terms of (higher-form) symmetry. The characterisation of the SM group is given by two indexes $(p,k)$, the first giving the group $G_p$ and the second the compositeness degree that underpins the problem of charge quantisation in a $U(1)$ theory. 

The results in this paper can be split into an IR half consisting of characterising the SM and its unbroken groups with theory tools and an UV part with the study of embeddings of the SM group and its predictions for $(p,k)$. 

On the IR front we introduced the charge operator $Q_6$ and its eigenvalue $\mod 6$ dubbed hexality ($n_6$) as well as a magnetic equivalent dubbed $n_6^m$. The choice of $p$ in $G_p$ leads to an electric 1-form $Z_{6/p}^{el\,(1)}$ and magnetic 1-form $Z_{p}^{mag\,(1)}$ symmetries and hexality and $n_6^m$ characterise the transfomation under each group of electric and magnetic representations. In addition allowed values of $n_6$ and $n_6^m$, which are correlated, when plotted against the $U(1)$ charges present a `stairway' structure whose periodicity determines $p$. In the magnetic spectrum $n_6^m$ is useful as a finer grading that topology is insensitive to, one can write the SM group as in eq.~\eqref{GpKp} to study topology while the action of the group $K_p$ in the denominator helps understand the structure of the group as illustrated in Fig.~\ref{fig:Kp}. We have shown how all stable monopole solutions have a flux aligned along $Q_6$ as an element of the Lie algebra.  For the electric spectrum on top of the 1-form discrete symmetry mentioned, we characterise the emergent electric 1-form symmetry $Z_{|6/p+6k|}^{emg(1)}$ in eq.~\eqref{eq:emergdef}, with charge $\hat n$ which is an integer $\mod |6/p+6k|$ given by $\hat n=(6/p+6k)(Q_{\textrm{em}}+n_c/3 \mod 1)$. The $p,k$ indexes can be extended down in energies to the unbroken groups $SU(3)_c\times U(1)_{\textrm{em}}$ and in the deep IR  $U(1)_{\textrm{em}}$ and we have shown the mapping between them as well as highlighting that cases with $k=0$ at the SM group level do lead to $k'\neq 0$ in $SU(3)_c\times U(1)_{\textrm{em}}$.

On the UV front we have studied the minimal non-Abelian embeddings for each $p$ and checked against naive expectations. We found that while embedding e.g. weak-isospin and hypercharge in a simple group does lead to $p=2$, there is another theory with the same group locally but with quotient $Z_3$ which returns instead $p=6$. On the other direction a simple group as $SU(7)$ does not necessarily imply $p=6$. The minimal groups here studied are Trinification and its quotient, 
Pati--Salam and its quotient, $SU(5)$, and to fill the $p=1$ gap we have proposed a new $SU(3)_c\times SU(2)_L \times SU(2)_Y$ model dubbed SU2Y, their correspondence with $p$ is displayed in Fig.~\ref{fig:MoneyPlot}. The SU2Y model, which itself could be taken in its quotient or universal cover from, has a wealth of fractionally charged particles but the new fermions cannot be heavier than $4\pi v$ as opposed to other models and while we did not study its phenomenology in detail, we provided its action, showed it is compatible with the SM mass spectrum, is gauge anomaly-free and renormalisable. 
\begin{figure}[h!]
\centering
\begin{center}
\begin{tikzpicture}[->,>=Stealth, node distance=1cm and 1.5cm]
\node (E6) at (0,4) {\( E_6 \)};
\node (Spin10) at (0,3) {\( \text{Spin}(10) \)};
\draw (E6) -- (Spin10);
\node (su4su2) at (-3,2) {\( \frac{\text{SU(4)} \times \text{SU(2)}^2}{\mathbb{Z}_2} \)};
\node (su5) at (0,2) {\( \text{SU(5)} \)};
\node (su3cube) at (3,2) {\( \frac{\text{SU(3)}^3}{\mathbb{Z}_3} \)};
\node (G6) at (0,1) {\( G_6 \)};
\draw (Spin10) -- (su4su2);
\draw (Spin10) -- (su5);
\draw (E6) -- (su3cube);
\draw (su4su2) -- (G6);
\draw (su5) -- (G6);
\draw (su3cube) -- (G6);
\end{tikzpicture}
\end{center}
\smallskip
\begin{center}
\begin{tikzpicture} [->,>=Stealth, node distance=1cm and 1.5cm]
\node (su4su22) at (-3,0) {\( \text{SU(4)} \times \text{SU(2)}^2 \)};
\node (G3) at (-3,-1) {\( G_3 \)};
\draw (su4su22) -- (G3);
\node (su3cube2) at (0,0) {\( \text{SU(3)}^3 \)};
\node (su3su22modZ2) at (3,0) {\( \frac{\text{SU(3)} \times \text{SU(2)}^2}{\mathbb{Z}_2} \)};
\node (G2) at (1.5,-1) {\( G_2 \)};
\draw (su3cube2) -- (G2);
\draw (su3su22modZ2) -- (G2);
\node (su3su22) at (6,0) {\( \text{SU(3)} \times \text{SU(2)}^2 \)};
\node (G1) at (6,-1) {\( G_1 \)};
\draw (su3su22) -- (G1);
\end{tikzpicture}
\end{center}
\caption{Atlas of UV embeddings of $G_p \,=\,\frac{SU(3)_c \times SU(2)_L\times U(1)_Y }{Z_p}$ comprising four distinct maps
for different values of $p\,=\,1,2,3,6$.}
 \label{fig:MoneyPlot}
\end{figure}
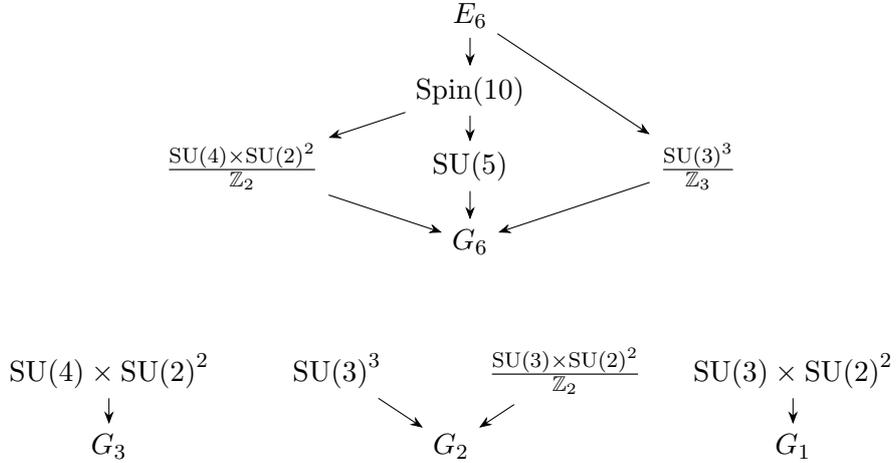 

The group $G_6$  can potentially descend from the $E_6$ Grand Unified Theory, as shown in the top map of Fig.~\ref{fig:MoneyPlot}, along the paths 
involving either the Pati--Salam model with the ${Z}_2$ quotient, the minimal $SU(5)$ GUT model, or the Trinification model with quotient
 ${Z}_3$. These three options can also be viewed as independent stand-alone UV embeddings of $G_6$ without the need of being further unified into a higher GUT theory. Note that the UV embeddings of the four groups $G_6$, $G_3$, $G_2$ and $G_1$ in Fig.~\ref{fig:MoneyPlot} fall into four distinct classes.

In addition, for all embedding groups mentioned, we showed how non-trivial compositeness $k\neq 0$ follows from placing $q_L$ in different representations of the embedding group and how the values obtained for possible hypercharge follow the pattern set by the choice of $p$. As an extra group which can lead to potentially any $(p,k)$ we revisited $SU(7)$ as discussed in the early GUT literature. The different embeddings offer a UV completion of Dirac monopoles in the form of `t~Hooft--Polyakov solutions characterised by their fluxes in the Lie algrebra of the encompassing group. This work identified the different monopoles (which are in fact unavoidable given we are embedding $U(1)_Y$ into non-Abelian group) in the UV and connected them to the known low energy spectrum in terms of $p,k$ again both for $k=0$ and the first few non-zero $k$'s. 
\medskip

As we already noted in the Introduction, one way to tell the Standard Model groups apart would be the discovery of particles with novel electric and magnetic charge quantisation conditions.
LHC constraints on the existence of particles with new fractional electric charges, such as~\cite{ATLAS:2019gqq,CMS:2024eyx}, 
and their collider signatures have been recently discussed 
in Ref.~\cite{Koren:2024xof} constraining their mass range for most cases to be above few hundred GeV, or above 2~TeV, depending on 
whether or not the new particle is charged under $SU(3)_c$. The model SU2Y here presented does indeed provide, given its EW scale tied spectrum, a specific target for LHC and future collider searches and will be ruled out (or in) just by present and future collider data.

 Since the lightest fractionally charged particle $\chi$ is necessarily stable, there are strong cosmological constraints~\cite{Langacker:2011db} on its relic abundance, 
with the strongest bound~\cite{Dunsky:2018mqs}, $\Omega_\chi /\Omega_{\textrm{DM}} \lesssim 10^{-10} - 10^{-16}$, arising from the observational constraints on 
$\chi$ from cosmic ray experiments and monopole searches.
There are also strong constraints from searches for fractionally charged
particles on Earth and in meteorites reviewed in~\cite{Perl:2009zz}.
All these constraints can be accommodated  by cosmological inflation at the scale below $M_\chi$
to dilute $\Omega_\chi$.
The resulting bound on the reheating temperature
$T_{\textrm{BBN}} \,<\, T_{\textrm{r}} \,< \, M_\chi/r$ where $M_\chi$ is the mass of the stable fractionally charged particle and $r \simeq 65$~\cite{Dunsky:2018mqs}.
The lower bound on $T_{\textrm{r}}$ is to ensure presence of
Standard Model plasma at the BBN temperatures~0.7~MeV, which is the only known requirement on limiting how low the inflation scale 
can be~\cite{Hannestad:2004px,deSalas:2015glj}.

There is also an ongoing experimental program to find magnetic monopoles either directly
  in high-energy collisions at the LHC (for TeV-scale monopoles)~\cite{MoEDAL:2019ort}, or to identify ionisation and Cherenkov radiation for both: high- and low-scale monopoles, as they traverse detectors~\cite{IceCube:2014xnp, IceCube:2021eye, NOvA:2020qpg}. Monopoles can also catalyse fast proton decay or synthesis via Callan--Rubakov processes
which provides distinctive observational signatures~\cite{Super-Kamiokande:2012tld, Candela:2025gwp}.   

It is the combination of data from multiple experimental fronts and the theory characterisation of the different possibilities (which this paper aims to contribute to) that has the potential to answer the fundamental question, which is the true gauge group of Nature?

\acknowledgments
We would like to thank Mohamed Anber and Martin Bauer for useful discussions.



\bibliographystyle{JHEP}
\bibliography{biblio.bib}

\providecommand{\href}[2]{#2}\begingroup\raggedright\begin{thebibliography}{10}

\bibitem{Anber:2021upc}
M.M.~Anber and E.~Poppitz, \emph{{Nonperturbative effects in the Standard Model with gauged 1-form symmetry}}, \href{https://doi.org/10.1007/JHEP12(2021)055}{\emph{JHEP} {\bfseries 12} (2021) 055} [\href{https://arxiv.org/abs/2110.02981}{{\ttfamily 2110.02981}}].

\bibitem{Langacker:2011db}
P.~Langacker and G.~Steigman, \emph{{Requiem for an FCHAMP? Fractionally CHArged, Massive Particle}}, \href{https://doi.org/10.1103/PhysRevD.84.065040}{\emph{Phys. Rev. D} {\bfseries 84} (2011) 065040} [\href{https://arxiv.org/abs/1107.3131}{{\ttfamily 1107.3131}}].

\bibitem{Hannestad:2004px}
S.~Hannestad, \emph{{What is the lowest possible reheating temperature?}}, \href{https://doi.org/10.1103/PhysRevD.70.043506}{\emph{Phys. Rev. D} {\bfseries 70} (2004) 043506} [\href{https://arxiv.org/abs/astro-ph/0403291}{{\ttfamily astro-ph/0403291}}].

\bibitem{deSalas:2015glj}
P.F.~de~Salas, M.~Lattanzi, G.~Mangano, G.~Miele, S.~Pastor and O.~Pisanti, \emph{{Bounds on very low reheating scenarios after Planck}}, \href{https://doi.org/10.1103/PhysRevD.92.123534}{\emph{Phys. Rev. D} {\bfseries 92} (2015) 123534} [\href{https://arxiv.org/abs/1511.00672}{{\ttfamily 1511.00672}}].

\bibitem{Hucks:1990nw}
J.~Hucks, \emph{{Global structure of the standard model, anomalies, and charge quantization}}, \href{https://doi.org/10.1103/PhysRevD.43.2709}{\emph{Phys. Rev. D} {\bfseries 43} (1991) 2709}.

\bibitem{Tong:2017oea}
D.~Tong, \emph{{Line Operators in the Standard Model}}, \href{https://doi.org/10.1007/JHEP07(2017)104}{\emph{JHEP} {\bfseries 07} (2017) 104} [\href{https://arxiv.org/abs/1705.01853}{{\ttfamily 1705.01853}}].

\bibitem{Wilson:1974sk}
K.G.~Wilson, \emph{{Confinement of Quarks}}, \href{https://doi.org/10.1103/PhysRevD.10.2445}{\emph{Phys. Rev. D} {\bfseries 10} (1974) 2445}.

\bibitem{tHooft:1977nqb}
G.~'t~Hooft, \emph{{On the Phase Transition Towards Permanent Quark Confinement}}, \href{https://doi.org/10.1016/0550-3213(78)90153-0}{\emph{Nucl. Phys. B} {\bfseries 138} (1978) 1}.

\bibitem{Aharony:2013hda}
O.~Aharony, N.~Seiberg and Y.~Tachikawa, \emph{{Reading between the lines of four-dimensional gauge theories}}, \href{https://doi.org/10.1007/JHEP08(2013)115}{\emph{JHEP} {\bfseries 08} (2013) 115} [\href{https://arxiv.org/abs/1305.0318}{{\ttfamily 1305.0318}}].

\bibitem{Gaiotto:2014kfa}
D.~Gaiotto, A.~Kapustin, N.~Seiberg and B.~Willett, \emph{{Generalized Global Symmetries}}, \href{https://doi.org/10.1007/JHEP02(2015)172}{\emph{JHEP} {\bfseries 02} (2015) 172} [\href{https://arxiv.org/abs/1412.5148}{{\ttfamily 1412.5148}}].

\bibitem{Dirac:1931kp}
P.A.M.~Dirac, \emph{{Quantised singularities in the electromagnetic field}}, \href{https://doi.org/10.1098/rspa.1931.0130}{\emph{Proc. Roy. Soc. Lond. A} {\bfseries 133} (1931) 60}.

\bibitem{Wu:1975es}
T.T.~Wu and C.N.~Yang, \emph{{Concept of Nonintegrable Phase Factors and Global Formulation of Gauge Fields}}, \href{https://doi.org/10.1103/PhysRevD.12.3845}{\emph{Phys. Rev. D} {\bfseries 12} (1975) 3845}.

\bibitem{Brennan:2023mmt}
T.D.~Brennan and S.~Hong, \emph{{Introduction to Generalized Global Symmetries in QFT and Particle Physics}},  \href{https://arxiv.org/abs/2306.00912}{{\ttfamily 2306.00912}}.

\bibitem{Cordova:2022ruw}
C.~Cordova, T.T.~Dumitrescu, K.~Intriligator and S.-H.~Shao, \emph{{Snowmass White Paper: Generalized Symmetries in Quantum Field Theory and Beyond}},  in \emph{{Snowmass 2021}}, 5, 2022 [\href{https://arxiv.org/abs/2205.09545}{{\ttfamily 2205.09545}}].

\bibitem{McGreevy:2022oyu}
J.~McGreevy, \emph{{Generalized Symmetries in Condensed Matter}}, \href{https://doi.org/10.1146/annurev-conmatphys-040721-021029}{\emph{Ann. Rev. Condensed Matter Phys.} {\bfseries 14} (2023) 57} [\href{https://arxiv.org/abs/2204.03045}{{\ttfamily 2204.03045}}].

\bibitem{Bhardwaj:2023kri}
L.~Bhardwaj, L.E.~Bottini, L.~Fraser-Taliente, L.~Gladden, D.S.W.~Gould, A.~Platschorre et~al., \emph{{Lectures on generalized symmetries}}, \href{https://doi.org/10.1016/j.physrep.2023.11.002}{\emph{Phys. Rept.} {\bfseries 1051} (2024) 1} [\href{https://arxiv.org/abs/2307.07547}{{\ttfamily 2307.07547}}].

\bibitem{Alonso:2024pmq}
R.~Alonso, D.~Dimakou and M.~West, \emph{{Fractional-charge hadrons and leptons to tell the Standard Model group apart}}, \href{https://doi.org/10.1016/j.physletb.2025.139354}{\emph{Phys. Lett. B} {\bfseries 863} (2025) 139354} [\href{https://arxiv.org/abs/2404.03438}{{\ttfamily 2404.03438}}].

\bibitem{Koren:2024xof}
S.~Koren and A.~Martin, \emph{{Fractionally charged particles at the energy frontier: The SM gauge group and one-form global symmetry}}, \href{https://doi.org/10.21468/SciPostPhys.18.1.004}{\emph{SciPost Phys.} {\bfseries 18} (2025) 004} [\href{https://arxiv.org/abs/2406.17850}{{\ttfamily 2406.17850}}].

\bibitem{Cordova:2023}
S.H.~C.~Córdova and L.~Wang, \emph{{Axion Domain Walls, Small Instantons, and Non‑Invertible Symmetry Breaking}},  \href{https://arxiv.org/abs/2309.05636}{{\ttfamily 2309.05636}}.

\bibitem{tHooft:1974kcl}
G.~'t~Hooft, \emph{{Magnetic Monopoles in Unified Gauge Theories}}, \href{https://doi.org/10.1016/0550-3213(74)90486-6}{\emph{Nucl. Phys. B} {\bfseries 79} (1974) 276}.

\bibitem{Polyakov:1974ek}
A.M.~Polyakov, \emph{{Particle Spectrum in Quantum Field Theory}}, {\emph{JETP Lett.} {\bfseries 20} (1974) 194}.

\bibitem{Gross:1972pv}
D.J.~Gross and R.~Jackiw, \emph{{Effect of anomalies on quasirenormalizable theories}}, \href{https://doi.org/10.1103/PhysRevD.6.477}{\emph{Phys. Rev. D} {\bfseries 6} (1972) 477}.

\bibitem{Weinberg:1996kr}
S.~Weinberg, \emph{{The quantum theory of fields. Vol. 2: Modern applications}}, Cambridge University Press (8, 2013), \href{https://doi.org/10.1017/CBO9781139644174}{10.1017/CBO9781139644174}.

\bibitem{Witten:1982fp}
E.~Witten, \emph{{An SU(2) Anomaly}}, \href{https://doi.org/10.1016/0370-2693(82)90728-6}{\emph{Phys. Lett. B} {\bfseries 117} (1982) 324}.

\bibitem{Wang:2018qoy}
J.~Wang, X.-G.~Wen and E.~Witten, \emph{{A New SU(2) Anomaly}}, \href{https://doi.org/10.1063/1.5082852}{\emph{J. Math. Phys.} {\bfseries 60} (2019) 052301} [\href{https://arxiv.org/abs/1810.00844}{{\ttfamily 1810.00844}}].

\bibitem{Davighi:2019rcd}
J.~Davighi, B.~Gripaios and N.~Lohitsiri, \emph{{Global anomalies in the Standard Model(s) and Beyond}}, \href{https://doi.org/10.1007/JHEP07(2020)232}{\emph{JHEP} {\bfseries 07} (2020) 232} [\href{https://arxiv.org/abs/1910.11277}{{\ttfamily 1910.11277}}].

\bibitem{ATLAS:2019gqq}
{\scshape ATLAS} collaboration, \emph{{Search for heavy charged long-lived particles in the ATLAS detector in 36.1 fb$^{-1}$ of proton-proton collision data at $\sqrt{s} = 13$ TeV}}, \href{https://doi.org/10.1103/PhysRevD.99.092007}{\emph{Phys. Rev. D} {\bfseries 99} (2019) 092007} [\href{https://arxiv.org/abs/1902.01636}{{\ttfamily 1902.01636}}].

\bibitem{CMS:2024eyx}
{\scshape CMS} collaboration, \emph{{Search for Fractionally Charged Particles in Proton-Proton Collisions at s=13{\,}{\,}TeV}}, \href{https://doi.org/10.1103/PhysRevLett.134.131802}{\emph{Phys. Rev. Lett.} {\bfseries 134} (2025) 131802} [\href{https://arxiv.org/abs/2402.09932}{{\ttfamily 2402.09932}}].

\bibitem{tHooft:1976rip}
G.~'t~Hooft, \emph{{Symmetry Breaking Through Bell-Jackiw Anomalies}}, \href{https://doi.org/10.1103/PhysRevLett.37.8}{\emph{Phys. Rev. Lett.} {\bfseries 37} (1976) 8}.

\bibitem{Belavin:1975fg}
A.A.~Belavin, A.M.~Polyakov, A.S.~Schwartz and Y.S.~Tyupkin, \emph{{Pseudoparticle Solutions of the Yang-Mills Equations}}, \href{https://doi.org/10.1016/0370-2693(75)90163-X}{\emph{Phys. Lett. B} {\bfseries 59} (1975) 85}.

\bibitem{tHooft:1976snw}
G.~'t~Hooft, \emph{{Computation of the Quantum Effects Due to a Four-Dimensional Pseudoparticle}}, \href{https://doi.org/10.1103/PhysRevD.14.3432}{\emph{Phys. Rev. D} {\bfseries 14} (1976) 3432}.

\bibitem{Kuzmin:1985mm}
V.A.~Kuzmin, V.A.~Rubakov and M.E.~Shaposhnikov, \emph{{On the Anomalous Electroweak Baryon Number Nonconservation in the Early Universe}}, \href{https://doi.org/10.1016/0370-2693(85)91028-7}{\emph{Phys. Lett. B} {\bfseries 155} (1985) 36}.

\bibitem{Glashow:1984gc}
A.~De~R\'ujula, H.~Georgi and S.L.~Glashow, \emph{{Trinification of All Elementary Particle Forces}},  in \emph{{Fifth Workshop on Grand Unification}}, 7, 1984.

\bibitem{Babu:1985gi}
K.S.~Babu, X.-G.~He and S.~Pakvasa, \emph{{Neutrino Masses and Proton Decay Modes in $SU(3)\times SU(3)\times SU(3)$ Trinification}}, \href{https://doi.org/10.1103/PhysRevD.33.763}{\emph{Phys. Rev. D} {\bfseries 33} (1986) 763}.

\bibitem{Hetzel:2015bla}
J.~Hetzel and B.~Stech, \emph{{Low-energy phenomenology of trinification: an effective left-right-symmetric model}}, \href{https://doi.org/10.1103/PhysRevD.91.055026}{\emph{Phys. Rev. D} {\bfseries 91} (2015) 055026} [\href{https://arxiv.org/abs/1502.00919}{{\ttfamily 1502.00919}}].

\bibitem{Georgi:1974sy}
H.~Georgi and S.L.~Glashow, \emph{{Unity of All Elementary Particle Forces}}, \href{https://doi.org/10.1103/PhysRevLett.32.438}{\emph{Phys. Rev. Lett.} {\bfseries 32} (1974) 438}.

\bibitem{Li:1981un}
L.F.~Li and F.~Wilczek, \emph{{Price of Fractionally Charged Particles in a Unified Model}}, \href{https://doi.org/10.1016/0370-2693(81)91148-5}{\emph{Phys. Lett. B} {\bfseries 107} (1981) 64}.

\bibitem{Farhi:1979zx}
E.~Farhi and L.~Susskind, \emph{{A Technicolored G.U.T.}}, \href{https://doi.org/10.1103/PhysRevD.20.3404}{\emph{Phys. Rev. D} {\bfseries 20} (1979) 3404}.

\bibitem{Umemura:1981bw}
I.~Umemura and K.~Yamamoto, \emph{{Symmetry Breaking Patterns in an SU(7) Grand Unified Model}}, \href{https://doi.org/10.1143/PTP.66.1430}{\emph{Prog. Theor. Phys.} {\bfseries 66} (1981) 1430}.

\bibitem{Corrigan:1976wk}
E.~Corrigan and D.I.~Olive, \emph{{Color and Magnetic Monopoles}}, \href{https://doi.org/10.1016/0550-3213(76)90525-3}{\emph{Nucl. Phys. B} {\bfseries 110} (1976) 237}.

\bibitem{vanBeest:2023mbs}
M.~van Beest, P.~Boyle~Smith, D.~Delmastro, R.~Mouland and D.~Tong, \emph{{Fermion-Monopole Scattering in the Standard Model}},  \href{https://arxiv.org/abs/2312.17746}{{\ttfamily 2312.17746}}.

\bibitem{Dokos:1979vu}
C.P.~Dokos and T.N.~Tomaras, \emph{{Monopoles and Dyons in the SU(5) Model}}, \href{https://doi.org/10.1103/PhysRevD.21.2940}{\emph{Phys. Rev. D} {\bfseries 21} (1980) 2940}.

\bibitem{Gardner:1983uu}
C.L.~Gardner and J.A.~Harvey, \emph{{Stable Grand Unified Monopoles With Multiple Dirac Charge}}, \href{https://doi.org/10.1103/PhysRevLett.52.879}{\emph{Phys. Rev. Lett.} {\bfseries 52} (1984) 879}.

\bibitem{Preskill:1984gd}
J.~Preskill, \emph{{Magnetic Monopoles}}, \href{https://doi.org/10.1146/annurev.ns.34.120184.002333}{\emph{Ann. Rev. Nucl. Part. Sci.} {\bfseries 34} (1984) 461}.

\bibitem{Lubkin:1963zz}
E.~Lubkin, \emph{{Geometric definition of gauge invariance}}, \href{https://doi.org/10.1016/0003-4916(63)90194-5}{\emph{Annals Phys.} {\bfseries 23} (1963) 233}.

\bibitem{Brandt:1979kk}
R.A.~Brandt and F.~Neri, \emph{{Stability Analysis for Singular Nonabelian Magnetic Monopoles}}, \href{https://doi.org/10.1016/0550-3213(79)90211-6}{\emph{Nucl. Phys. B} {\bfseries 161} (1979) 253}.

\bibitem{Coleman:1982cx}
S.R.~Coleman, \emph{{The Magnetic Monopole Fifty Years Later}},  in \emph{{Les Houches Summer School of Theoretical Physics: Laser-Plasma Interactions}}, pp.~461--552, 6, 1982.

\bibitem{Khoze:2024hlb}
V.V.~Khoze, \emph{{Monopoles and fermions in the Standard Model}}, \href{https://doi.org/10.1007/JHEP09(2024)146}{\emph{JHEP} {\bfseries 09} (2024) 146} [\href{https://arxiv.org/abs/2405.18689}{{\ttfamily 2405.18689}}].

\bibitem{Dunsky:2018mqs}
D.~Dunsky, L.J.~Hall and K.~Harigaya, \emph{{CHAMP Cosmic Rays}}, \href{https://doi.org/10.1088/1475-7516/2019/07/015}{\emph{JCAP} {\bfseries 07} (2019) 015} [\href{https://arxiv.org/abs/1812.11116}{{\ttfamily 1812.11116}}].

\bibitem{Perl:2009zz}
M.L.~Perl, E.R.~Lee and D.~Loomba, \emph{{Searches for fractionally charged particles}}, \href{https://doi.org/10.1146/annurev-nucl-121908-122035}{\emph{Ann. Rev. Nucl. Part. Sci.} {\bfseries 59} (2009) 47}.

\bibitem{MoEDAL:2019ort}
{\scshape MoEDAL} collaboration, \emph{{Magnetic Monopole Search with the Full MoEDAL Trapping Detector in 13 TeV pp Collisions Interpreted in Photon-Fusion and Drell-Yan Production}}, \href{https://doi.org/10.1103/PhysRevLett.123.021802}{\emph{Phys. Rev. Lett.} {\bfseries 123} (2019) 021802} [\href{https://arxiv.org/abs/1903.08491}{{\ttfamily 1903.08491}}].

\bibitem{IceCube:2014xnp}
{\scshape IceCube} collaboration, \emph{{Search for non-relativistic Magnetic Monopoles with IceCube}}, \href{https://doi.org/10.1140/epjc/s10052-014-2938-8}{\emph{Eur. Phys. J. C} {\bfseries 74} (2014) 2938} [\href{https://arxiv.org/abs/1402.3460}{{\ttfamily 1402.3460}}].

\bibitem{IceCube:2021eye}
{\scshape IceCube} collaboration, \emph{{Search for Relativistic Magnetic Monopoles with Eight Years of IceCube Data}}, \href{https://doi.org/10.1103/PhysRevLett.128.051101}{\emph{Phys. Rev. Lett.} {\bfseries 128} (2022) 051101} [\href{https://arxiv.org/abs/2109.13719}{{\ttfamily 2109.13719}}].

\bibitem{NOvA:2020qpg}
{\scshape NOvA} collaboration, \emph{{Search for slow magnetic monopoles with the NOvA detector on the surface}}, \href{https://doi.org/10.1103/PhysRevD.103.012007}{\emph{Phys. Rev. D} {\bfseries 103} (2021) 012007} [\href{https://arxiv.org/abs/2009.04867}{{\ttfamily 2009.04867}}].

\bibitem{Super-Kamiokande:2012tld}
{\scshape Super-Kamiokande} collaboration, \emph{{Search for GUT monopoles at Super\textendash{}Kamiokande}}, \href{https://doi.org/10.1016/j.astropartphys.2012.05.008}{\emph{Astropart. Phys.} {\bfseries 36} (2012) 131} [\href{https://arxiv.org/abs/1203.0940}{{\ttfamily 1203.0940}}].

\bibitem{Candela:2025gwp}
P.M.~Candela, V.V.~Khoze and J.~Turner, \emph{{Monopoles at Future Neutrino Detectors}},  \href{https://arxiv.org/abs/2504.14918}{{\ttfamily 2504.14918}}.

\end{thebibliography}\endgroup






\end{document}